\documentclass[12pt,letterpaper,english,aps,preprint,double-spaced,nofootinbib,superscriptaddress]{revtex4}
\usepackage[FIGTOPCAP]{subfigure}
\usepackage{color}
\usepackage{graphicx}
\usepackage{amssymb}  
\usepackage{babel}
\makeatletter


\begin{document}  
\global\long\def\oalphas{O(\alpha_{s})}
\global\long\def\met{\not\!\! E_{T}}

\preprint{ANL-HEP-PR-09-94}
\preprint{EFI-09-22}
\preprint{MSUHEP-091025} 

\title{Next-to-leading order QCD corrections to s-channel single top quark
production and decay at the LHC}

\author{Sarah Heim}
\email{heimsara@msu.edu}
\affiliation{Department of Physics $\&$ Astronomy, Michigan State University,
East Lansing, MI, 48824, USA}

\author{Qing-Hong Cao}
\email{caoq@hep.anl.gov}
\affiliation{High Energy Physics Division, Argonne National Laboratory, Argonne, IL, 60439, USA}
\affiliation{Enrico Fermi Institute, University of Chicago, Chicago, IL, 60637, USA}

\author{Reinhard Schwienhorst}
\email{schwier@pa.msu.edu}
\affiliation{Department of Physics $\&$ Astronomy, Michigan State University,
East Lansing, MI, 48824, USA}

\author{C.-P. Yuan}
\email{yuan@pa.msu.edu}
\affiliation{Department of Physics $\&$ Astronomy, Michigan State University,
East Lansing, MI, 48824, USA}

\begin{abstract}
We present a study of electroweak production of top and antitop quarks
in the s-channel mode at the LHC, including next-to-leading order
(NLO) quantum chromodynamics (QCD) corrections to the production and
decay of the single (anti)top quark. The spin is preserved in
production and decay by using the narrow width approximation for the
(anti)top quark. We show the effect of different $O(\alpha_{s})$
contributions on the inclusive cross section and various kinematic distributions 
at parton level after imposing relevant
kinematic cuts to select s-channel single top quark events. We also
discuss several possibilities for measuring the top quark polarization. 
\end{abstract}
\maketitle

\section{Introduction}

Recent results at the Tevatron $p\bar{p}$ collider have confirmed
the existence of electroweak single top quark production%
~\cite{Abazov:2006gd,Abazov:2008kt,Aaltonen:2008sy,Aaltonen:2009jj,Abazov:2009ii}.
While the Tevatron can be considered a $t\bar{t}$ factory, measurements
of single top quark properties are statistics limited. This limitation
does not exist at the LHC. The main mode of top quark production at
the LHC is still strong interaction top quark pair production, but
the number of produced single top quark events will be large enough
for precision measurements.

Single top quark events are of considerable importance for probing
the Standard Model of particle physics (SM). 
As the top quark decays via the weak interaction before it can hadronize,
it is possible to measure its polarization. In single top quark events,
the top quark is coupled to the bottom quark with an amplitude proportional
to the Cabibbo-Kobayashi-Maskawa (CKM) matrix element $V_{tb}$, so
that a value for $V_{tb}$ can be obtained by measuring the single
top quark production cross section.

\begin{figure*}
\includegraphics[scale=0.7]{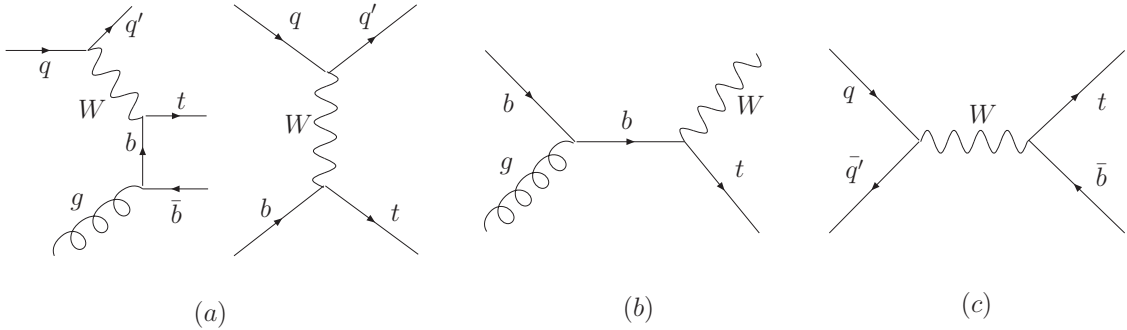} 
\caption{Representative Feynman diagrams of the three single top quark production modes: t-channel
(a), associated production (b) and s-channel (c). 
\label{fig:feynman}}
\end{figure*}

Electroweak single top quark production at the LHC occurs in three
different modes (cf. Fig.~\ref{fig:feynman}). The process with the largest cross section is the
t-channel exchange of a virtual $W$ boson ($bq\rightarrow tq'$ and
$b\bar{q}'\rightarrow t\bar{q}$), also referred to as $W$-gluon
fusion, followed by associated production of a top quark and a $W$
boson ($bg\rightarrow tW^{-}$) and the s-channel decay of a virtual
$W$ boson ($q\bar{q}'\rightarrow W^{*}\rightarrow t\bar{b}$).

The top quark is the heaviest known elementary particle, which makes
it an excellent candidate for new physics searches. The cross sections
of the three single top quark production modes are sensitive to different
kinds of new physics~\cite{Hikasa:1998wx,Han:1998tp,Tait:2000sh,Liu:2004bb,
Cakir:2005rf,Guasch:2006hf,Carlson:1994bg,Tait:1996dv,Espriu:2001vj,
Cao:2006wk,Cao:2007ea,AguilarSaavedra:2008gt,Abazov:2006aj,Abazov:2008vj,Aaltonen:2009qu,
Abazov:2007ev,Aaltonen:2008qr,Abazov:2008sz,Abazov:2009ky,Abazov:2008rn}.
The single top quark s-channel cross section
is especially sensitive to additional bosons, which makes it a very important channel in spite
of its comparably small cross section. The t-channel provides good
means to search for flavor changing neutral currents (FCNC)
and the associated production cross section changes with a modified
$Wtb$ or $Wqq'$ coupling.
Both t-channel and $Wt$ associated production are sensitive to the
bottom quark parton distribution function (PDF). A detailed knowledge
of the properties of single top quark production is necessary, as
it is an important background for several Higgs boson production modes.
The $Wt$ associated production serves as background to Higgs boson searches
in the decay channel $H\rightarrow WW$~\cite{Moretti:1997ng} and
also to processes involving a charged Higgs boson like $bg\rightarrow t H^{\pm}$
and $H^{\pm}\rightarrow\tau\nu$~\cite{Odagiri:1999fs}.

In order to extract the single top quark signal from the large QCD
and $W+jets$ backgrounds, but also in cases in which single top quark
production is a background itself, accurate theoretical predictions
including higher order QCD corrections are needed. The NLO QCD
corrections to the single top quark production have been carried out in
Refs.~ \cite{Smith:1996ij,Stelzer:1995mi,Bordes:1994ki,Zhu:2001hw,Harris:2002md,
Sullivan:2004ie,Cao:2008af,Frixione:2008yi,Campbell:2009ss,Campbell:2009gj}.
Furthermore, the complete NLO QCD calculations including both single top
quark production and decay have been carried out for the Tevatron in
several studies~ \cite{Campbell:2004ch,Cao:2004ky,Cao:2004ap,Cao:2005pq,Frixione:2005vw,Alioli:2009je}.
To a lesser extend, single top quark production is also affected
by electroweak corrections and possibly virtual
supersymmetric effects~\cite{Li:1996ir,Zhang:2006cx,Beccaria:2008av}.
In this paper we use the full NLO QCD calculations for single top quark
production at the LHC in order to present a detailed phenomenological
analysis, focusing on signal cross sections and kinematical distributions
at parton level after imposing simple kinematic cuts.

In contrast to the Tevatron, the LHC is a $pp$ collider and we have
to analyze top and antitop quark production separately. In this paper,
we present distributions for top quark production alone where the
top-antitop quark differences are small and contrast them with the
results for the antitop quark where they are not.

In Sec.~\ref{sec:InclXS}, we first present the inclusive cross section
for s-channel single top quark production and discuss its dependence
on the center of mass energy of the collider ($E_{c.m.}$), 
the top quark mass ($m_t$) and 
renormalization and factorization
scales. We also evaluate PDF uncertainties. In Sec.~\ref{sec:Acceptance},
we examine the effect of various kinematic cuts. Kinematical distributions
of final state objects and spin correlations are discussed in Sec.~\ref{sec:EventDistr}.
Our conclusion follows in Sec.~\ref{sec:Conclusions}.

\section{Cross Section (Inclusive Rate)\label{sec:InclXS}}

In this section, we show the inclusive production rates for s-channel
single top quark production and discuss their dependence on $E_{c.m.}$, $m_t$ 
and factorization
and renormalization scales. We present numerical
results for s-channel single top quark events considering the leptonic
decay of the $W$ boson from the top quark decay at the LHC (a $pp$
collider). Unless otherwise specified, we use the NLO parton distribution
function (PDF) set CTEQ6.6M~\cite{Pumplin:2002vw,Tung:2006tb,Lai:2007dq,Pumplin:2007wg,Nadolsky:2008zw},
defined in the $\overline{MS}$ scheme, and the NLO (2-loop) running
coupling $\alpha_{s}$ with $\Lambda_{\overline{MS}}$ provided by
the PDFs. For the CTEQ6.6M PDFs, $\Lambda_{\overline{MS}}^{(4)}=0.326$ GeV
 for four active quark flavors. 
For the numerical evaluation, we choose the following set of SM input parameters:
$G_{\mu}=1.16637\times10^{-5}~\mathrm{GeV}^{-2}$, $M_{W}=80.413$ GeV, $M_{Z}=91.187$ GeV,
and  $\alpha_s(M_Z)=0.1186$.  The square of the weak gauge coupling is 
$g^{2}=4\sqrt{2}M_{W}^{2}G_{\mu}$.

In this study we focus on the electron leptonic decay of the $W$
boson from the top quark only, but for muon leptons the analysis
procedure would be analogue. Including the $\oalphas$ corrections
to $W\rightarrow\bar{q}q'$, the branching ratio for the decay of
the $W$ boson into leptons is $Br(W\rightarrow l^{+}\nu)=0.108$%
~\cite{Cao:2004yy}. If not otherwise specified, the top quark mass is chosen
to be $m_{t}=175~\rm{GeV}$~\cite{Collaboration:2009eq}, the center of mass energy of the collisions $E_{c.m.} = $ 14 TeV and we will choose
the renormalization scale ($\mu_{R}$) as well as the factorization
scale ($\mu_{F}$) to be equal to $m_{t}$. In the current section we present inclusive cross sections, which include all $W$ boson decay modes.

In order to calculate NLO QCD differential cross sections we adopt
the one-cutoff phase space slicing (PSS) method~\cite{Giele:1991vf,Giele:1993dj,Keller:1998tf}
with a cutoff parameter $s_{min}=5~\rm{GeV}^2$.  

\subsection{Inclusive cross section}
As in our previous studies \cite{Cao:2004ap,Cao:2005pq}, we divide
the higher-order QCD corrections into three separate gauge invariant
sets: corrections to the initial particles (INIT), corrections to
the final state (FINAL), and corrections to the top quark decay (SDEC).
The explicit diagrams and definitions for the different corrections
can be found in Ref.~\cite{Cao:2004ky}. For $E_{c.m.}$ = 14 TeV and $m_t$ = 175 GeV,
Table~\ref{tab:inclusive}
shows the inclusive cross sections for top and antitop quark production
as well as the individual $\oalphas$ contributions. The effects
of the finite widths of top quark and $W$ boson have been
included. The total NLO s-channel single top quark production cross section agrees with Ref.~\cite{Smith:1996ij},
but updated values for the electroweak parameters are used. 
\begin{table*}
\begin{centering}
\begin{ruledtabular}
\begin{tabular}{lcccc}
& \multicolumn{2}{c}{Top} & \multicolumn{2}{c}{Antitop}\tabularnewline
 & Cross section  & Fraction of & Cross section  & Fraction of\tabularnewline
 & {[}pb{]} & NLO ($\%$) & {[}pb{]} & NLO ($\%$)\tabularnewline
\hline
Born level & 4.42  & 72.55 & 2.70  & 72.07\tabularnewline
\hline 
INIT 	   & 1.18  & 19.38 & 0.73  & 19.48\tabularnewline
FINAL      & 0.80  & 13.16 & 0.51  & 13.57\tabularnewline
SDEC       & -0.31 & -5.09 & -0.19 & -5.11\tabularnewline
\hline 
$\oalphas$ sum & 1.67 & 27.44 & 1.04 & 27.93\tabularnewline
\hline 
NLO        & 6.09  & 100   & 3.74  & 100\tabularnewline
\end{tabular}
\end{ruledtabular}
\par\end{centering}

\caption{Inclusive single top quark production cross sections for different
subprocesses, for top quark production (left) and antitop quark production
(right). $E_{c.m.} =$ 14 TeV and $m_t = 175~\rm{GeV}$. 
\label{tab:inclusive}}
\end{table*}

As can be seen in Table~\ref{tab:inclusive}, the LO cross section
for antitop quark production ($\bar{u}d\rightarrow\bar{t}b$) is 39\%
smaller than the cross section for top quark production ($u\bar{d}\rightarrow t\bar{b}$).
This is due to the difference in parton densities of the colliding protons. While in both cases
the antiquark is from the quark sea of one of the incoming protons,
the probability that it collides with an up quark from the other proton
is higher than the probability for a collision with a down quark.

The $\oalphas$ corrections increase the cross section by 38\%
for both top and antitop quark production. The largest contribution
to the $\oalphas$ corrections comes from the initial state,
due to the enhancement of collinear physics and the large phase space
for additional parton radiation. As expected, and as also observed
at the Tevatron, the correction to the top quark decay is small \cite{Cao:2004ap,Cao:2005pq}.
\begin{table*}
\begin{centering}
\begin{ruledtabular}
\begin{tabular}{lccc}
 		 &  		 & Top   & Antitop \tabularnewline
$m_{t}$ {[}GeV{]}  & $E_{\rm c.m.}$ {[}TeV{]}  & Cross section {[}pb{]}  & Cross section {[}pb{]}\tabularnewline
\hline 
 	      & 14   & 6.09  & 3.74 \tabularnewline
175    & 10    & 3.96  & 2.28 \tabularnewline

 		  & 7   & 2.45  & 1.30 \tabularnewline
\hline
 	      & 14    & 6.43  & 3.96 \tabularnewline
172.5 & 10    & 4.19  & 2.42 \tabularnewline
		  & 7   & 2.58  & 1.38 \tabularnewline
\hline 
 	      & 14    & 6.76  & 4.18 \tabularnewline 
170    & 10    & 4.42  & 2.56 \tabularnewline
 		  & 7   & 2.73  & 1.47 \tabularnewline
\end{tabular}
\end{ruledtabular}
\par\end{centering}

\caption{Inclusive single top quark production cross sections for top quark
production (left) and antitop quark production (right) at $E_{c.m.} =$
 14, 10 and 7 TeV and three different $m_t$. 
\label{tab:cross}}
\end{table*}
\begin{figure*}
\subfigure[]{\includegraphics[scale=0.3]{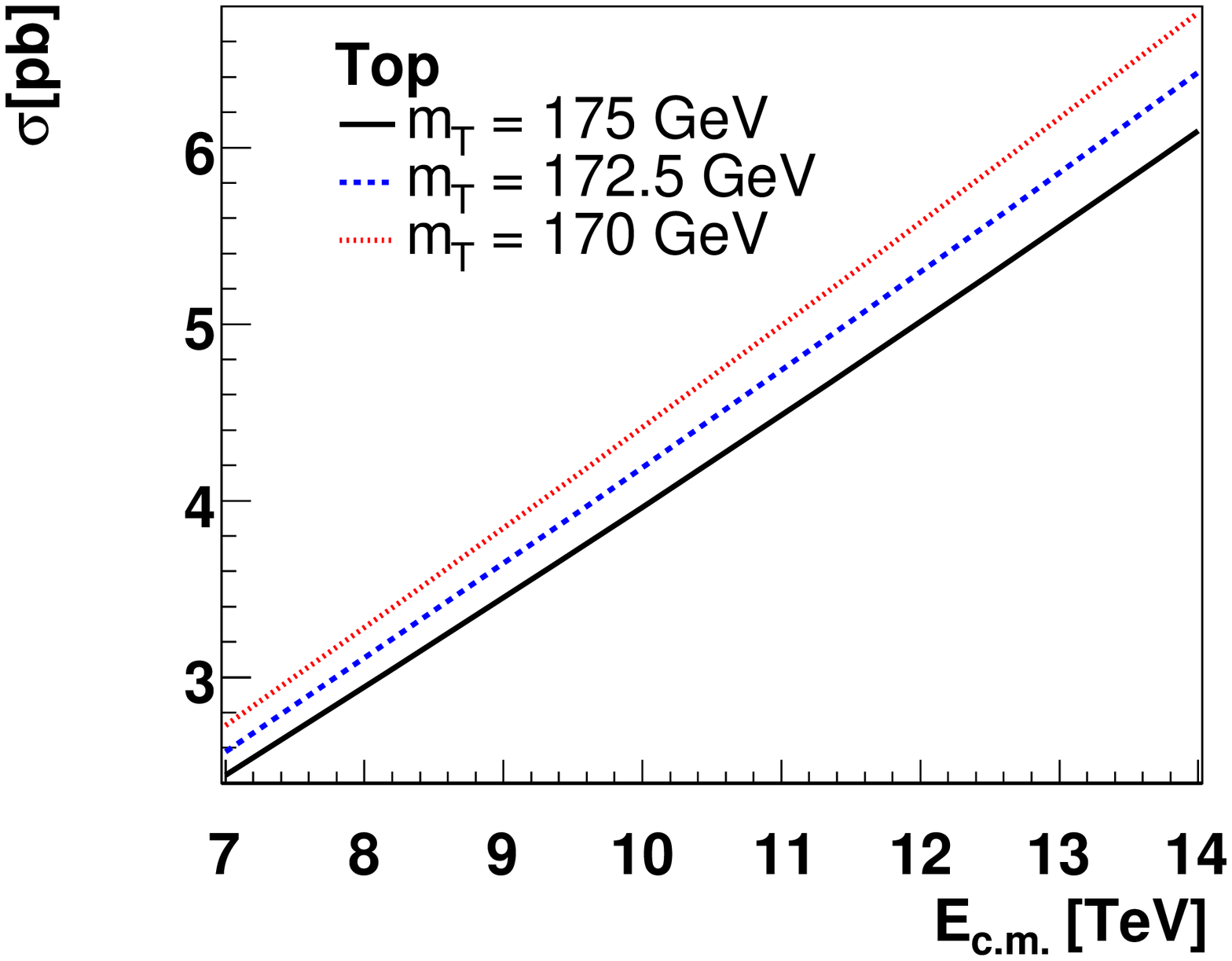}}\subfigure[]{\includegraphics[scale=0.3]{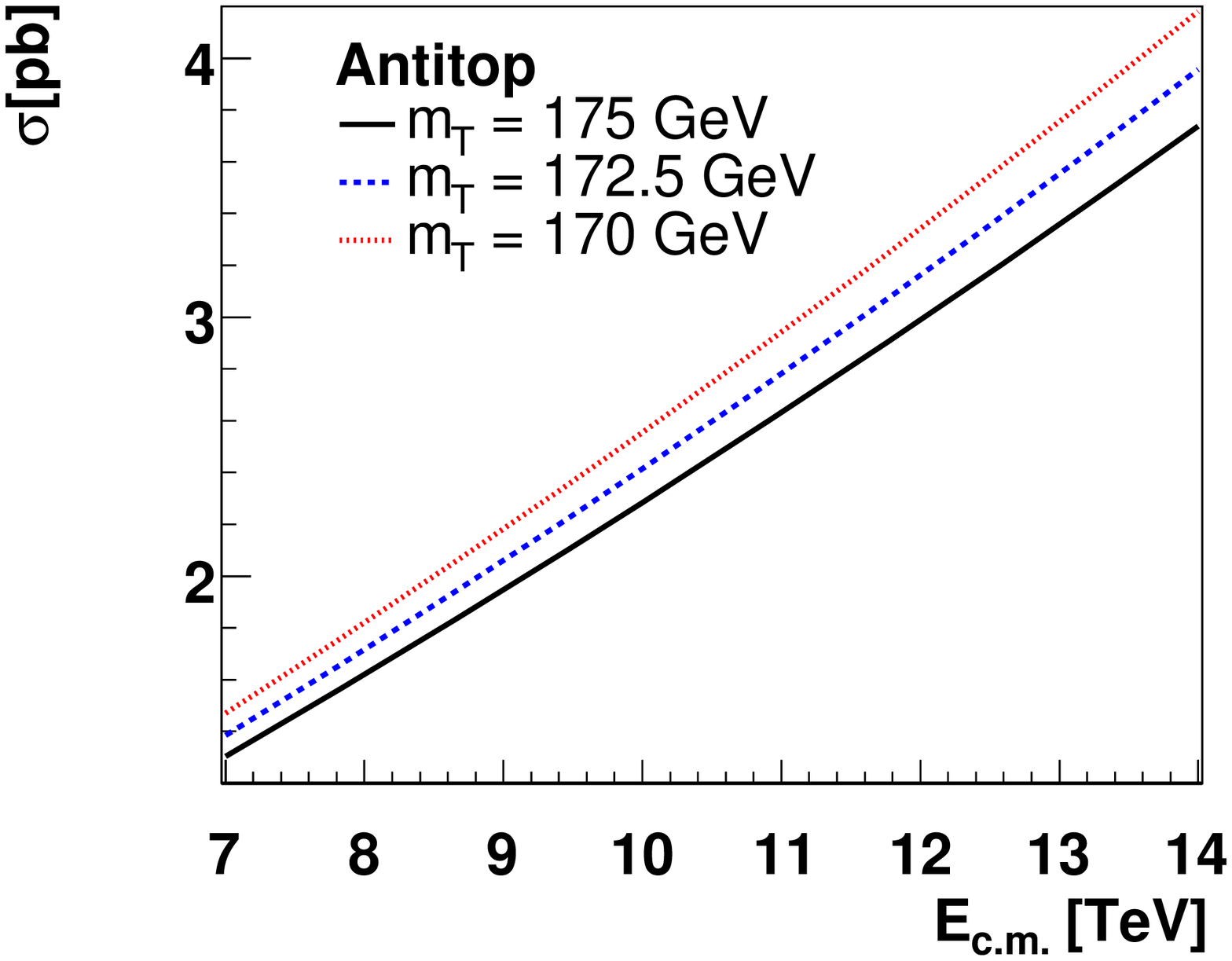}}\caption{$E_{c.m.}$ dependence of the single top (a) and antitop (b) quark production cross section for $m_{T} = $ 175, 172.5, 170 GeV. \label{fig:cmcrosssection3masses}}
\end{figure*} 

The obtained inclusive cross sections at the LHC for s-channel
single top quark events (considering the $W$ boson decay branching
ratio and using $m_{t}=175~\rm{GeV}$, $172.5~\rm{GeV}$ and $170~\rm{GeV}$) are shown
in Table~\ref{tab:cross} and Fig.~\ref{fig:cmcrosssection3masses} for top and antitop quark production at different $E_{c.m.}$.
For the LHC injection energy, the cross sections become very small, for $m_t =175~\rm{GeV}$ and $E_{c.m} = 900~\rm{GeV}$, the inclusive cross section for top (antitop) quark production is 0.018 (0.005) pb.

\begin{figure*}
\subfigure[]{\includegraphics[scale=0.28]{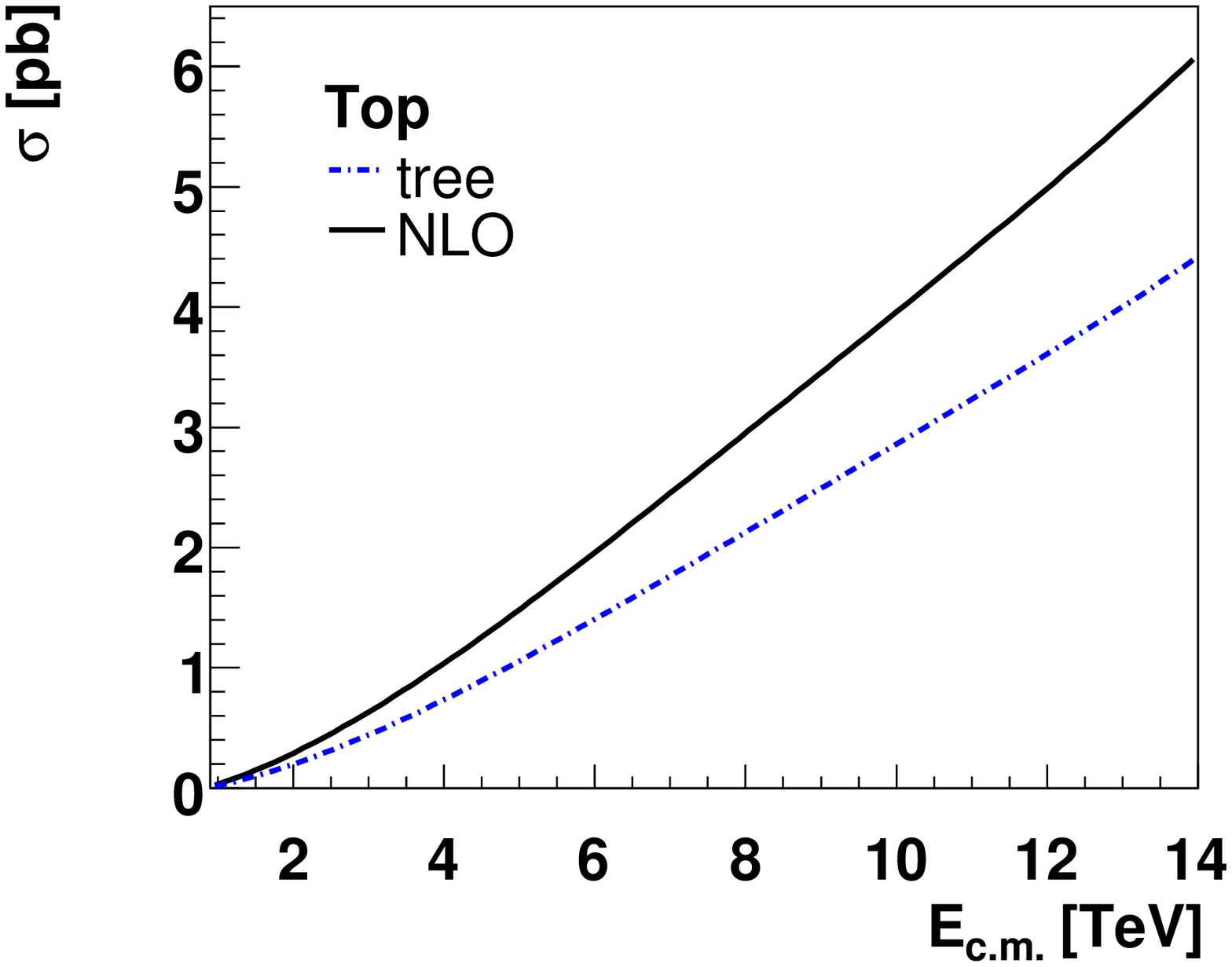}}\subfigure[]{\includegraphics[scale=0.28]{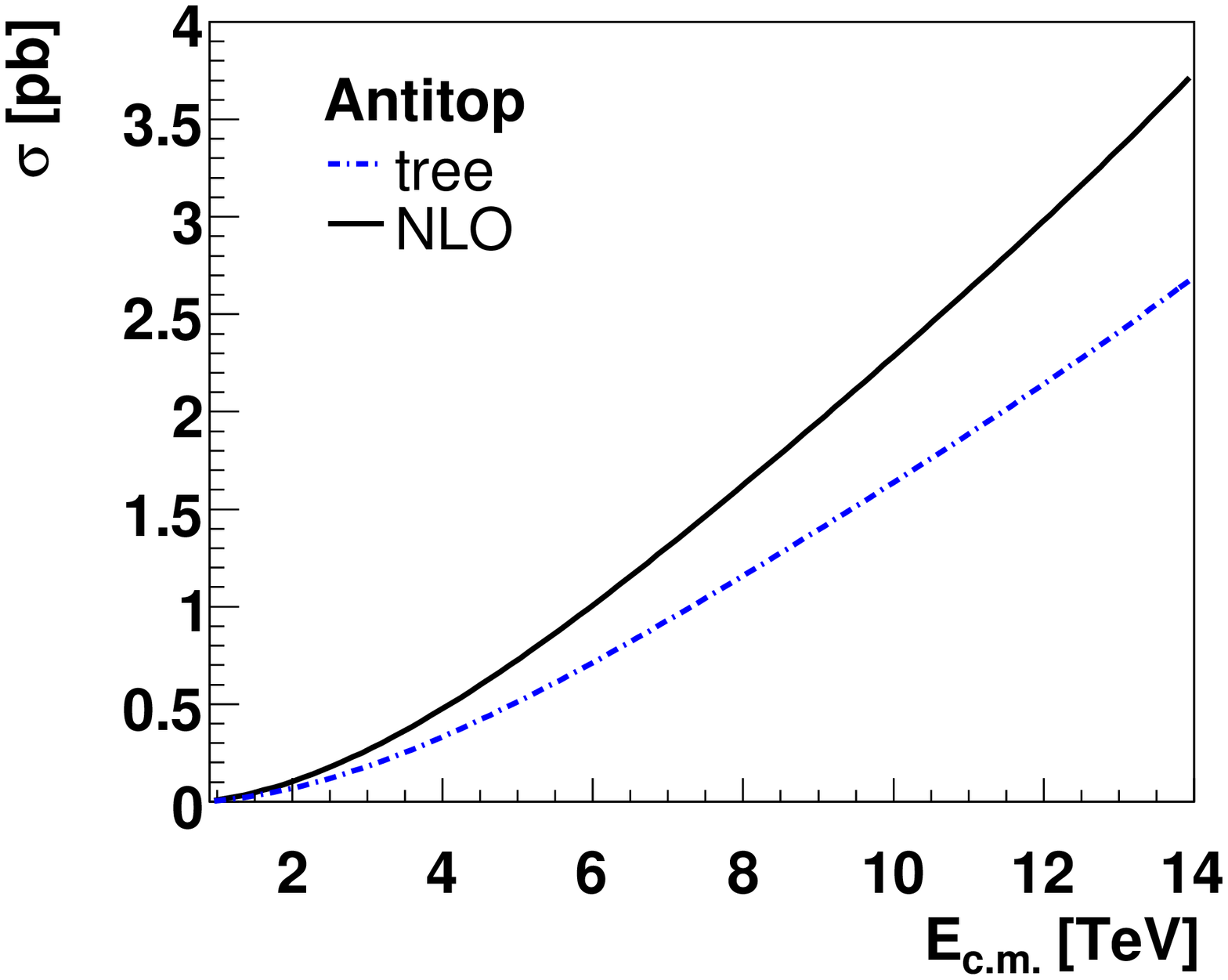}}\subfigure[]{\includegraphics[scale=0.28]{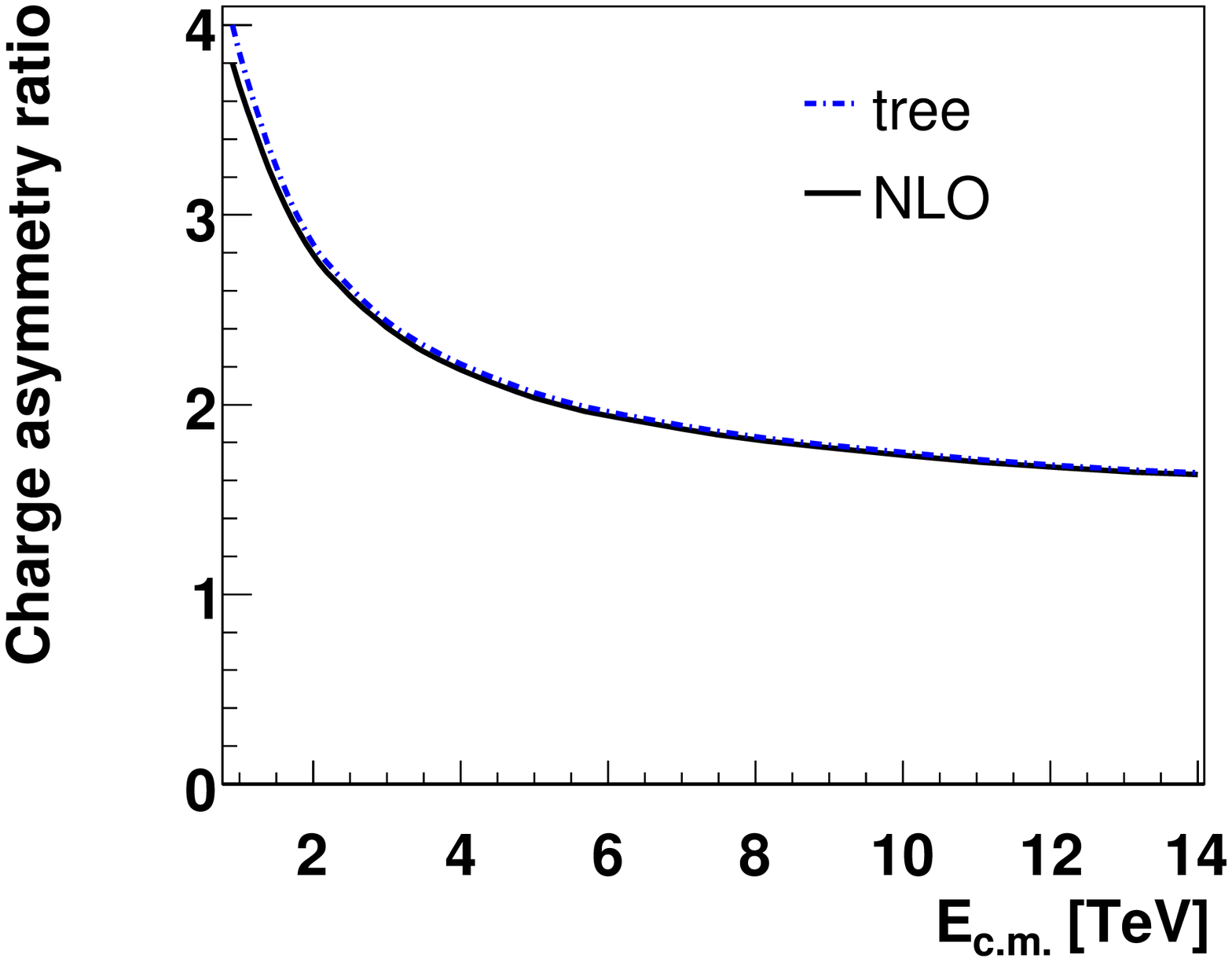}}\caption{$E_{c.m.}$ dependence of single top (a) and antitop (b) quark production cross section at NLO and leading order (LO) for $m_{T} = $ 175 GeV and ratio of cross sections for top/antitop quark production (c). \label{fig:cmcrosssection}}
\end{figure*} 

For $m_t$ = 175 GeV, Fig.~\ref{fig:cmcrosssection} compares the $E_{c.m.}$ dependence of NLO and
 LO cross sections, for top and antitop quark, and for the charge asymmetry ratio ($\mathcal{R}$). $\mathcal{R}$ is defined as
the ratio of top over antitop quark production cross sections. Different from the Tevatron, the single top quark production at the LHC (a $pp$ collider) has a large charge asymmetry, leading to a difference in the numbers of top
versus antitop quarks produced. Such an asymmetry is preserved in the charge of 
charged leptons from the top quark decay and will be measured at the LHC.
It can be seen that for both top and antitop quark, with rising $E_{c.m.}$, the cross section
 grows faster at NLO than at LO. This is mainly due to the higher momentum of the sea quarks
and gluons in the colliding protons which increases the initial state corrections. The ratio of top/antitop quark production decreases with higher energy as the PDF difference between up and down quark looses significance.

Without losing generality, in the remaining paper we consider $E_{\rm c.m.}=$ 14 TeV only.

\subsection{Top quark mass dependence and theoretical uncertainties}

In order to predict the cross section for single top quark production
as precisely as possible, we need to understand how it depends on variations
of the input parameters, such as $m_t$ dependence,  
scale dependences, and PDF uncertainties.  

\begin{figure*}
\subfigure[]{\includegraphics[scale=0.28]{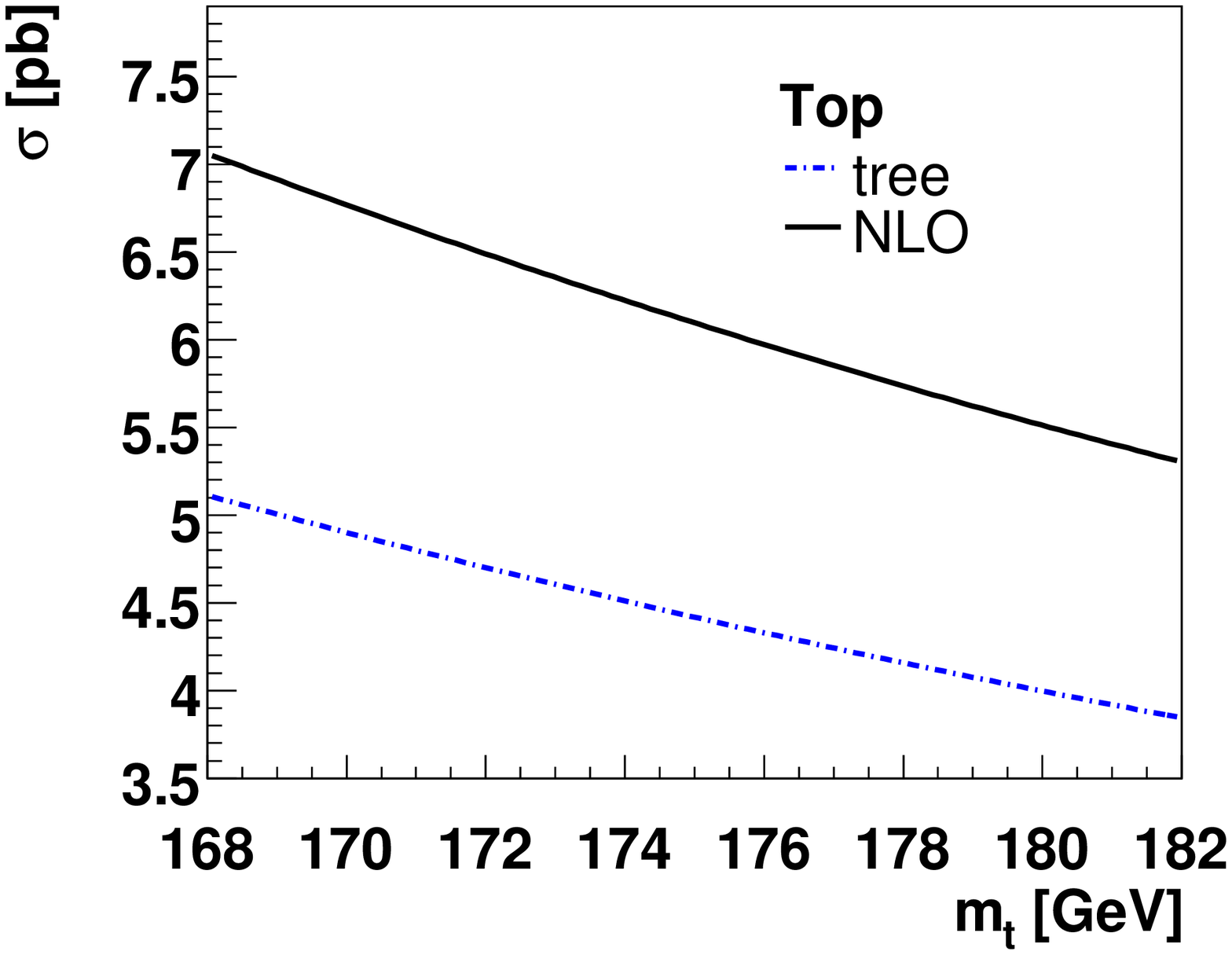}}%
\subfigure[]{\includegraphics[scale=0.28]{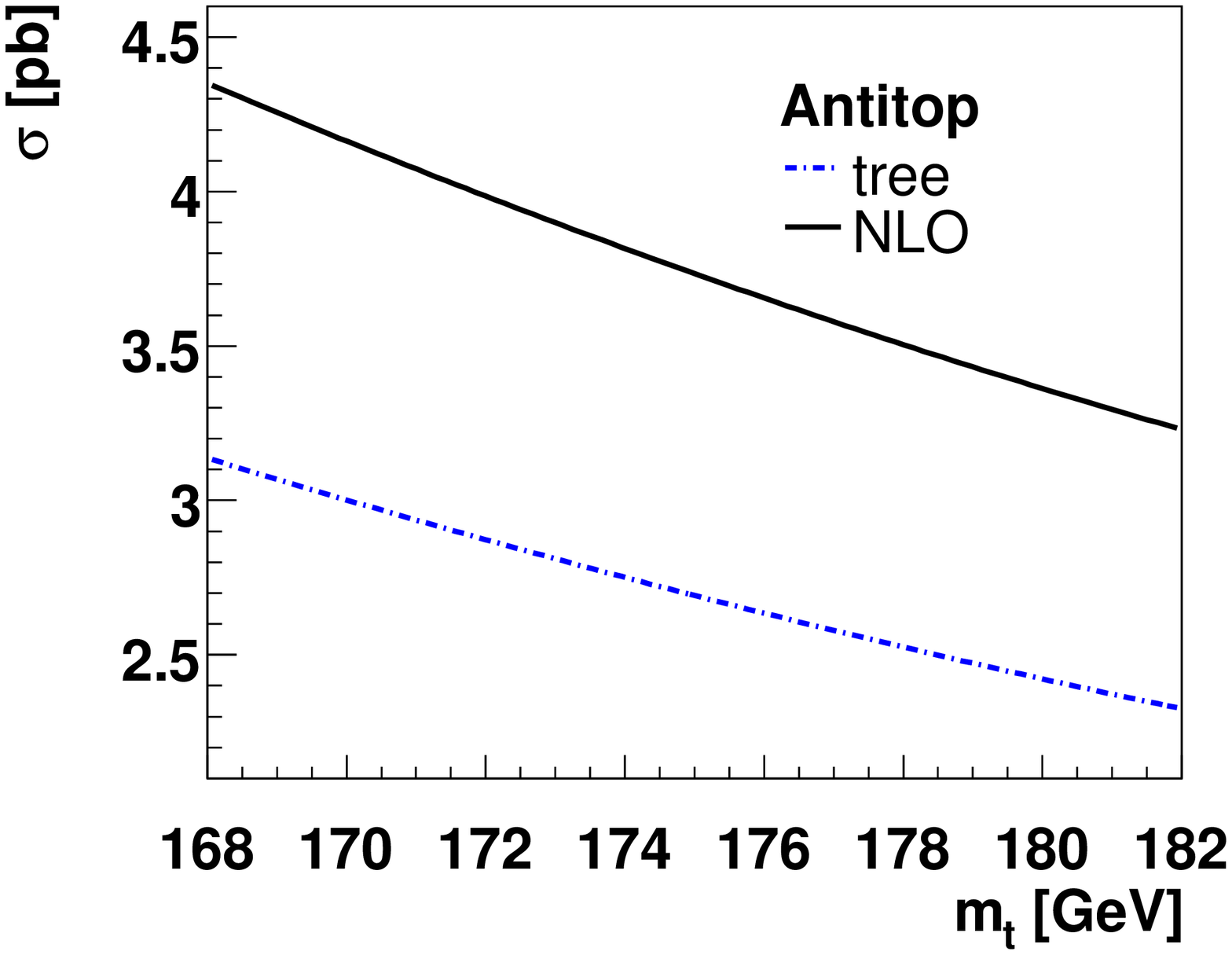}}%
\subfigure[]{\includegraphics[scale=0.28]{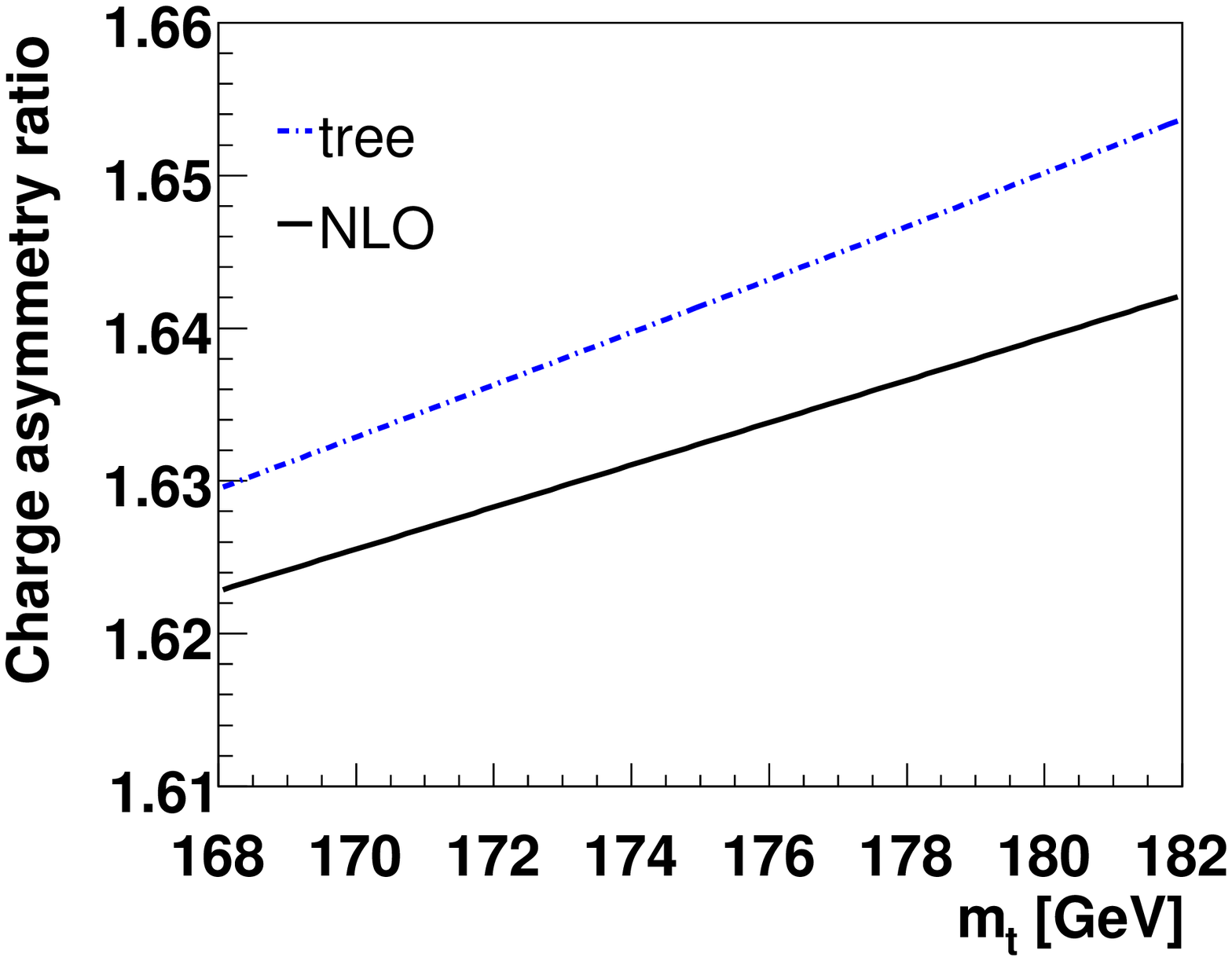}}
\caption{$m_t$ dependence of single top (a) and antitop (b) quark production cross
section and ratio of
cross sections for top/antitop quark production (c), $E_{c.m.} = $ 14 TeV.  
\label{fig:crossSectionMass}} 
\end{figure*}

The Tevatron has accomplished to reduce the
uncertainty of the newest world average $m_t$ to
$1.3$ GeV \cite{Tevatron:2009ec}, so it is of interest to see how the cross
section varies in this range. 
It can be seen in Fig.~\ref{fig:crossSectionMass}, that a variation
of $m_t$ at 175 GeV of $\pm$ 5 GeV changes the cross
section of single top quark production by about $\mp$ 10\%. For the
current world average of 173.1 $\pm$ 1.3 GeV \cite{Tevatron:2009ec}, the
predicted numerical result for the s-channel single top quark production cross
section is 6.36 pb $\pm$ 0.19 pb, where the error of 3\% is due to
mass uncertainty only. For antitop quark production the calculation
yields 3.90 pb $\pm$ 0.12 pb, also with an error of 3\%.

In Fig.~\ref{fig:crossSectionMass}~(c) the charge asymmetry ratios 
 are $\sim 1.6$ at LO and NLO in the region of $168~\rm{GeV} < m_t < 182~\rm{Ge
V}$.
Such a large charge asymmetry will be detectable at the LHC.
We also note that the ratio curves at LO and NLO both rise with increasing $m_t
$, which can be understood as follows. 
If we set $x_1 \approx x_2 = x$ where $x_{1,2}$ is the fraction of incoming 
proton 
energy carried by the parton, we obtain the average $x$ as $\left<x\right> 
\approx m_t/\sqrt{s}$ from $\hat{s}=x_1 x_2 s$, 
where $\sqrt{s}=E_{c.m.}$ and $\sqrt{\hat s}$ is the invariant mass of the incoming partons.
The heavier the top quark, the 
larger $\left<x\right>$.
Since the down quark PDF peaks at lower $x$ value than the up quark PDF, the antitop quark production rate decreases faster than the top quark production rate with increasing $x$ (i.e. increasing $m_t$). 
Therefore, it yields the increasing ratio curves in Fig.~\ref{fig:crossSectionMass}~(c).
The $\oalphas$ corrections, involving more production channels, only distort this picture slightly. 
For the remainder of this paper we use $m_t =$ 175 GeV.

\begin{figure*}
\subfigure[]{\includegraphics[scale=0.28]{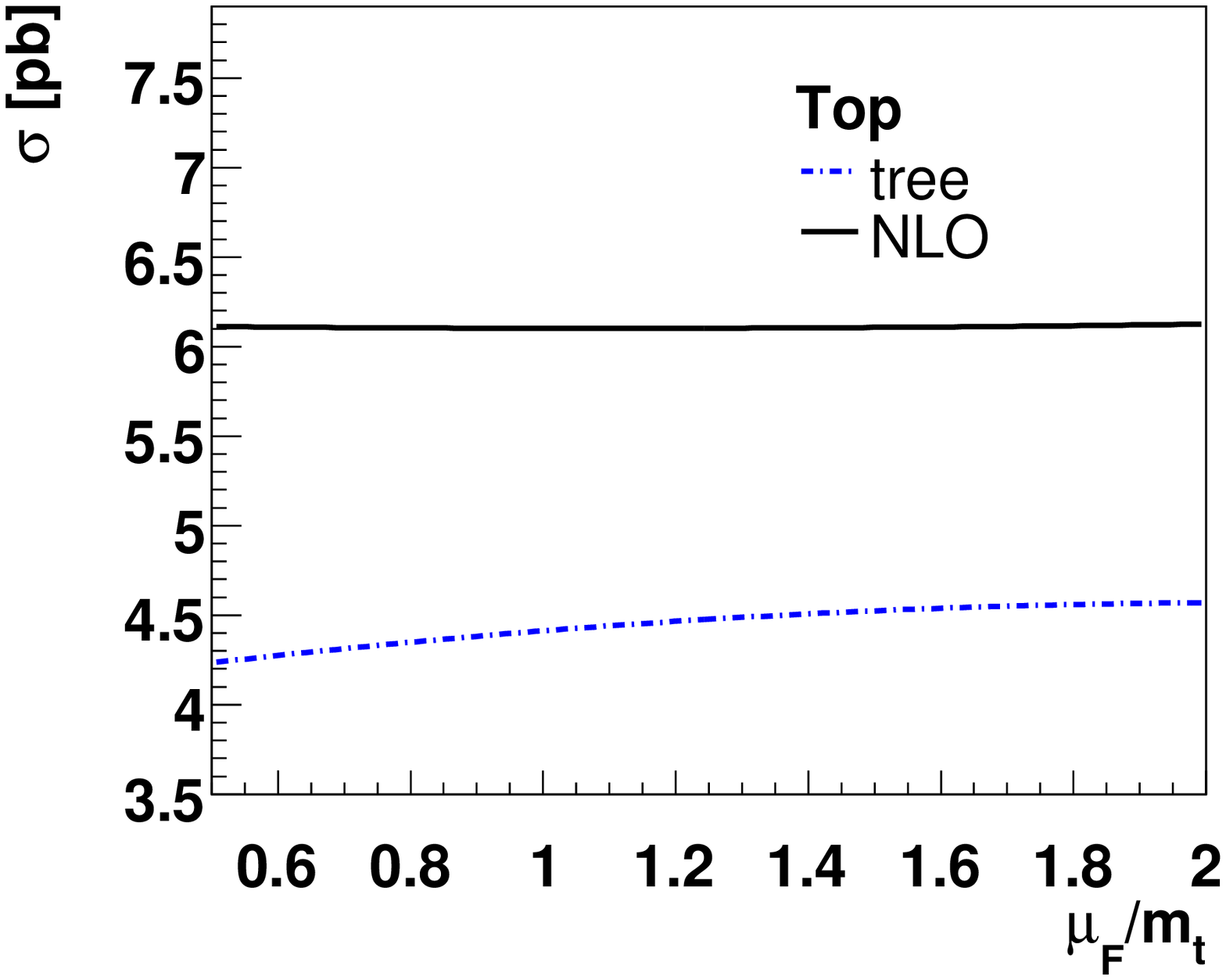}}%
\subfigure[]{\includegraphics[scale=0.28]{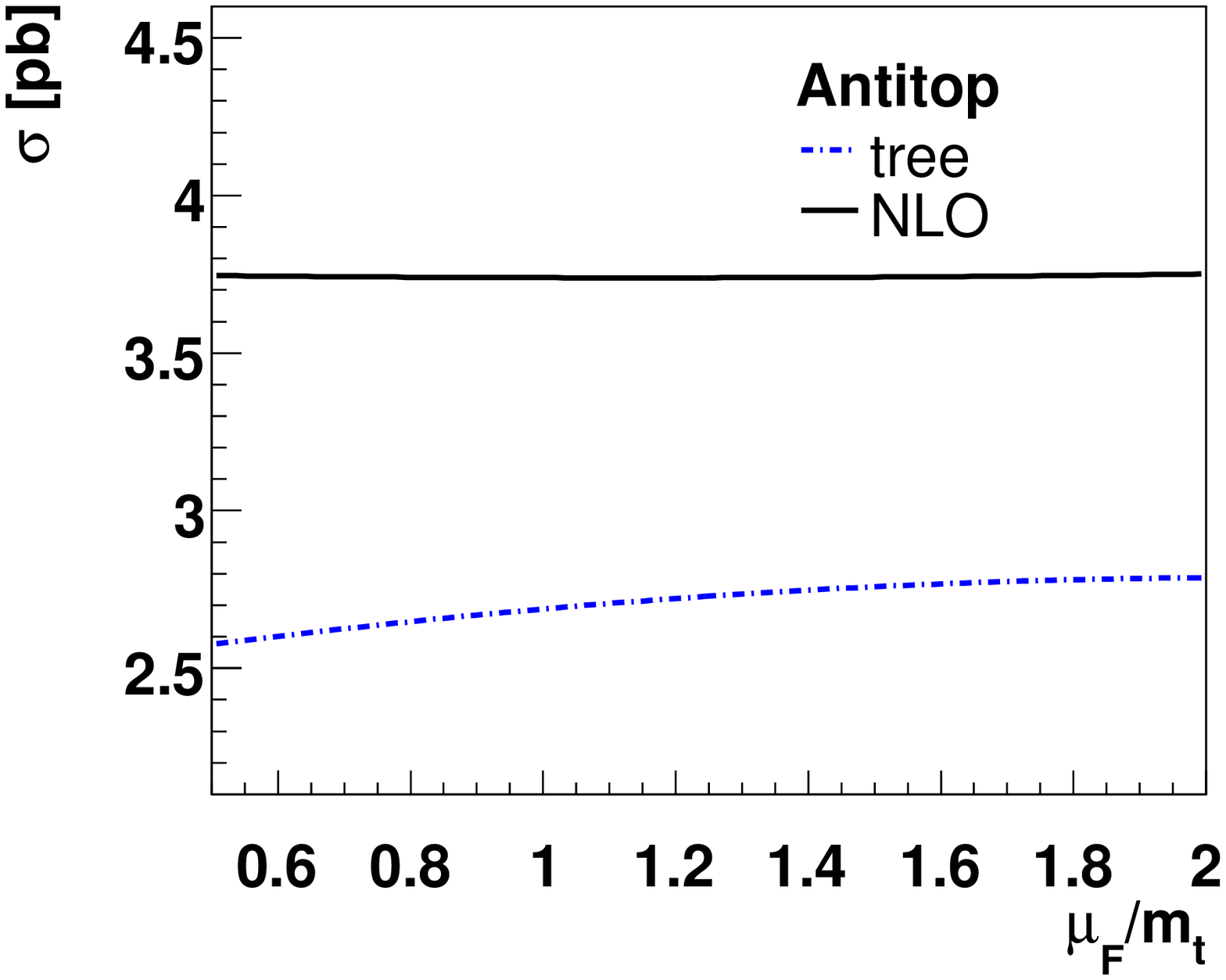}}%
\subfigure[]{\includegraphics[scale=0.28]{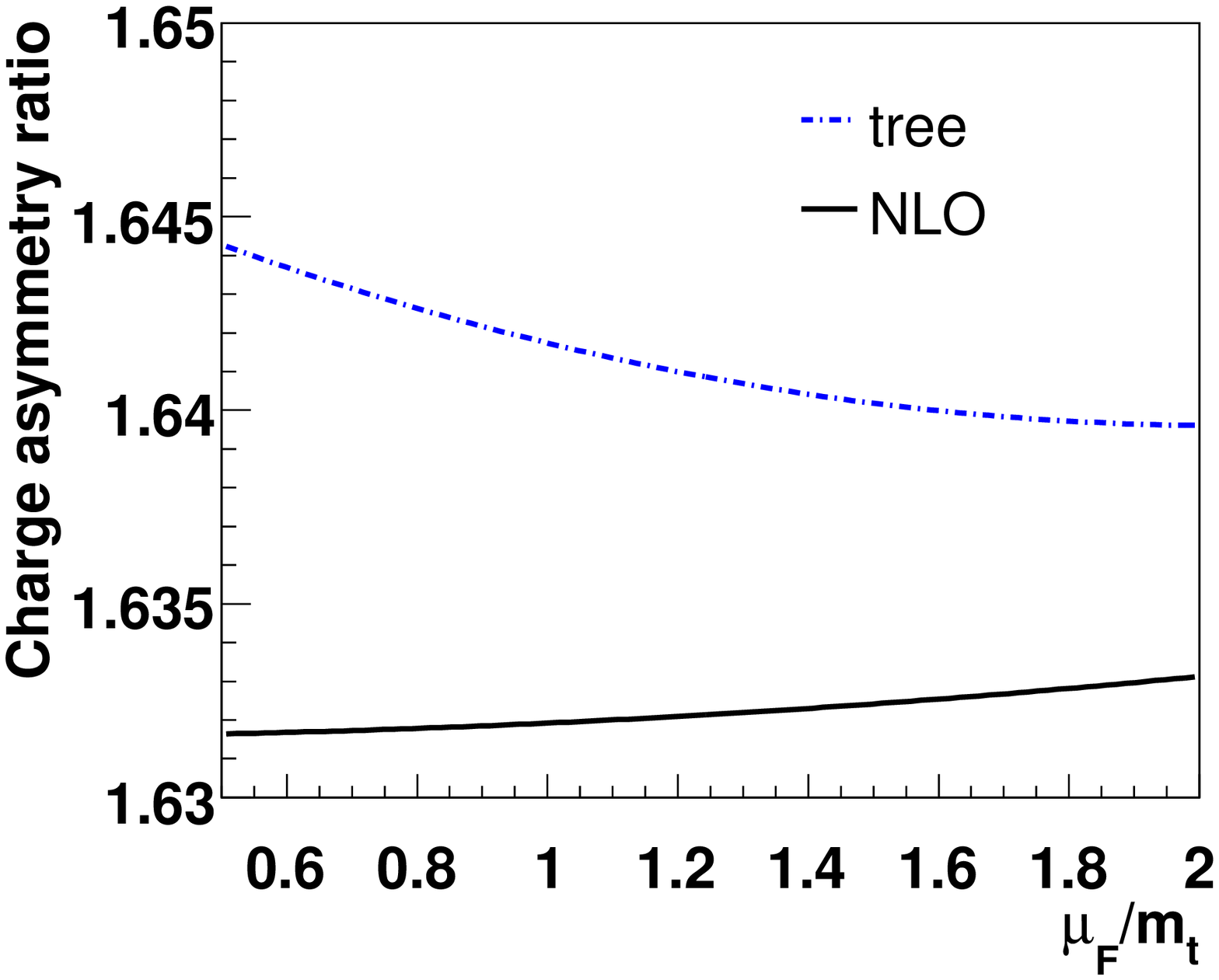}}      

\caption{Scale dependence of single top (a) and antitop (b) quark production cross section and
ratio of cross sections for top/antitop quark production (c). $E_{c.m.} =$ 14 TeV and $m_t =$ 175 GeV. \label{fig:crossSectionScale}}
\end{figure*}

Besides $m_t$ dependence, single top quark production also suffers from scale dependence, a theoretical uncertainty originated from the unknown higher order corrections.
There are two kinds of uncertainties: 
One is the renormalization scale $\mu_{R}$ which is used for redefining 
the bare parameters in terms of the renormalized parameters, the other is 
the factorization scale $\mu_{F}$ which is introduced in order to absorb 
the collinear divergence into the PDFs. 
Although both $\mu_{R}$ and $\mu_{F}$ are only introduced for technical reasons
and our predictions for the cross section should not depend on them,
we see such dependences, as we only work at first order in perturbation theory.
In principle, $\mu_R$ and $\mu_F$ are two independent theoretical parameters. For simplicity, we choose 
$\mu_{R}=\mu_{F}=\mu_{0}=m_t$. We vary $\mu_{0}$ by a factor 2 to estimate the size of 
higher order quantum corrections. 
Figure~\ref{fig:crossSectionScale}~(a,~b) shows that the $\oalphas$
corrections reduce the scale dependence of the single top and antitop
quark production cross section. In Fig.~\ref{fig:crossSectionScale}~(c) we
see that the charge asymmetry ratio falls at LO with rising $\mu_{F}$, 
but increases at NLO. 
This difference comes from additional constituent processes available at NLO.

Another theoretical uncertainty comes from the PDFs. 
The uncertainty of the CTEQ6.6M PDFs is given in Ref.~\cite{Nadolsky:2008zw}. 
At NLO the PDF uncertainty is found to be roughly about 3\% for 
both top and antitop quark productions. Such a
small uncertainty is due to relatively large $x$ values ($x \sim m_t/\sqrt{s} \sim 0.01$)
typical for s-channel single top quark production, where PDF uncertainties 
are generally small~\cite{Nadolsky:2008zw}.

\section{Single top quark acceptance studies\label{sec:Acceptance}}

The $W$ boson from the top quark can decay into jets or leptons.
We only consider the leptonic decay mode as the hadronic decay mode
is very difficult to observe experimentally, due to the large QCD
background. Thus, the experimental signature of an s-channel single
top quark event at NLO is the following: one charged lepton, missing transverse
energy ($\met$) and two or three jets. As we are discussing single top
quark production at parton level here, we only approximate the kinematic
acceptance of the detector, and do not consider other detector effects
such as $b$-tagging efficiency or jet energy resolution. For the discussion
of the effects of gluon radiation, jets have to be defined as infrared-safe
observables. For this study we use the cone-jet algorithm \cite{Alitti:1990aa},
as explained in Ref.~\cite{Cao:2004ap} with cone-size $R=0.4$ \cite{O'Neil:2002ks}.
In this section we furthermore consider $R=1.0$ for reference. The
same $R$-separation is also applied to the separation between
leptons and jets.

We consider two sets of kinematic cuts on the final state objects,
a `loose' and a `tight' cut set~\cite{Cristinziani:2008er}: 
\begin{eqnarray}
\label{eqnarray:cuts}
p_{T}^{\ell}\ge30\,{\rm~GeV} & , & \left|\eta_{\ell}\right|\le2.5, \nonumber     \\ 
\met\ge20\,{\rm~GeV} &,& \nonumber   \\ 
p_{T}^{j}\ge p_{T}^{min} &,& \left|\eta_{j}\right|\le\eta_{j}^{max},   \\    \nonumber 
\Delta R_{\ell j}\ge R_{cut} &,& \Delta R_{jj}\ge R_{cut}.
\end{eqnarray}
For both sets of cuts, each event is required to have one lepton and
at least 2 jets passing all selection criteria. The cut on the separation
is chosen to be $R_{cut}=$ 0.4 (and for reference 1.0). The `loose' set
of cuts requires the jets to have a transverse momentum of at least
$p_{T}^{min}=30$ GeV, and a pseudo-rapidity of at most $\eta_{j}^{max}$
= 5. At least one of the jets has to come from a $b$ quark. The `tight'
set requires two $b$ jets and all jets to have at least $p_{T}^{min}=50$ GeV 
and at most $\eta_{j}^{max}$ = 2.5. 

\begin{table*}
\begin{centering}
\begin{ruledtabular}
\begin{tabular}{lc|ccccc|ccccc}
\multicolumn{2}{c|}{} & \multicolumn{5}{c|}{Top} & \multicolumn{5}{c}{Antitop}\tabularnewline
\multicolumn{2}{c|}{$\sigma$ {[}pb{]}} %
                   & LO     & NLO    & INIT   & FINAL  & SDEC    & LO     & NLO    & INIT   & FINAL  & SDEC \tabularnewline
\hline
(a)  & 2 \& 3 jet  & 1.42  & 2.02  & 0.57  & 0.19  & -0.16  & 0.90  & 1.28  & 0.37  & 0.12  & -0.10 \tabularnewline
     & 3 jet       &        & 0.81  & 0.62  & 0.15  & 0.04   &        & 0.50  & 0.38  & 0.09  & 0.02 \tabularnewline
\hline 
(b)  & 2 \& 3 jet  & 1.19  & 1.69  & 0.45  & 0.17  & -0.12  & 0.75  & 1.08  & 0.29  & 0.11  & -0.07 \tabularnewline
     & 3 jet       &        & 0.44  & 0.38  & 0.06  & 0.01   &        & 0.27  & 0.23  & 0.04  & 0.004 \tabularnewline
\hline 
(c) & 2 \& 3 jet   & 0.68  & 0.79  & 0.21  & 0.02  & -0.13  & 0.44  & 0.51  & 0.14  & 0.01  & -0.08 \tabularnewline
    & 3 jet        &        & 0.24  & 0.19  & 0.05  & 0.004   &        & 0.15  & 0.12  & 0.03  & 0.002 \tabularnewline
\end{tabular}
\end{ruledtabular}
\par\end{centering}
\caption{LO and NLO contributions of cross sections for inclusive two-jet events as well as for exclusive three-jet events, for different
cut scenarios. (a) `loose' cuts with $R_{cut}$ = 0.4, (b) `loose' cuts
with $R_{cut}$ = 1.0, (c) `tight' cuts with $R_{cut}$ = 0.4. $E_{c.m.} =$ 14 TeV and $m_t = $ 175 GeV.
\label{tab:total}}
\end{table*}

Table~\ref{tab:total} shows the cross sections for top and antitop quark production
in the s-channel single top quark mode, split up into the different LO and NLO contributions
 after applying the two sets of cuts. For the `loose' cut set the results are shown for two different values of $R_{cut}$ and for the `tight' cut set for $R_{
cut} = $ 0.4. The results for
 the `tight' cuts and $R_{cut} = $ 1.0 follow the same trend as the other
 numbers in the table.
For inclusive two-jet events and $R_{cut} = $ 0.4, the top quark acceptance for the `loose' cut set is around 33\% 
both at LO and NLO level, and for the `tight' cut set around 15\%
for the LO contribution and around 13\% for the NLO contributions.
For inclusive two-jet events and $R_{cut} = $ 0.4, the acceptance for antitop quarks is a bit higher than for top quarks: 34\% at LO and NLO for the `loose' cut set and 16\% (14\%) for 
the `tight' cut set at LO (NLO). The low acceptance is mainly due to the $p_{T}$ 
cuts on the leptons and jets. 
The lepton is a product of the $W$ boson decay and its
$p_{T}$ distribution peaks under 30 GeV, cf. Fig.~\ref{fig:ele}. 
Imposing a smaller $p_T$ cut on the lepton alone would not improve
the overall acceptance much because 
the $p_T$ distribution of the $b$ jet produced in association with
the (anti)top quark also peaks around 30 GeV (cf. Fig.~\ref{fig:bpt}). 
Hence, a low acceptance still follows from the $p_T$ cut on the $b$ jet.
A larger value for $R_{cut}$ reduces the cross section because
more events fail the lepton-jet separation cut. For inclusive two-jet events after applying the `loose' cuts, the reduction from $R_{cut} =$ 0.4 to $R_{cut} =$ 1.0 is about 17\% for top and antitop quark production, both at LO 
and NLO level.  
We further note that the acceptance for antitop quark production is slightly 
higher than
the acceptance for top quark production. This is due to the $\eta$ cuts on the lepton as
the lepton in antitop quark production remains in a more central 
rapidity region than the lepton 
in top quark production, cf.  Fig.~\ref{fig:ele}. This follows directly 
from the fact that the top quark is boosted more than the antitop quark.
The top quark, when produced,
receives large contributions from the valence up quark in the large $x$ 
regime, therefore it has large momentum along the moving direction of the
incoming up quark.
Such a boost effect is transferred to the top quark decay products, yielding 
the wide 
distribution of the lepton $\eta$.  

\begin{figure*}
\subfigure[]{\includegraphics[scale=0.3]{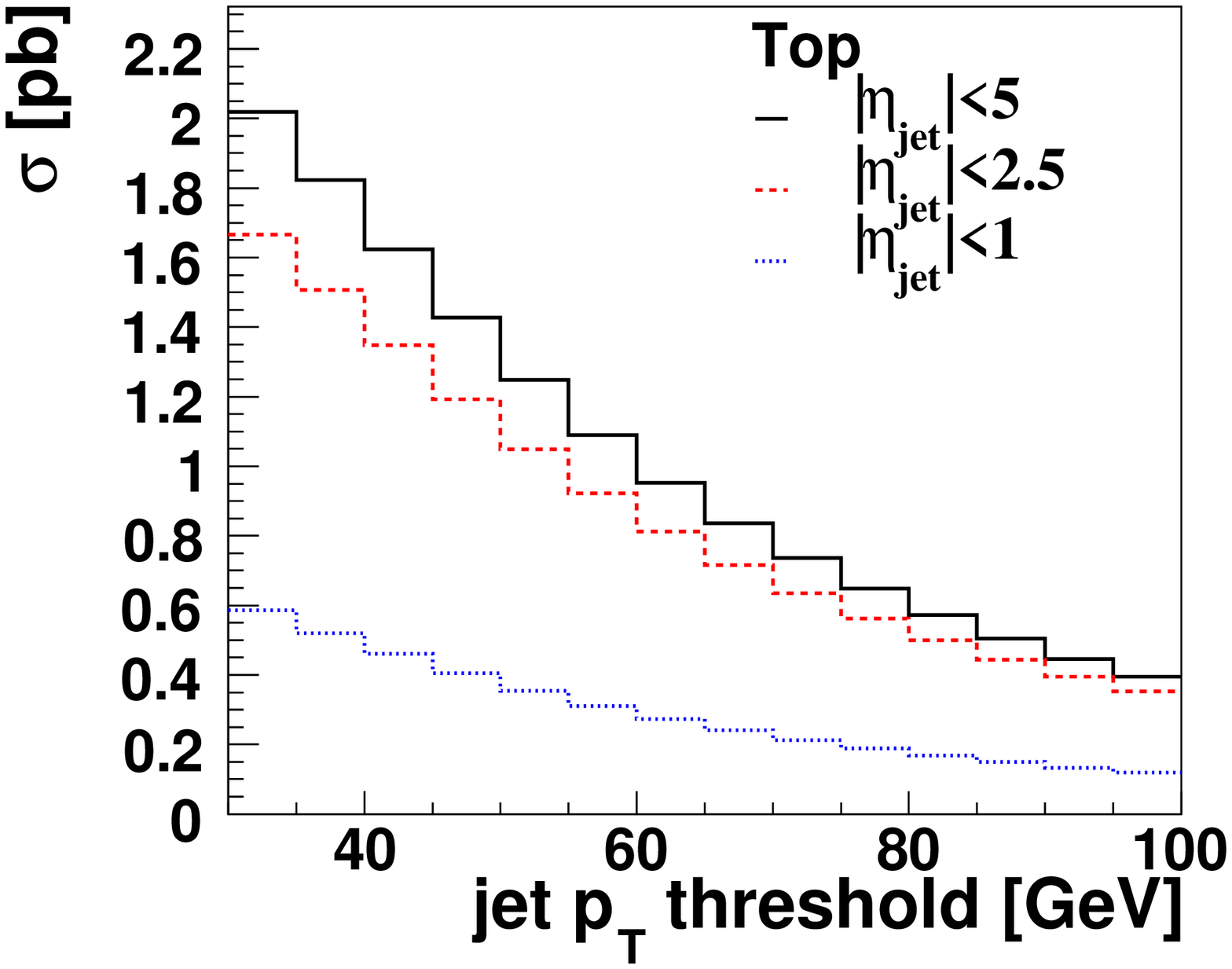}}%
\subfigure[]{\includegraphics[scale=0.3]{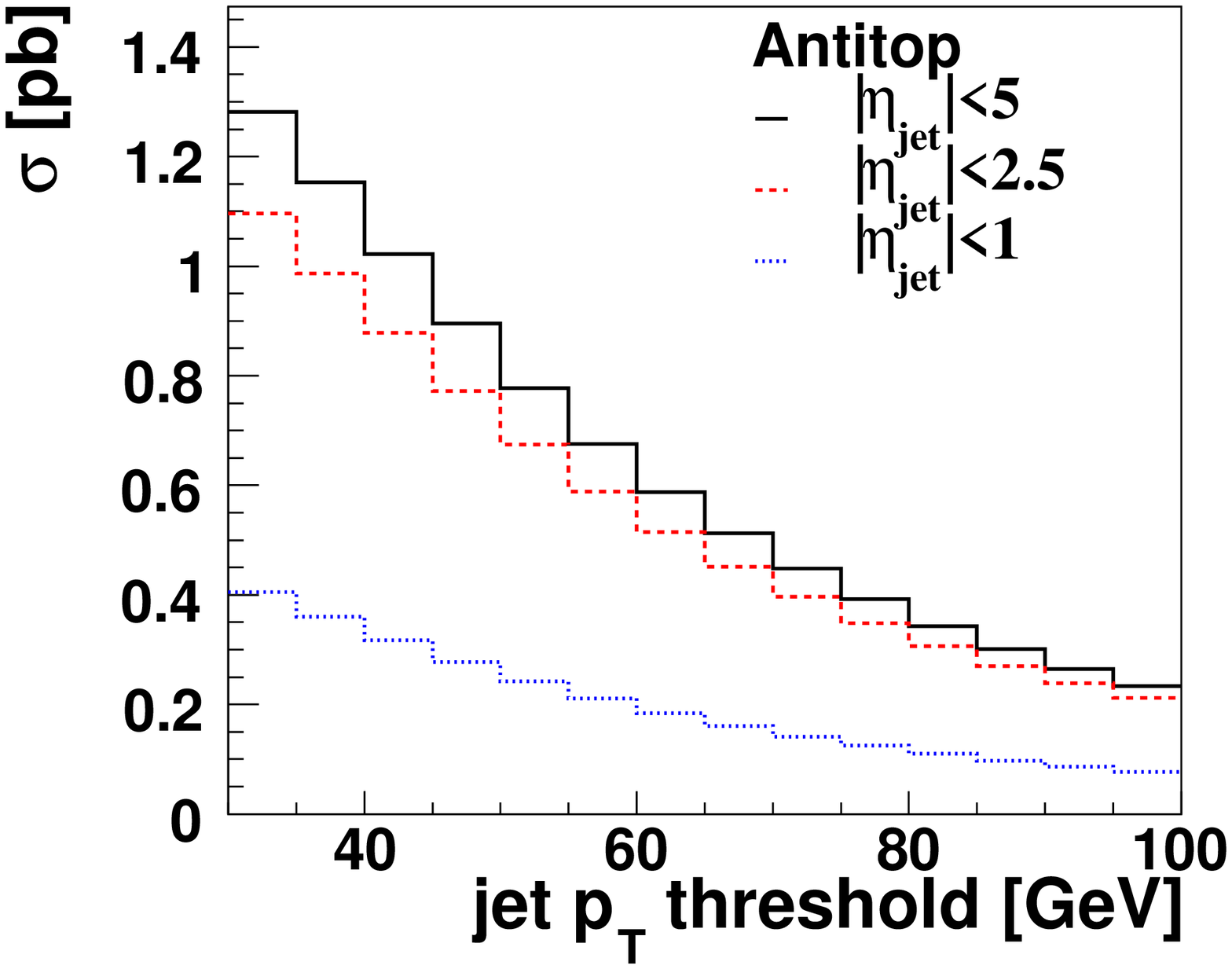}}%

\subfigure[]{\includegraphics[scale=0.3]{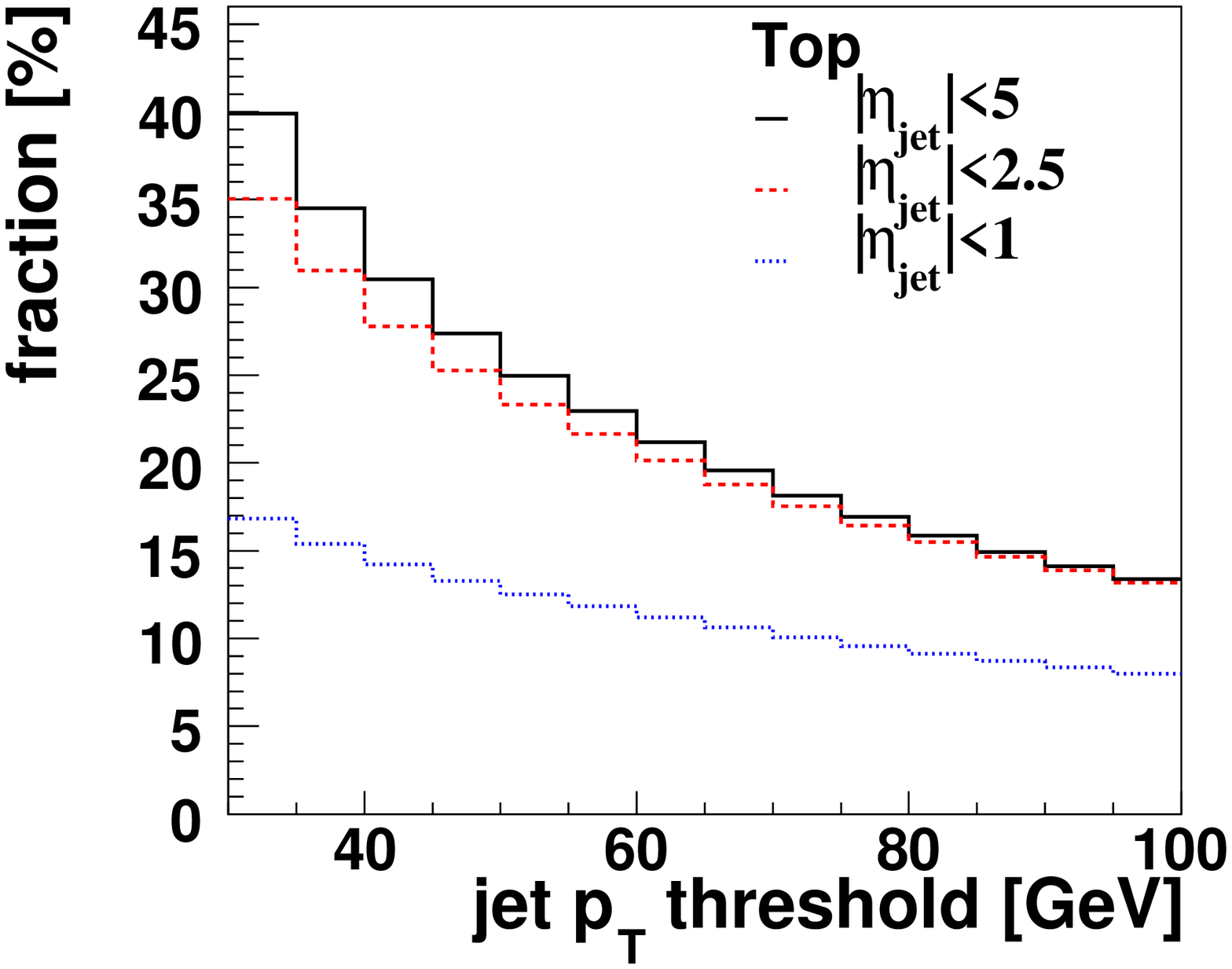}}
\subfigure[]{\includegraphics[scale=0.3]{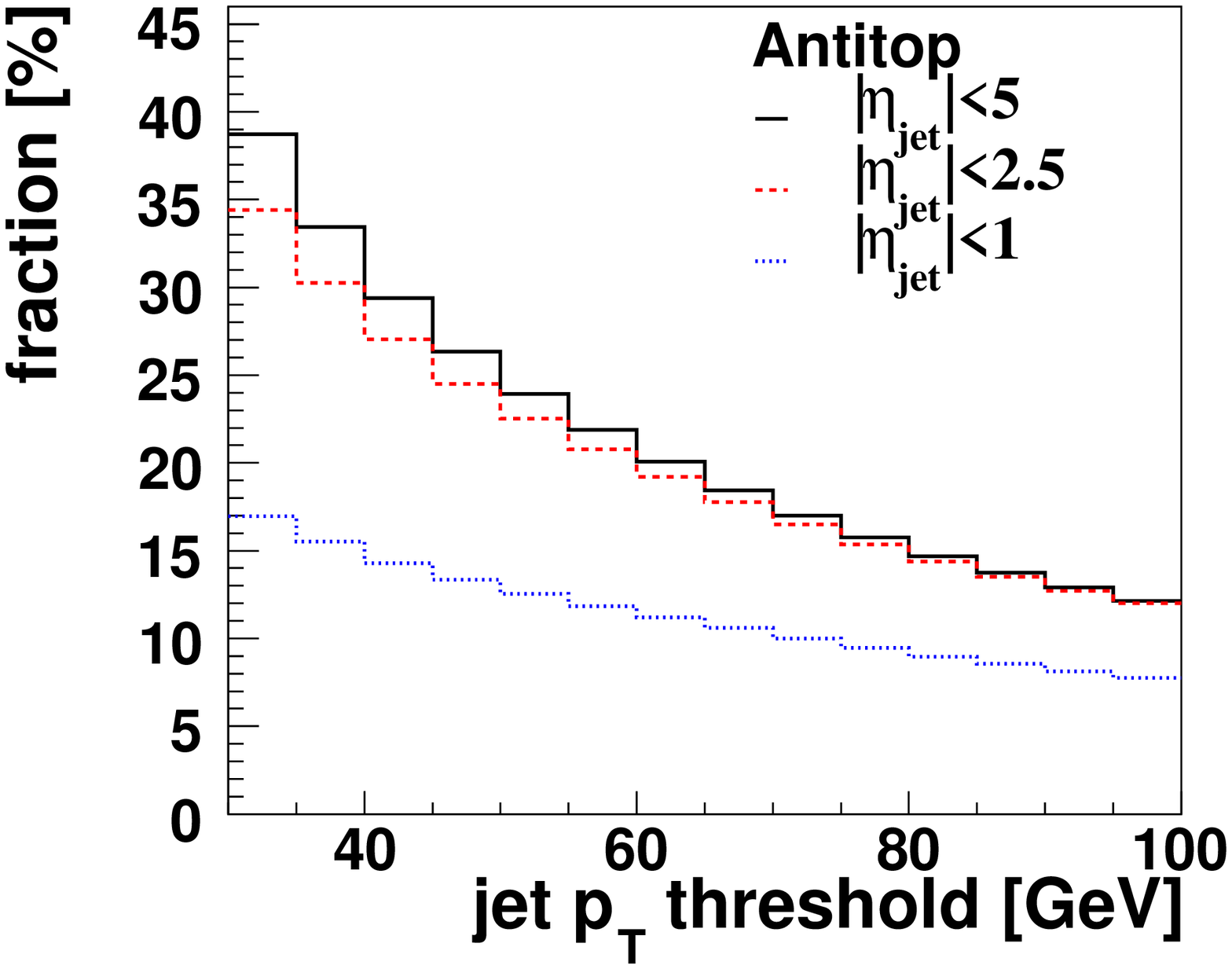}}

\caption{Single top and antitop quark production cross section and fraction of three-jet events at NLO for
varying jet $p_{T}$ cuts, after applying the `loose' cut set. (a, b)
Total cross section for inclusive two-jet events as a function
of the jet $p_{T}$ cut for three different jet $\eta$ cuts. (c, d)
Fraction of exclusive three-jet events as a function of the jet $p_{T}$
cut for three different jet $\eta$ cuts. 
\label{fig:jetPTEta}}
\end{figure*}

Figure~\ref{fig:jetPTEta}~(a, b) shows the NLO single (anti)top quark production
cross section after applying the `loose' cut set, for different $\eta_{j}^{max}$
as a function of the jet $p_{T}$ threshold. 
It can be seen that 
an $\eta_{j}^{max}$ of 2.5 as applied in our `tight' cut set does
not reduce the acceptance much compared to the $\eta_{j}^{max}$
= 5 of the `loose' cuts, but 
the reduction to very central regions
of the detector greatly decreases the acceptance. The fraction of
three-jet events is relatively large even up to high jet $p_{T}$
cuts, cf. Fig.~\ref{fig:jetPTEta}~(c, d). Due to collinear enhancement from INIT corrections,
the third jet has large pseudo-rapidity, so that the fraction of
events with an additional jet greatly decreases when only considering
central regions of the detector. In the following discussion of event distributions we will
use the `loose' cut set and $R_{cut}=0.4$. For the `tight' cut set, the results are similar. 

\section{Single top quark event distributions\label{sec:EventDistr}}

In this section we discuss some of the kinematic properties of final
state objects in s-channel single top quark events. As described in
Sec.~\ref{sec:Acceptance}, the final state consists of one charged
lepton, $\met$, two $b$ jets and possibly one additional jet due to
NLO corrections. It is experimentally impossible to determine which
of the $b$ jets is produced in association with the top quark and
which comes from the top quark decay. With the third jet from NLO
corrections, the situation becomes even more complicated, as the additional
jet has to be correctly identified to come either from the production
or decay of the top quark. A third jet coming from the SDEC correction 
should be included in the top quark reconstruction, while a third jet from INIT 
or FINAL corrections should not be. To find the best prescription for classifying the gluon/light quark
jet correctly, we first examine various kinematical distributions
of the final state particles. This includes a discussion of $b_{fin}$,
the $\bar{b}$ ($b$) jet that is produced in association with the
top (antitop) quark, and $b_{dec}$, which is the $b$ ($\bar{b}$)
jet from the top (antitop) quark decay. We then describe different
prescriptions for the top quark reconstruction. After showing the
superiority of the best-jet algorithm, we are able to reconstruct
the top quark and discuss its kinematic properties. The kinematical
distributions of the final state particles are examined after applying
the `loose' set of selection cuts as discussed in Eq.~(\ref{eqnarray:cuts}). 

\subsection{Final state object distributions}

\subsubsection{Leptons and missing transverse energy\label{sub:LepMET}}

\begin{figure*}
\subfigure[]{\includegraphics[scale=0.3]%
{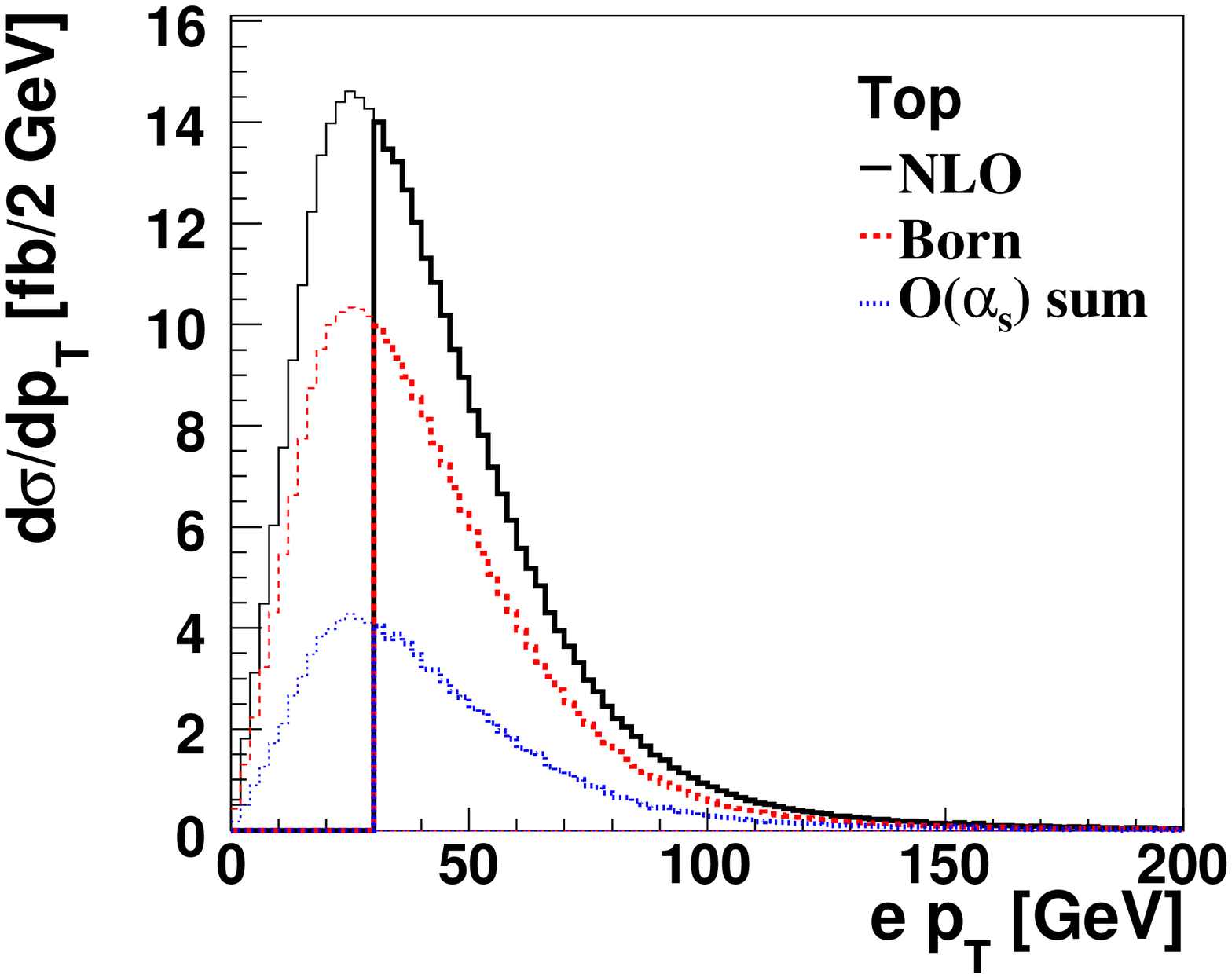}}%
\subfigure[]{\includegraphics[scale=0.3]%
{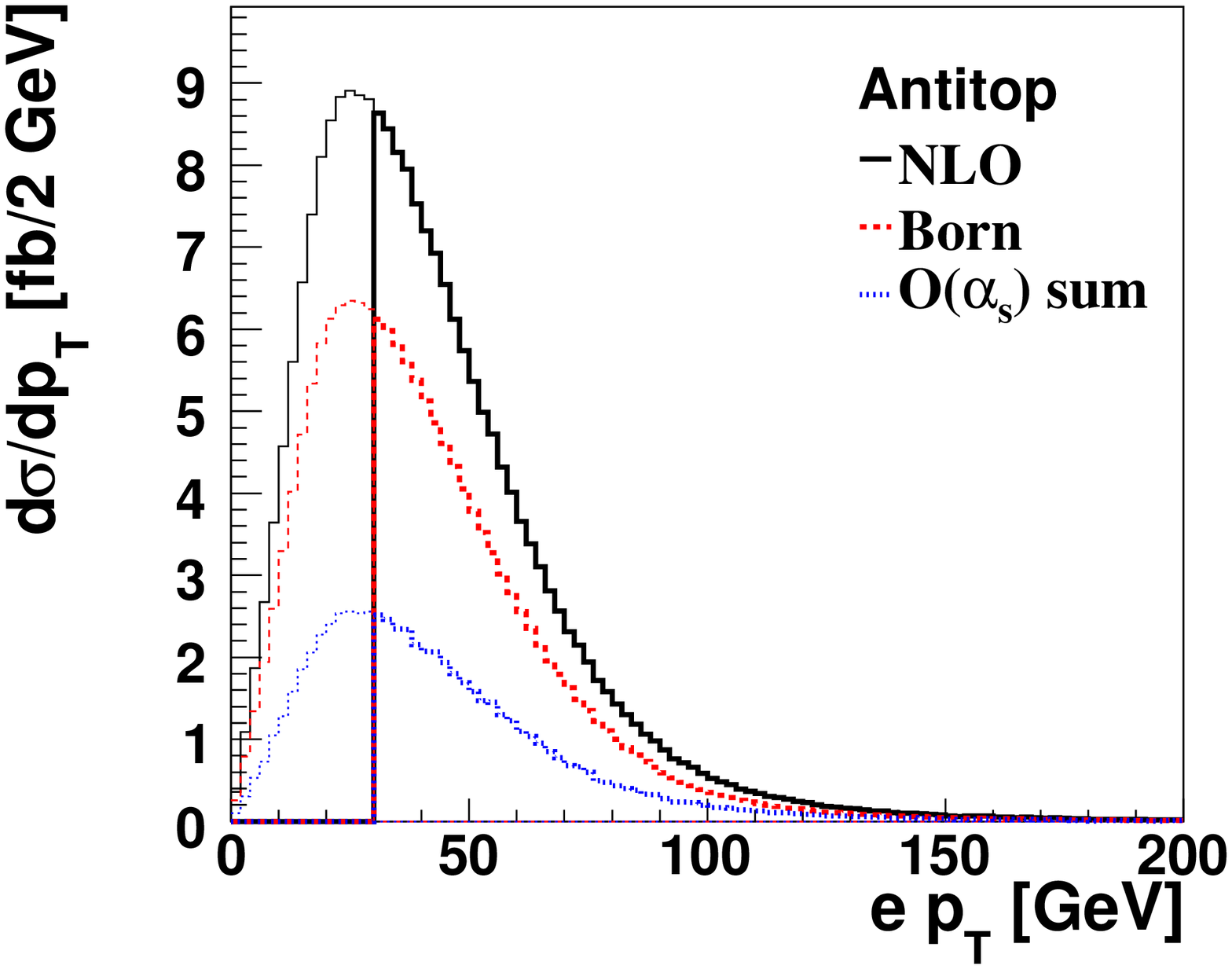}}
\subfigure[]{\includegraphics[scale=0.3]%
{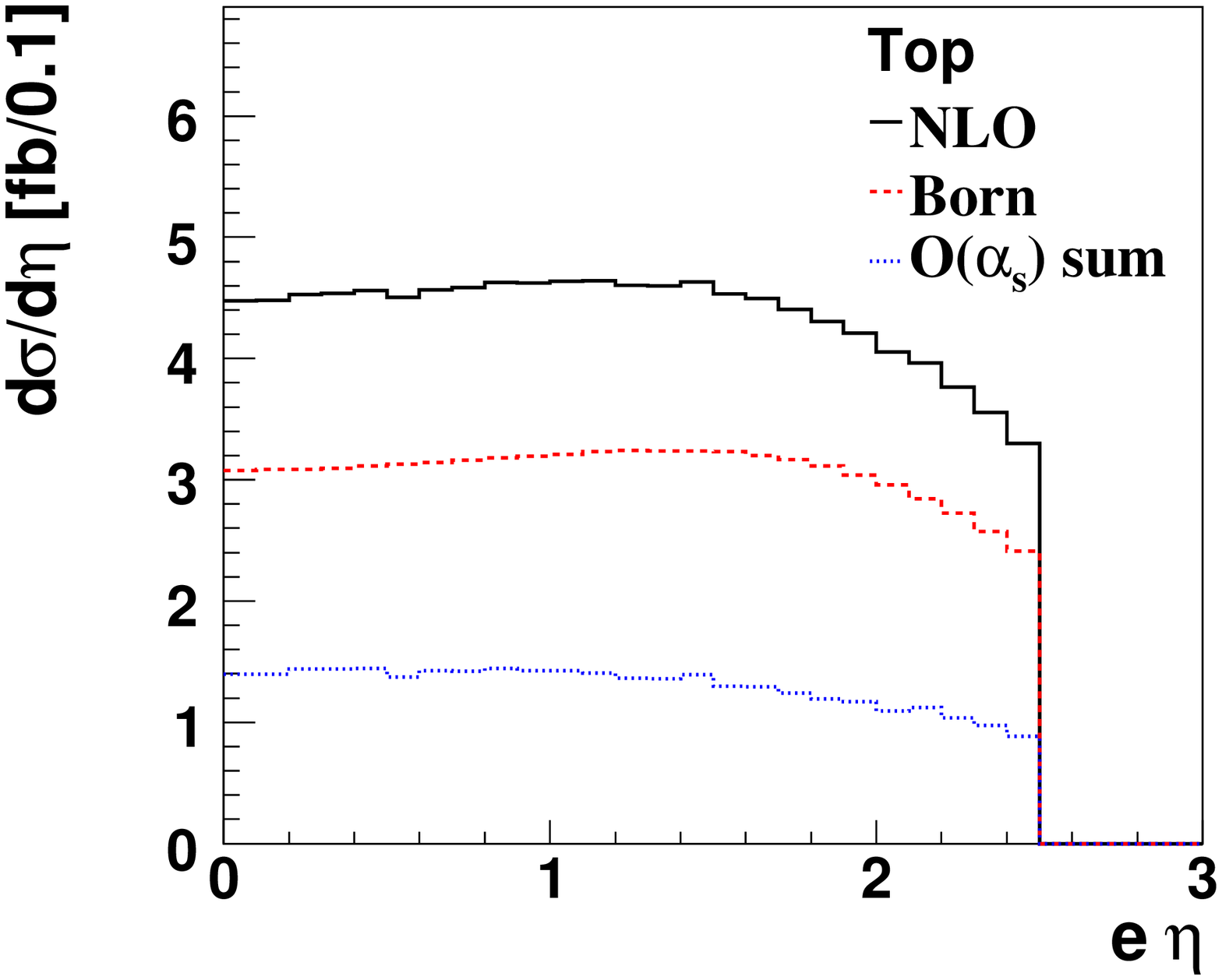}}%
\subfigure[]{\includegraphics[scale=0.3]%
{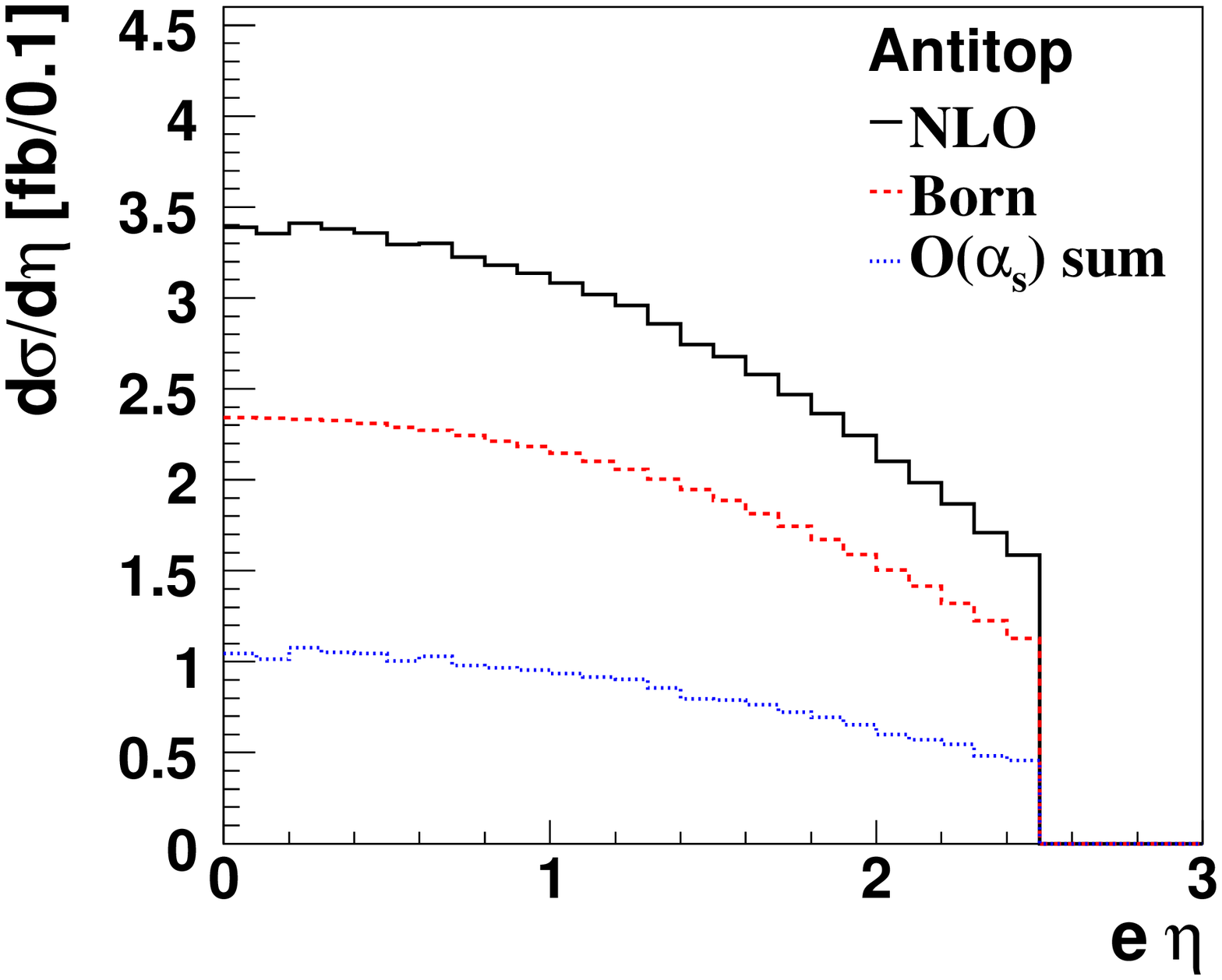}}
\subfigure[]{\includegraphics[scale=0.3]%
{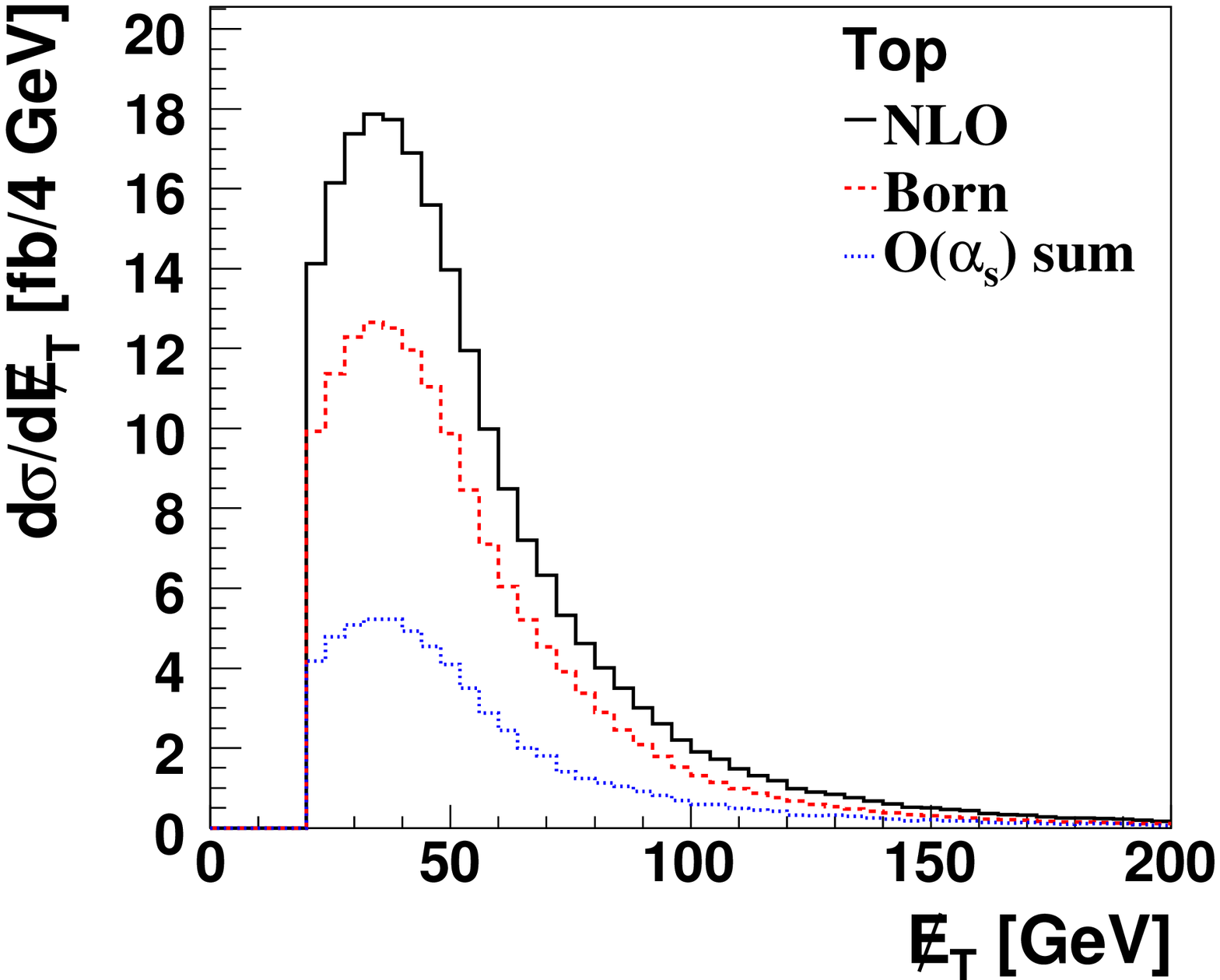}}
\subfigure[]{\includegraphics[scale=0.3]%
{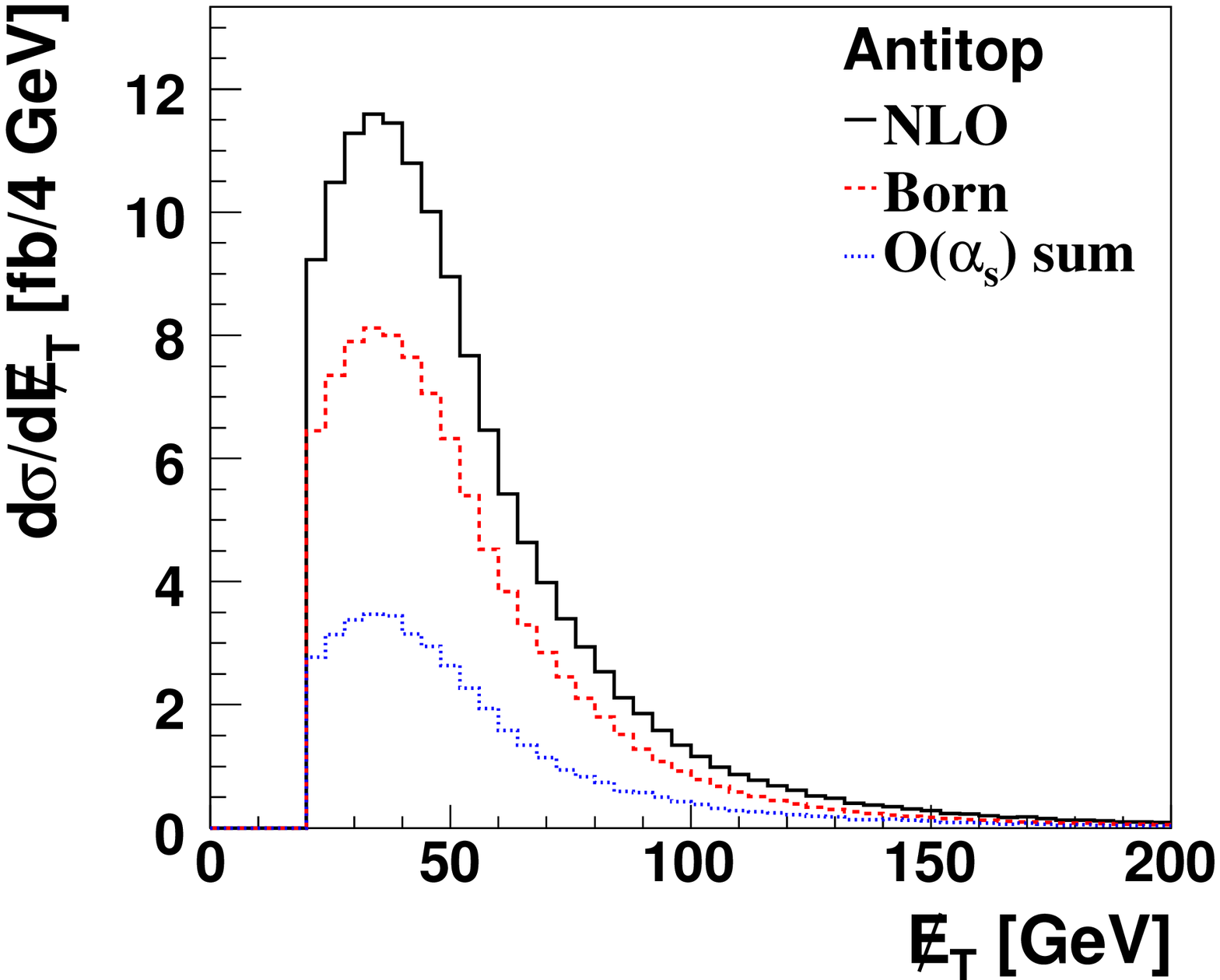}}

\caption{ $p_{T}$ (a, b) and $\eta$ (c, d) for the lepton and $\met$ (e, f)
after selection cuts, comparing LO to $O(\alpha_{s})$ corrections.
(a, c, e) show top quark events, (b, d, f) antitop quark events. In (a, b) the distributions are also shown before the lepton $p_T$ cut. 
\label{fig:ele}}
\end{figure*}

Figure~\ref{fig:ele} shows a comparison between the kinematical distributions
of lepton and $\met$ for both top and antitop quark production at LO
and with $\oalphas$ corrections. As the distributions are for
leptons and not quarks, the $\oalphas$ corrections do not change
the general form of the $p_{T}$ distributions. The shapes of the
distributions for the lepton $p_{T}$ look similar for top and
antitop quarks. 
Furthermore, the $\met$ distributions peak at about 32 GeV for both quark types. 
This is mainly because of a similar
phase space in the transverse direction and similar spin correlations.
 
\begin{figure*}
\subfigure[]{\includegraphics[scale=0.6]{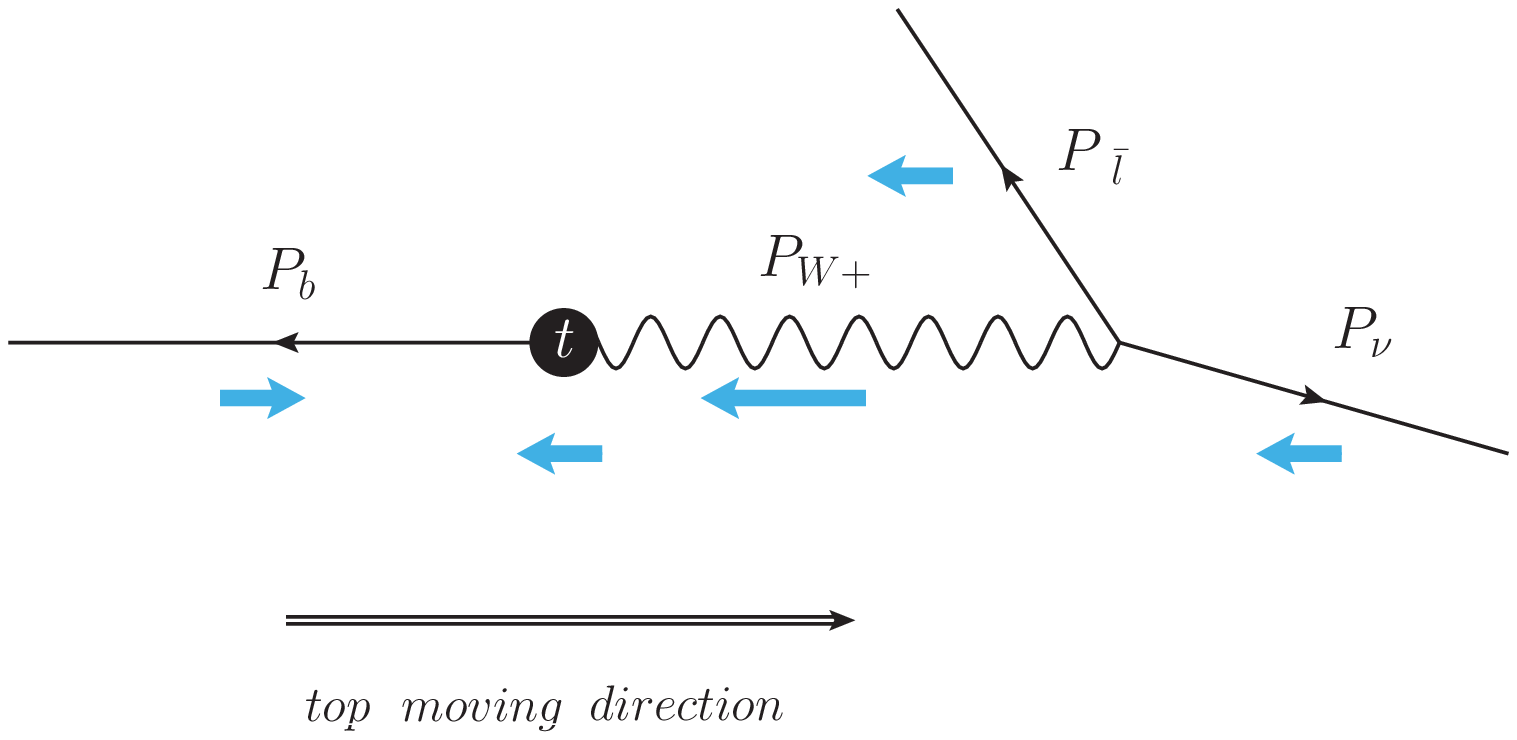}}%

\subfigure[]{\includegraphics[scale=0.6]{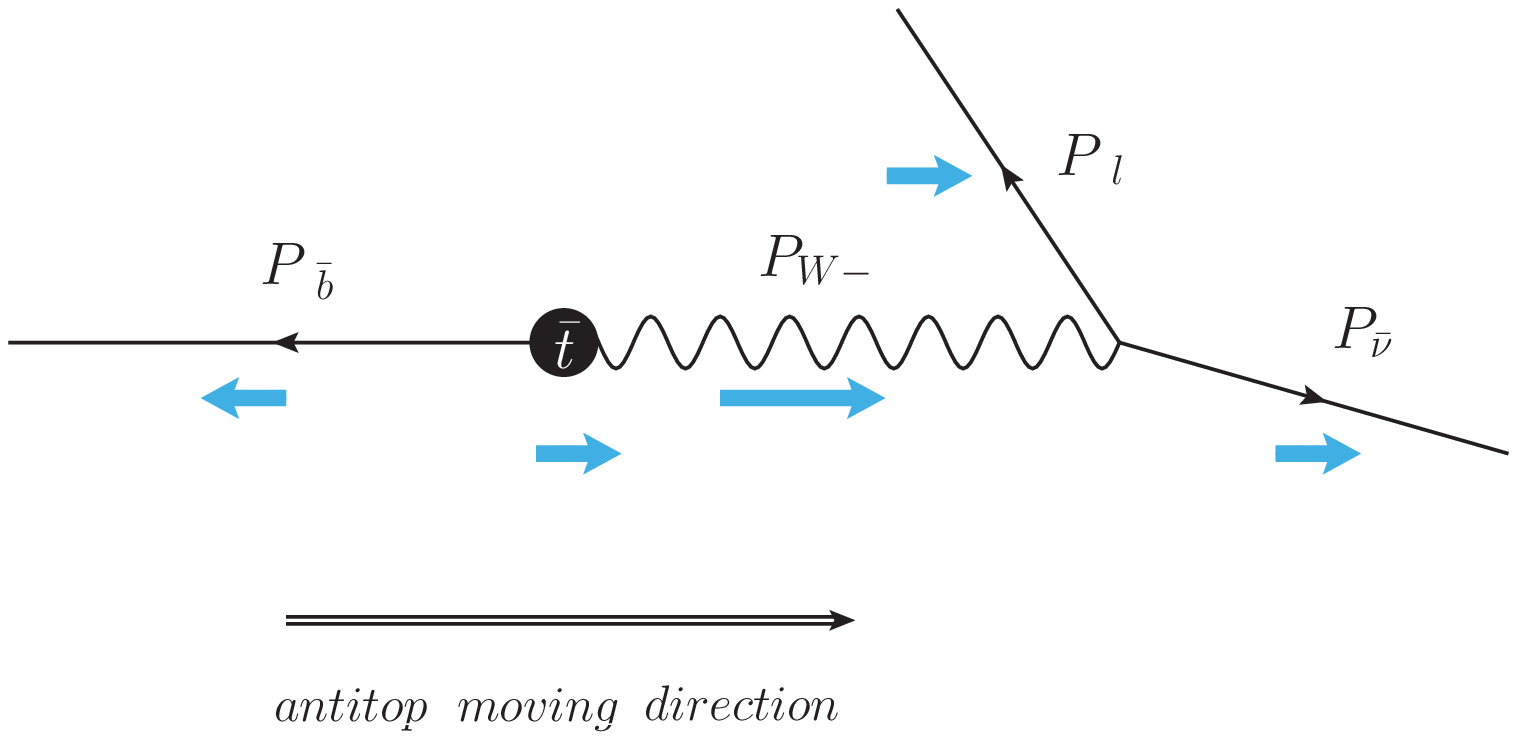}}%
\caption{Spin correlations for top (a) and antitop (b) quark decay in the (anti)top quark rest frame. The thin lines describe momenta while bold lines indicate spins.\label{fig:topdecay}}
\end{figure*}

It can also be seen that the $\met$ distribution peaks at a higher
energy than the lepton $p_{T}$ distribution. This is due to spin
correlations and the left-handedness of the weak interaction and illustrated in Fig.~\ref{fig:topdecay}: The
(anti)neutrino from the $W$ boson decays preferentially following the moving direction
of the (anti)top quark and is therefore boosted in comparison to the lepton,
which decays in the opposite direction when seen in the rest frame
of the $W$ boson. 
The lepton $\eta$ distributions are
the same for positive and negative values of $\eta$ as the $pp$
initial state of the LHC is parity symmetric. The NLO $\eta$ distribution
for the positron from the top quark decay peaks at around $\eta=1.4$.
This non-zero peak is due to LO and cannot be seen in the $\oalphas$
corrections. It originates from the longitudinal boost that the intermediate
$W$ boson ($W_{int}$)
 receives from the PDFs.
For example, a single top quark is preferentially produced through collision of an up valence quark
and a down antiquark from the quark sea. The valence quarks carry
a large momentum fraction of the incoming proton, while the sea
quarks carry a small fraction. As a result, the $(t\bar{b})$
system, which is equivalent to the $(u\bar{d})$ system at LO, 
is naturally boosted along the direction of the incoming valence quark.
This longitudinal boost is less strong for antitop quark production as
in the antitop quark case a down valence quark collides with an up
antiquark, with the down valence quark carrying a smaller momentum
fraction than the up valence quark. 
At NLO, the reacting parton in the initial state could be a sea (anti-)quark or gluon,  
which results in a smaller difference in momenta
between the colliding partons. Therefore, the lepton $\eta$ distributions
of the $\oalphas$ corrections peak around zero.

\subsubsection{$b_{fin}$ and $b_{dec}$ jets}

\begin{figure*}
\subfigure[]{\includegraphics[scale=0.3]%
{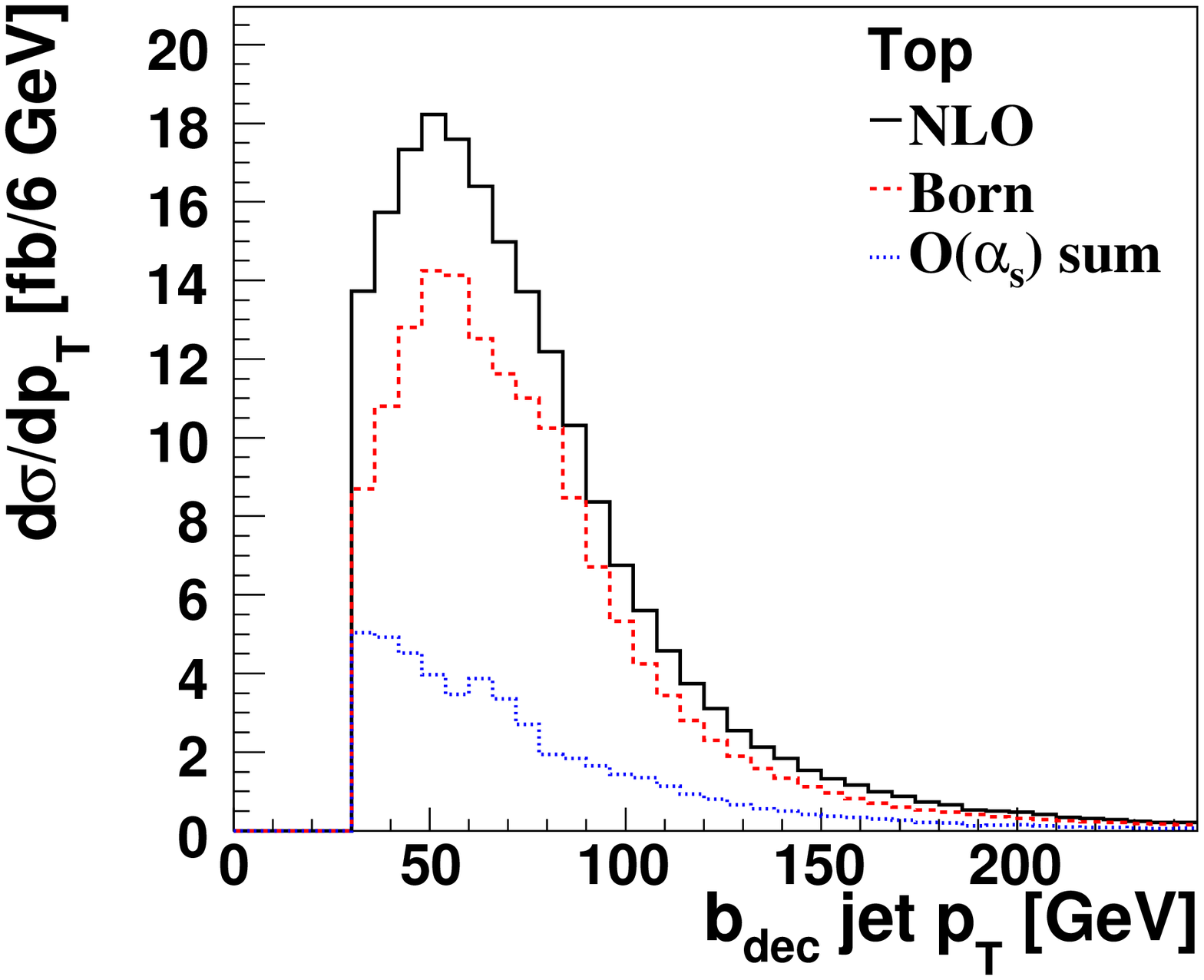}}%
\subfigure[]{\includegraphics[scale=0.3]%
{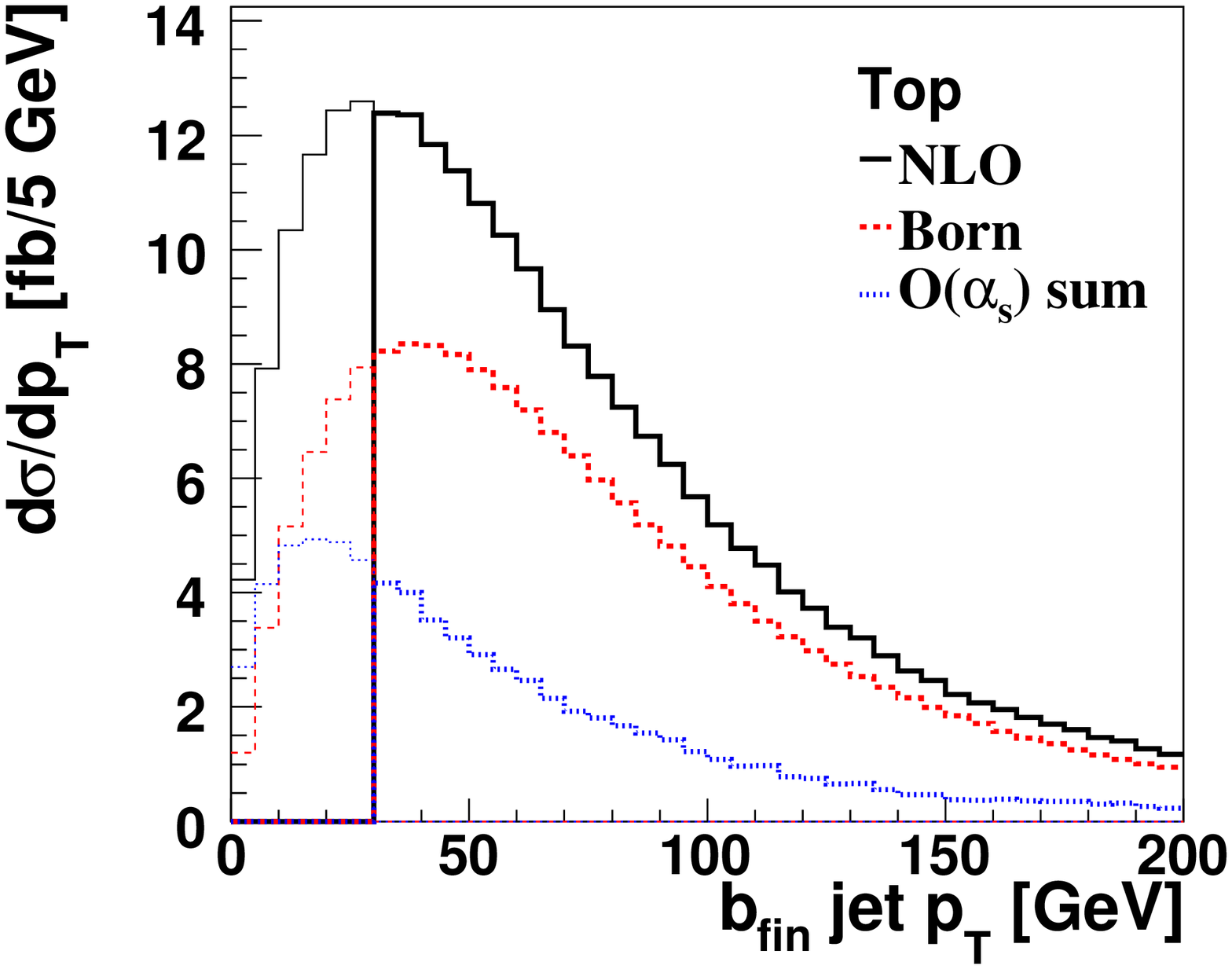}}

\subfigure[]{\includegraphics[scale=0.3]%
{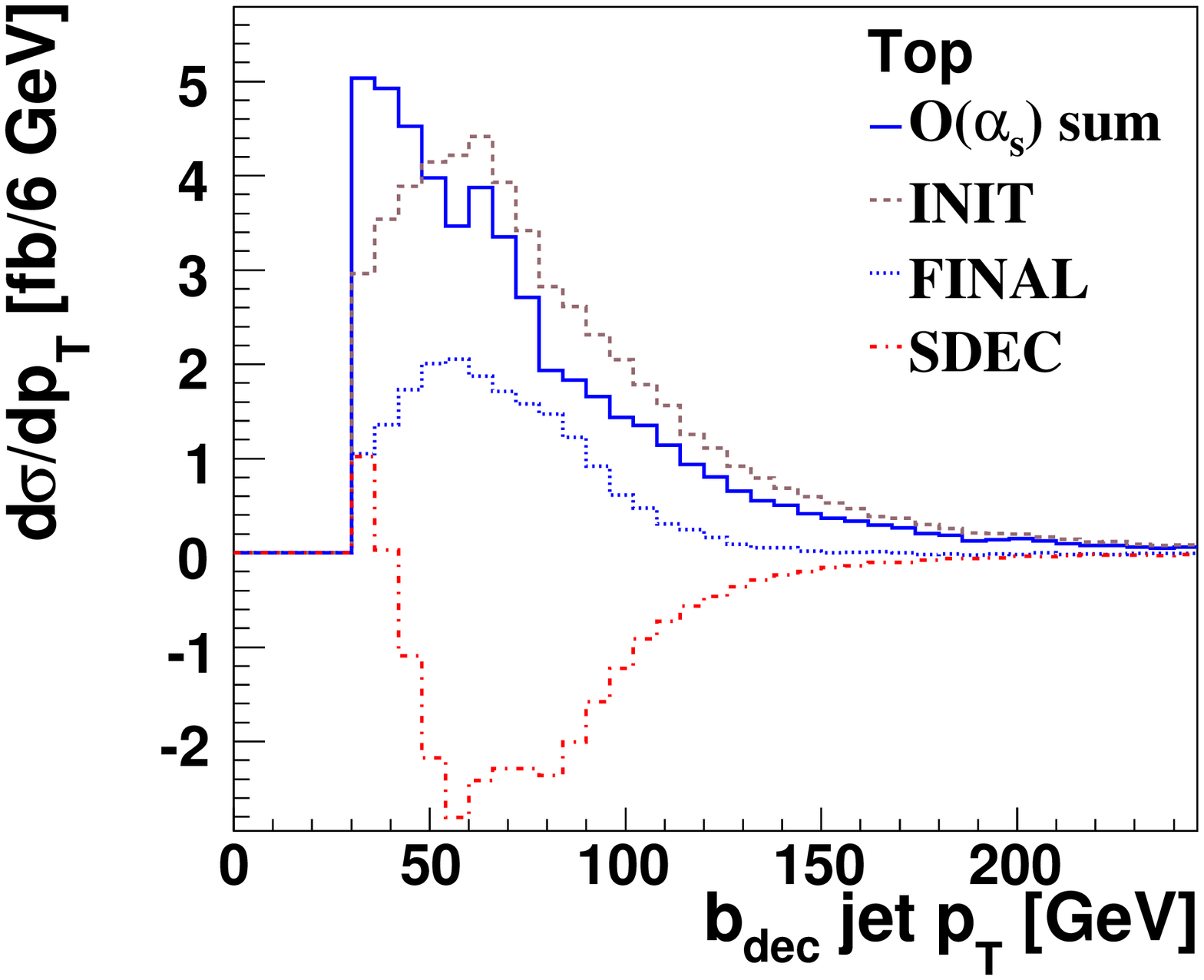}}
\subfigure[]{\includegraphics[scale=0.3]%
{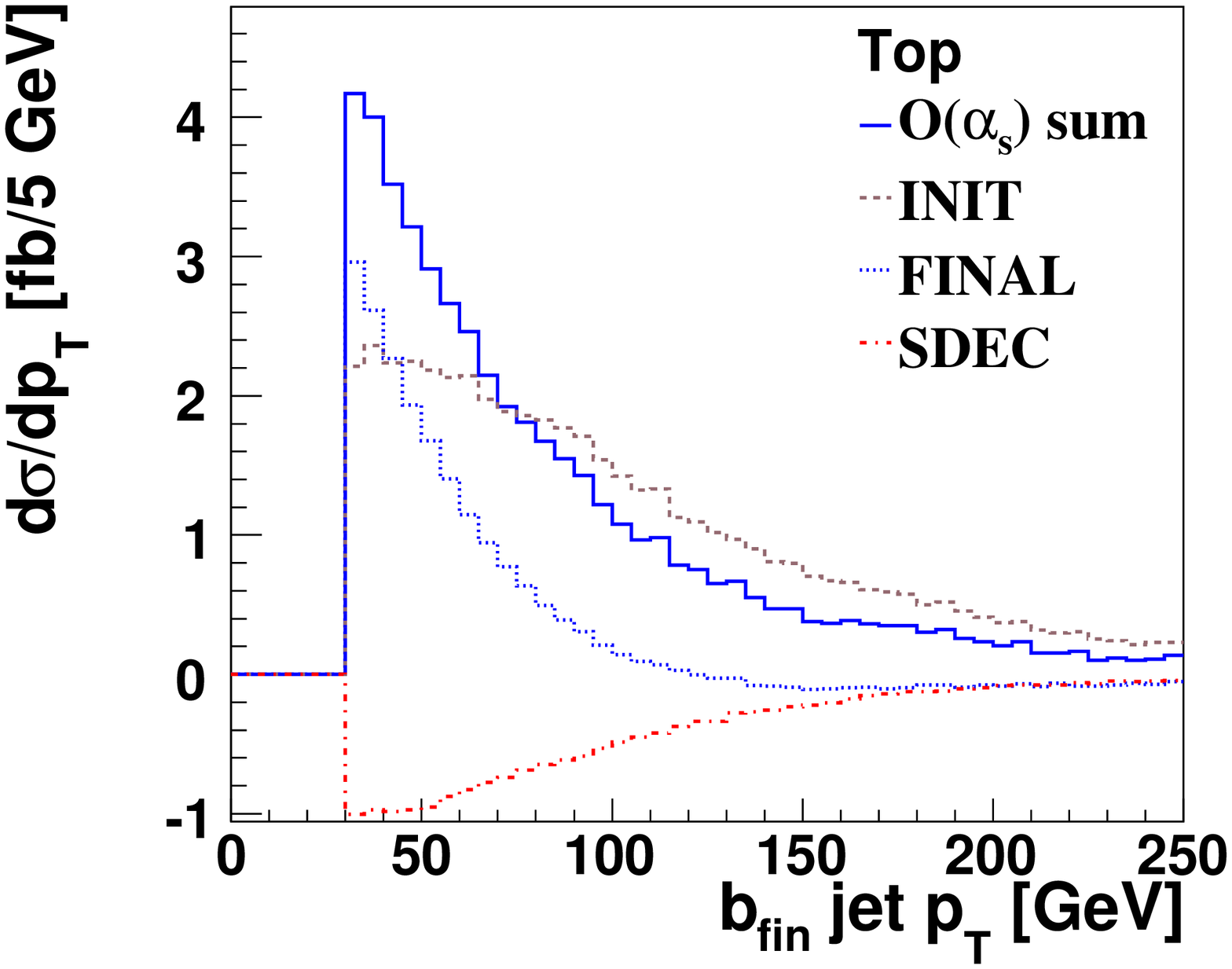}}

\caption{$p_{T}$ distributions for the $b_{dec}$ jet from the top quark decay
(a, c) and the $b_{fin}$ jet (b, d) after selection cuts [(b) is also shown before the $b_{fin}$ jet $p_{T}$ cut]. (a, b) show NLO,
Born and sum of $O(\alpha_{s})$ contributions, (c, d) individual $O(\alpha_{s})$
contributions. \label{fig:bpt}}

\end{figure*}

The comparison of the $p_T$ distributions of the $b_{dec}$ and $b_{fin}$
jets is shown in Fig.~\ref{fig:bpt} for single top quark production. It is experimentally not possible
to distinguish $b$ and $\bar{b}$ jets, but very instructive to consider
their distributions individually. 
The $p_T$ Born distribution of the $b_{dec}$ jet peaks at roughly 1/3 of $m_t$, as it is a top quark decay product, while that 
of the $b_{fin}$ jet peaks at lower $p_{T}$ and has a long tail
to high $p_{T}$ values, because it is produced in association with
the heavy top quark and has to balance the top quark $p_{T}$. 
The peak positions of both $b_{fin}$ and $b_{dec}$ jets are shifted to
a slightly lower value by the QCD corrections. This is mainly due to the 
FINAL (SDEC) contribution because the emitted gluon tends to be collinear to the
$b_{fin}$ ($b_{dec}$) jet in the FINAL (SDEC) correction. 
Furthermore, the NLO distribution is broadened due to the INIT contribution
which adds additional $p_{T}$ to the event by emitting a third jet in the initial
state. Note that among the three $\oalphas$ corrections the INIT contribution
dominates as it receives both soft and collinear enhancements at NLO.
 The soft gluon contributions 
in the FINAL and SDEC contributions are suppressed in comparison
to the INIT correction, due to the large top quark mass.
 The single antitop quark production exhibits very similar 
distributions and is not shown here. 

\begin{figure*}
\subfigure[]{\includegraphics[scale=0.3]%
{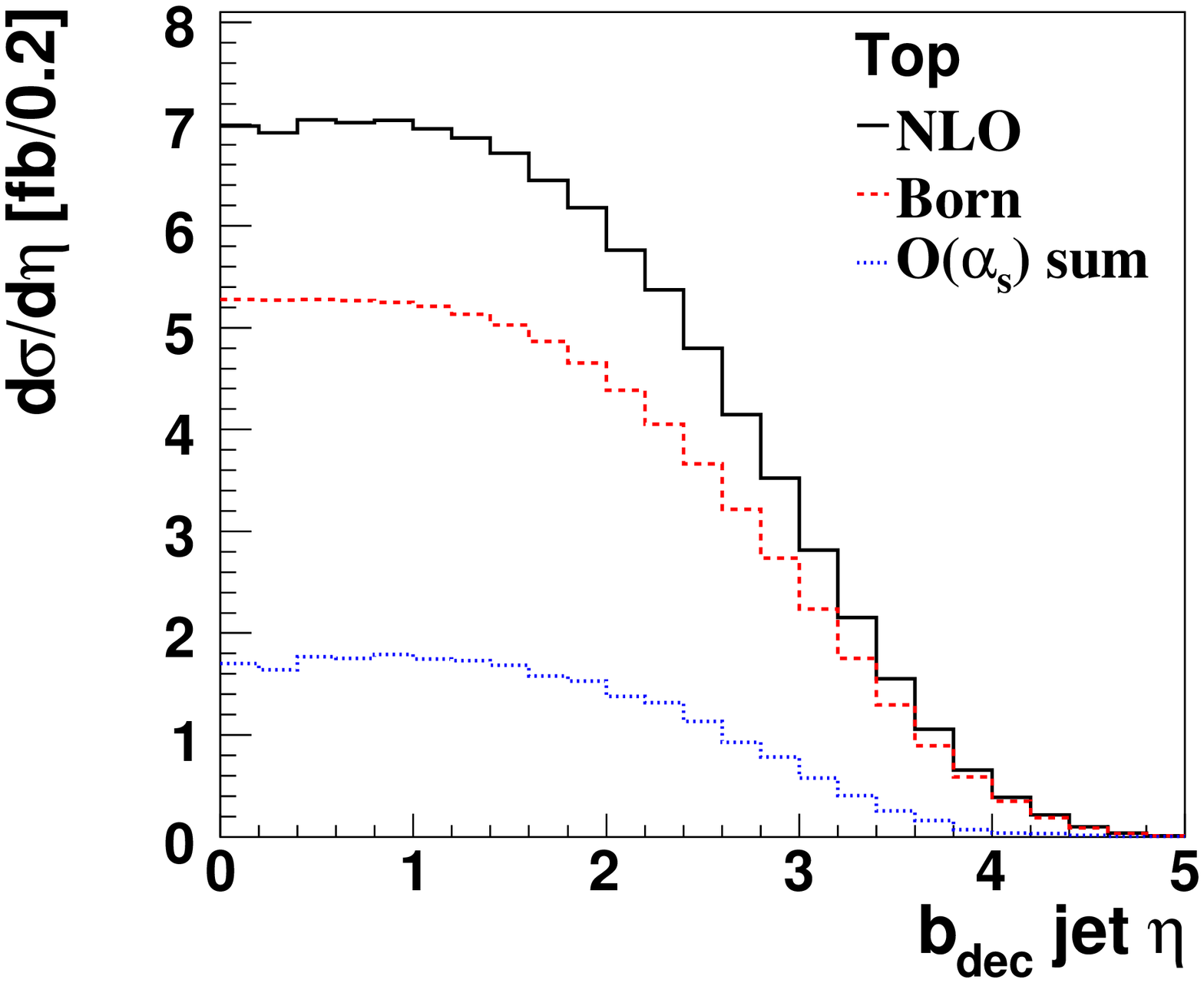}}%
\subfigure[]{\includegraphics[scale=0.3]%
{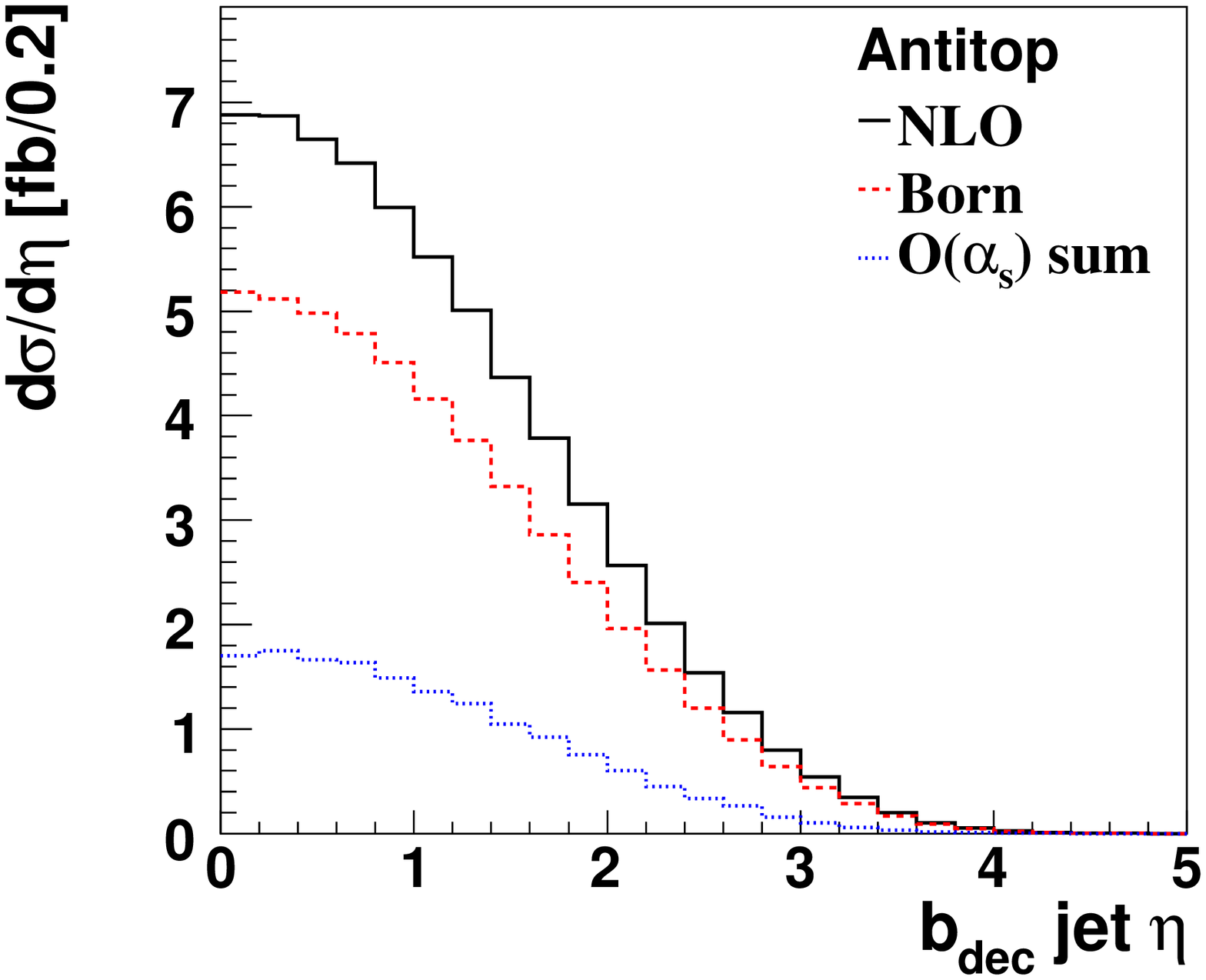}}%

\subfigure[]{\includegraphics[scale=0.3]%
{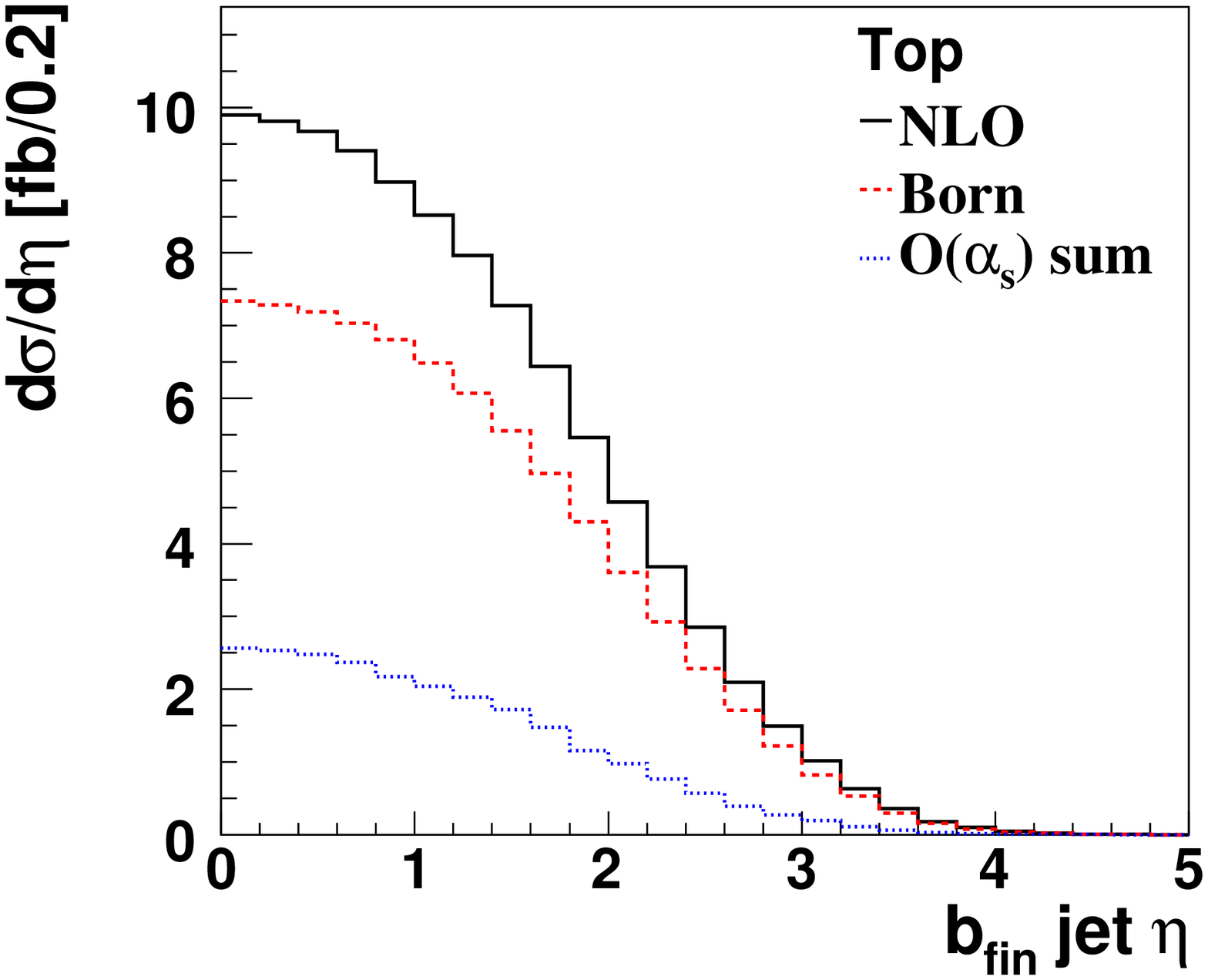}}
\subfigure[]{\includegraphics[scale=0.3]%
{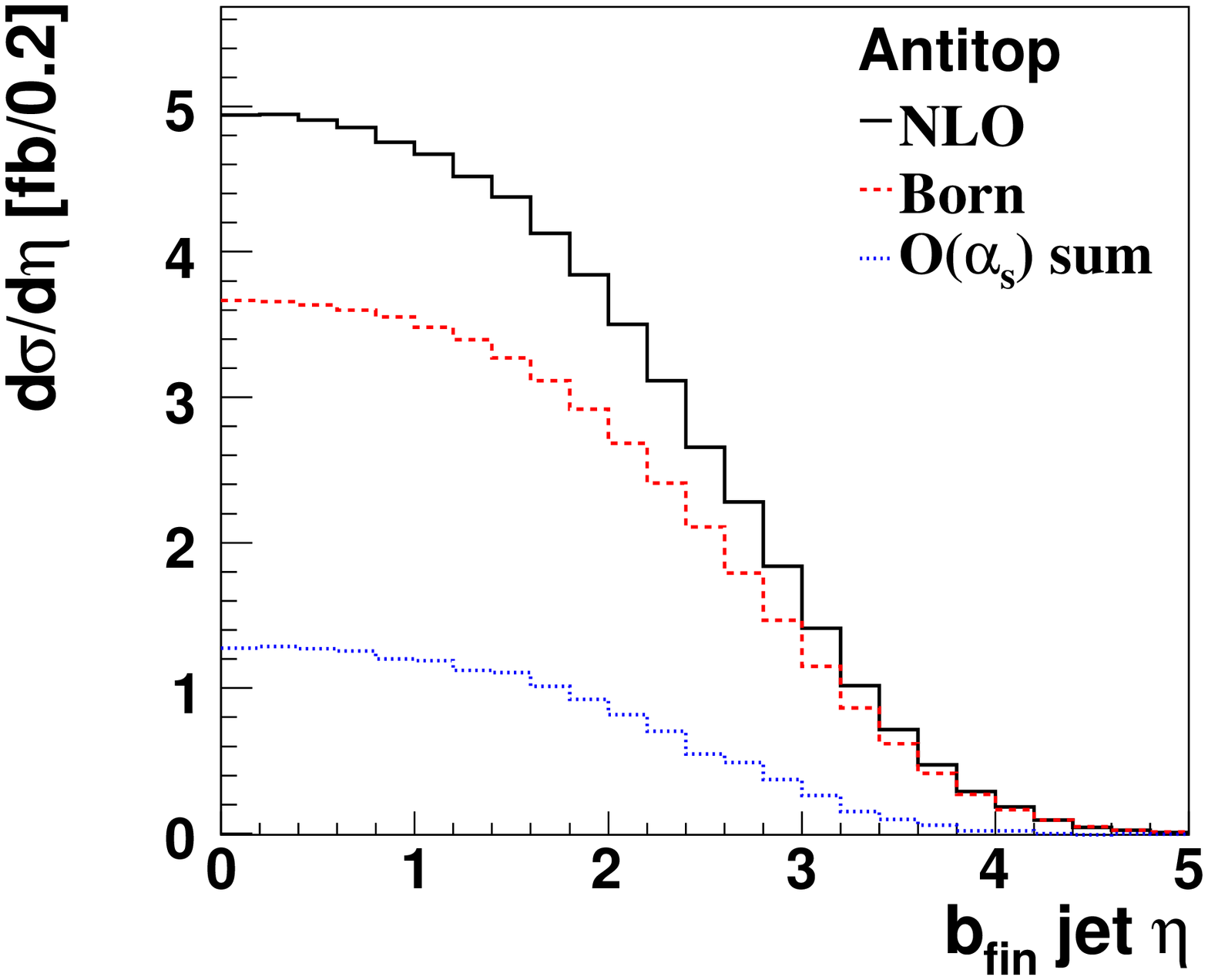}} 

\caption{$\eta$ distributions of the $b_{dec}$ jet from the top/antitop
quark decay (a/b) and the $b_{fin}$ jet produced in association with
the top/antitop quark (c/d) after applying the `loose' cut
set.
\label{fig:beta}}
\end{figure*}

\begin{figure}
\subfigure[]{\includegraphics[scale=0.5]%
{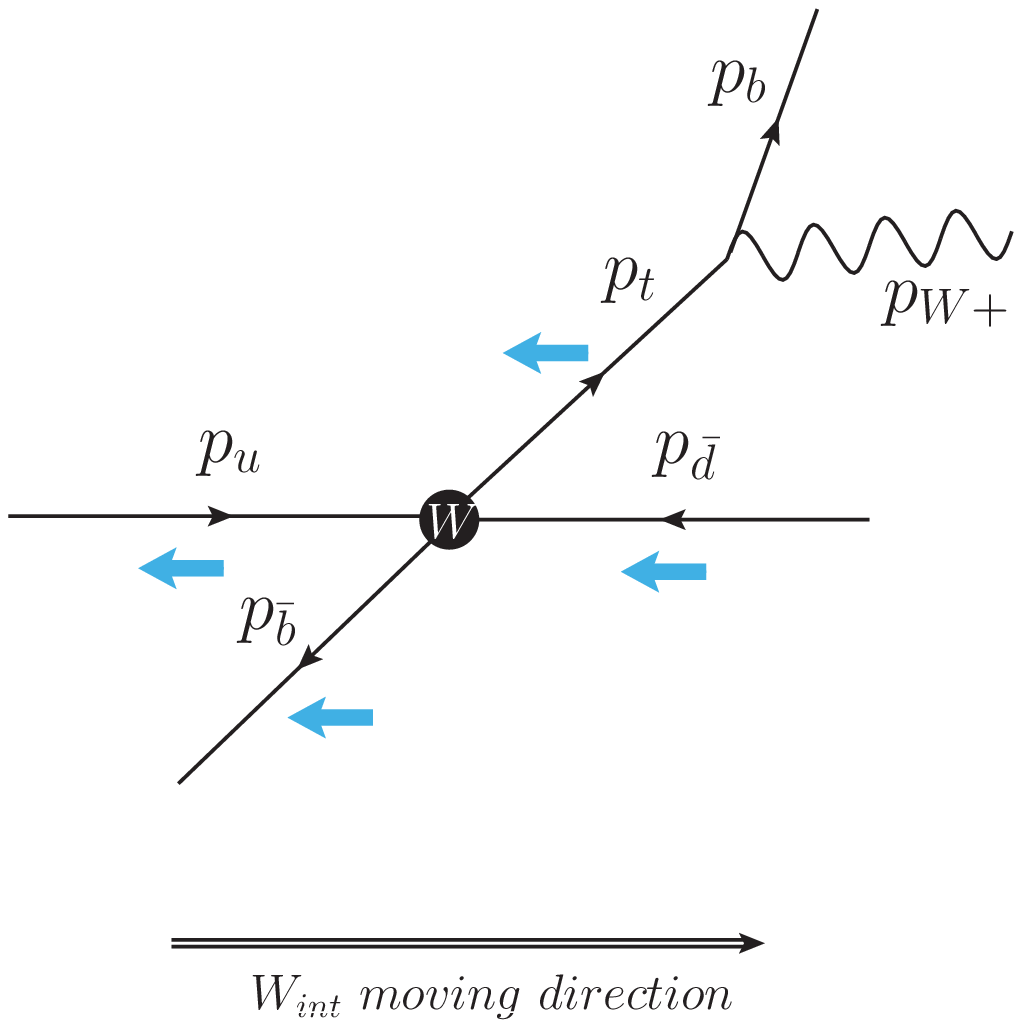}}
\hspace{20mm}
\subfigure[]{\includegraphics[scale=0.5]%
{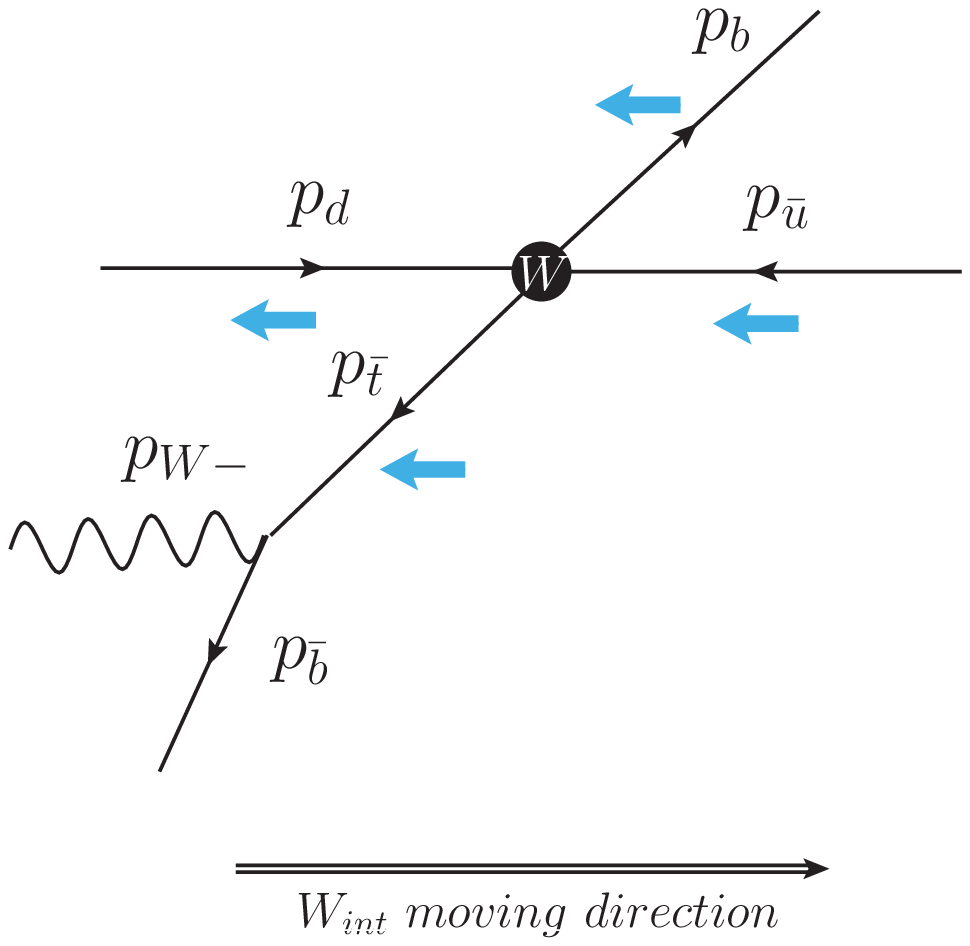}} 
\caption{ Pictorial illustration of spin correlations and boost effects
in top (a) and antitop (b) quark production in the c.m. frame of $W_{int}$. The thin lines denote
 particle momenta while the bold lines label spin directions.  
\label{fig:spinboost}}
\end{figure}
 
Figure~\ref{fig:beta} shows the $\eta$ distributions of the $b_{fin}$
and $b_{dec}$ jet for both single top and antitop quark
production. For top quark production the Born distribution of
the $b_{dec}$ jet $\eta$ is much broader than the Born distribution
of the $b_{fin}$ jet $\eta$. On the other hand, for antitop production 
the $b_{fin}$ jet exhibits a wider $\eta$ distribution than the $b_{dec}$ jet.
This is due to spin correlations and the boost of $W_{int}$ 
along the beam line.
Figure~\ref{fig:spinboost} shows single top and antitop quark production in the c.m. frame of $W_{int}$. 
Due to spin conservation, for top quark production, in the c.m. frame of $W_{int}$, the top quark mainly follows the 
direction of motion of the incoming up quark, 
while the $b_{fin}$ jet [$\bar{b}$ quark in Fig.~\ref{fig:spinboost} (a)] 
follows the direction of motion of the incoming antiquark.
Furthermore, in the top quark production process, $\bar b$ is right-handed and 
$b$ is left-handed due to the left-handed charged current interaction in the 
SM. 
Since the valence quarks carry a larger fraction of the proton momentum,
$W_{int}$ is boosted along the direction of motion of the up quark. The top quark and its decay product, the $b_{dec}$ jet, go in the same direction, and therefore have a larger longitudinal momentum and higher $\eta$ than the $b_{fin}$ jet, which, in the c.m. frame of $W_{int}$, follows the opposite direction and therefore receives a smaller longitudinal boost. 
For the antitop quark production it is the $b_{fin}$ jet [$b$ quark in Fig.~\ref{fig:spinboost} (b)]
and not the antitop quark that, in the c.m. frame of $W_{int}$, follows the direction of motion of the
incoming quark. Therefore the $\eta$ distribution of the $b_{fin}$
jet is wider than the same distribution of the $b_{dec}$ jet from
the antitop quark decay. As mentioned in Sec.~\ref{sub:LepMET},
the boost is smaller for antitop quark production, so the effect on
the $\eta$ width is less strong. 

The $O(\alpha_{s})$ corrections
shift the $b_{fin}$ jet produced in association with the top quark
to more central rapidities. This effect is not very strong and mainly
due to initial state corrections, which add to the $p_{T}$ of the
event while reducing the boost along the beam line direction. The shape of the
$\eta$ distribution of the $b_{dec}$ jet remains practically unchanged
by the $O(\alpha_{s})$ corrections. 

\begin{figure*}
\subfigure[]{\includegraphics[scale=0.3]%
{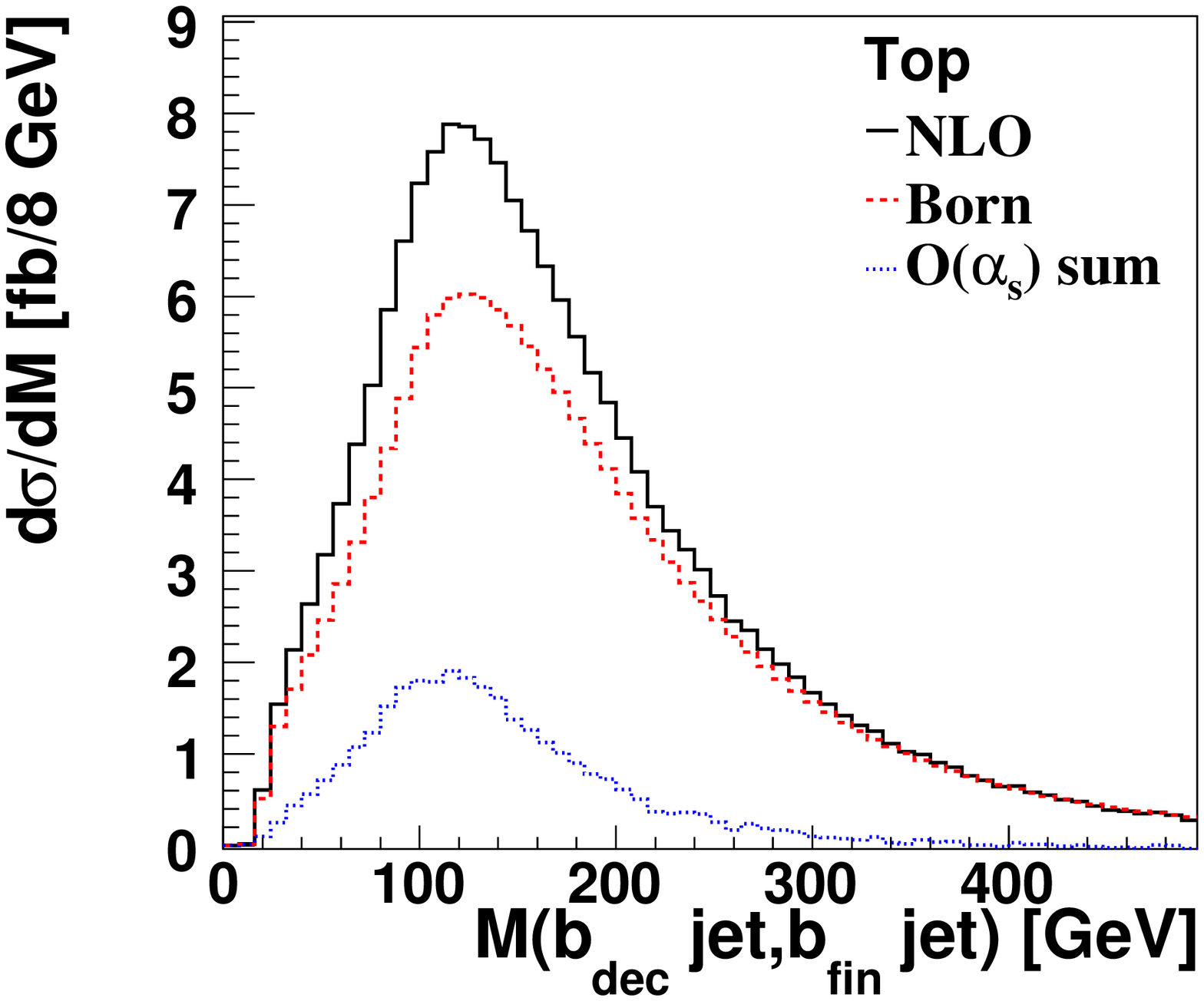}}%
\subfigure[]{\includegraphics[scale=0.3]%
{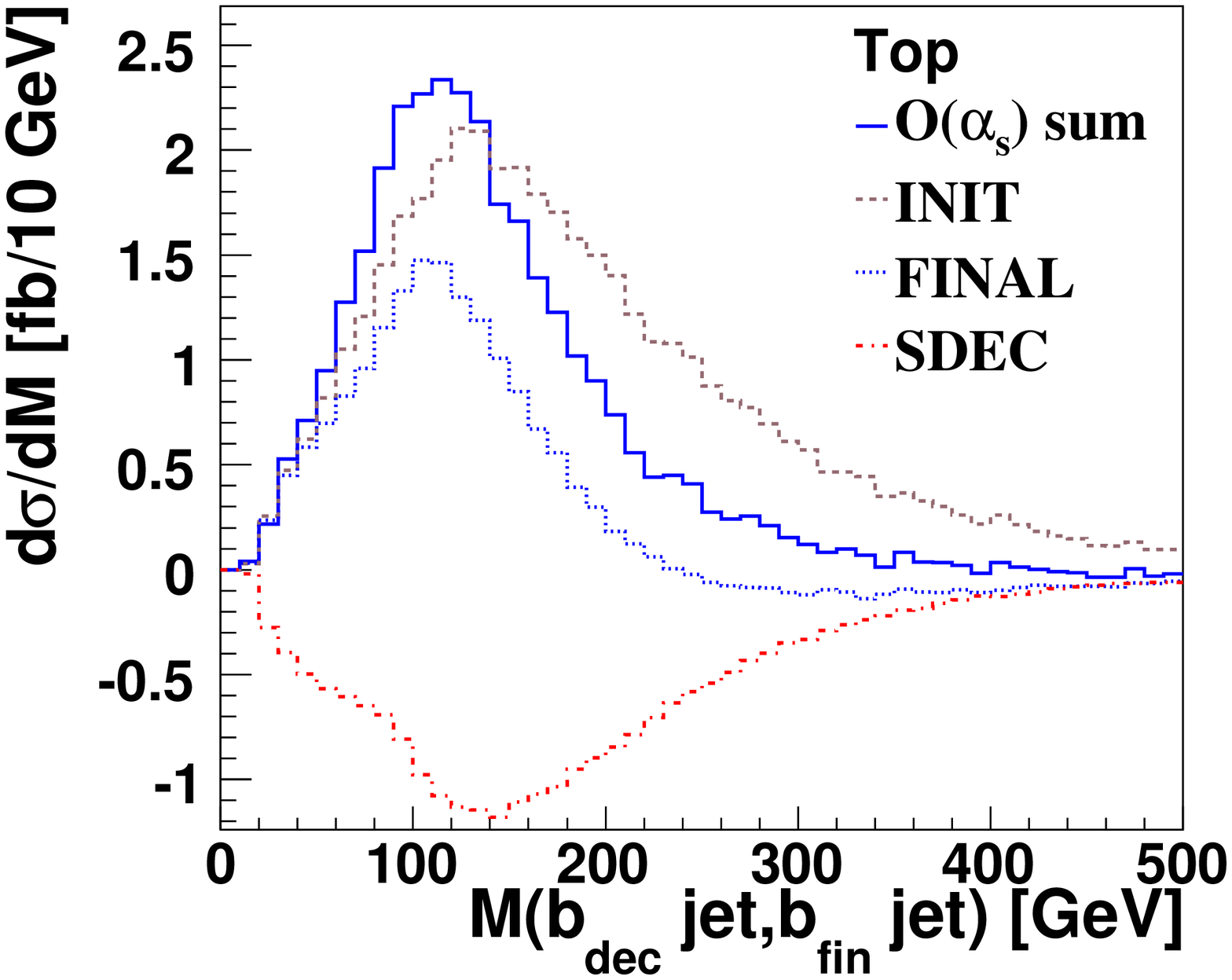}} 

\caption{Invariant mass of the ($b_{dec}$ jet, $b_{fin}$ jet) system after
selection cuts, comparing Born level to $O(\alpha_{s})$ corrections.
\label{fig:bbbarM}}
\end{figure*}

Single top quark production is an irreducible background to the Higgs boson search, 
e.g. $W^{\pm}H$ associated production with the subsequent Higgs boson decay $H\to b\bar{b}$. 
Even though $W^{\pm}H$ production is not the largest production channel at the LHC,
one has to combine its contribution with other channels in order to reach $5\sigma$
statistics significance. Thus it is crucial to have a good understanding 
of the SM backgrounds.
Since a light Higgs boson predominantly decays into a $b\bar{b}$ pair, 
we examine the impact of $\oalphas$ corrections on the invariant mass distribution
of the $b\bar{b}$ pair in single top quark production.
Figure~\ref{fig:bbbarM} plots the invariant mass distribution of the ($b_{dec}$ jet, $b_{fin}$
jet) system. It can be seen that the FINAL correction shifts the peak
of the invariant mass to slightly lower values. This is the case
if a third jet is produced in addition to the $b_{dec}$ jet and the
$b_{fin}$ jet. The INIT corrections tend to have peaks at higher
invariant mass, due to the additional $p_{T}$ they provide. The drop-off
at higher mass values is faster at NLO level. The invariant mass
distribution of the two $b$ jets for antitop quark events looks very
similar in spite of the differences in initial state PDFs.

\subsubsection{Invariant mass M($b_{dec}$ jet, lepton)}

\begin{figure*}
\subfigure[]{\includegraphics[scale=0.3]%
{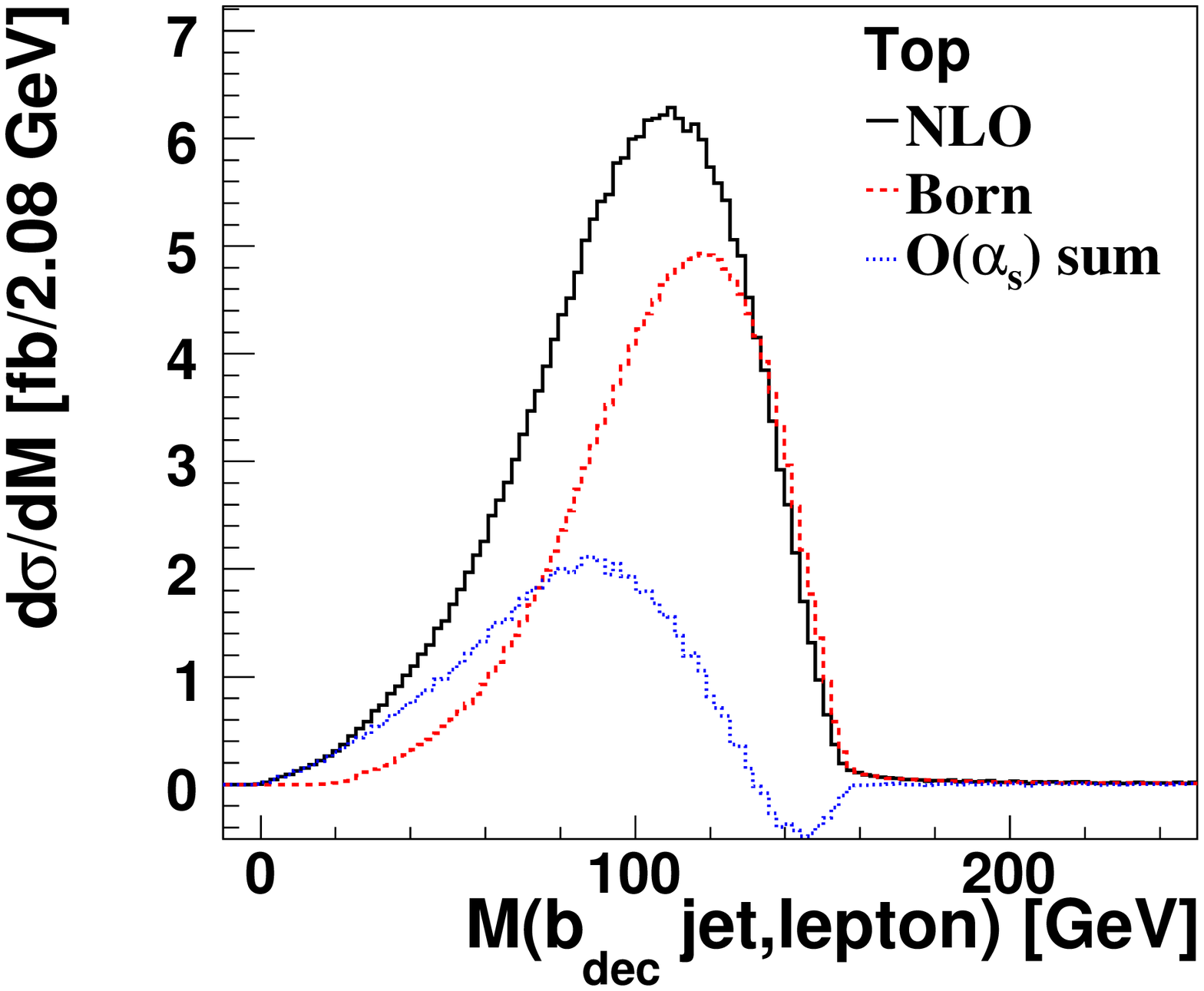}}%
\subfigure[]{\includegraphics[scale=0.3]%
{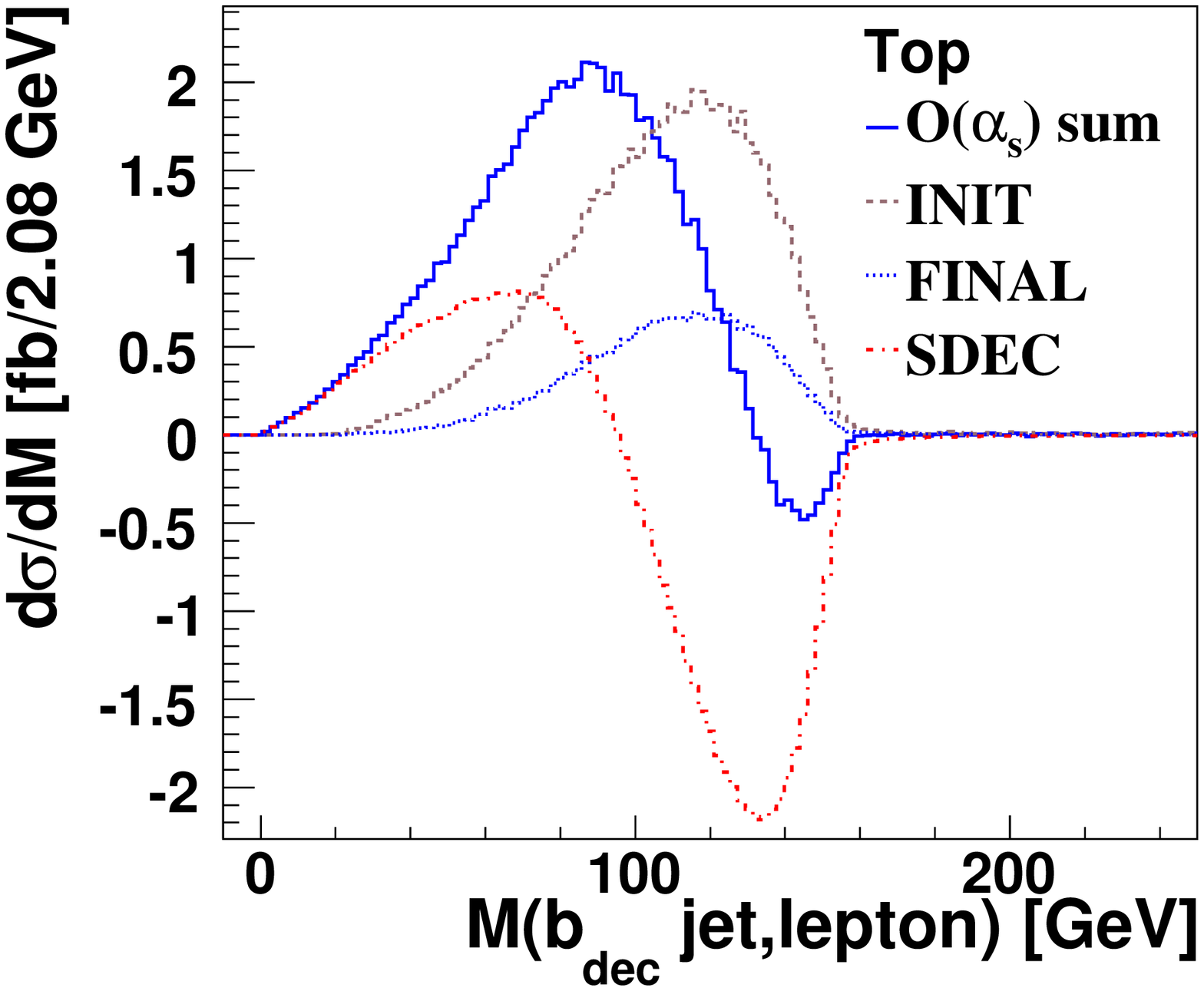}} 

\caption{Invariant mass of the ($b_{dec}$ jet, lepton) system after selection
cuts, comparing Born level to $O(\alpha_{s})$ corrections.\label{fig:beM}}

\end{figure*}

The top quark decays into a $b_{dec}$ jet and a $W$ boson, which
itself can decay leptonically. The invariant mass distribution of
the ($b_{dec}$ jet, lepton) system is characteristic of the decay
of the SM top quark and sensitive to the $W$-$t$-$b$ coupling 
(or the $W$ boson helicity)~\cite{Kao:1993ua,PhysRevD.61.055004},
$$
m_{b_{jet}\ell}^2 \approx \frac{1}{2}\left(m_t^2-m_W^2\right)
\left(1-\cos\theta^{\star}_{\ell}\right)
$$
where $\theta_{\ell}^{\star}$ is the polar angle of the charged lepton
in the rest frame of the $W$ boson which is defined in the rest frame of the top quark.
Figure~\ref{fig:beM} shows the drop-off behavior of the invariant
mass distribution at $\sqrt{m_{t}^{2}-m_{W}^{2}}~\approx~$155 GeV,
due to the kinematics of the event. It also shows that the SDEC correction
shifts the invariant mass peak to lower values, mainly because it
weakens the spin correlations. 

\subsubsection{$H_{T}$ distributions}

\begin{figure*}
\subfigure[]{\includegraphics[scale=0.3]%
{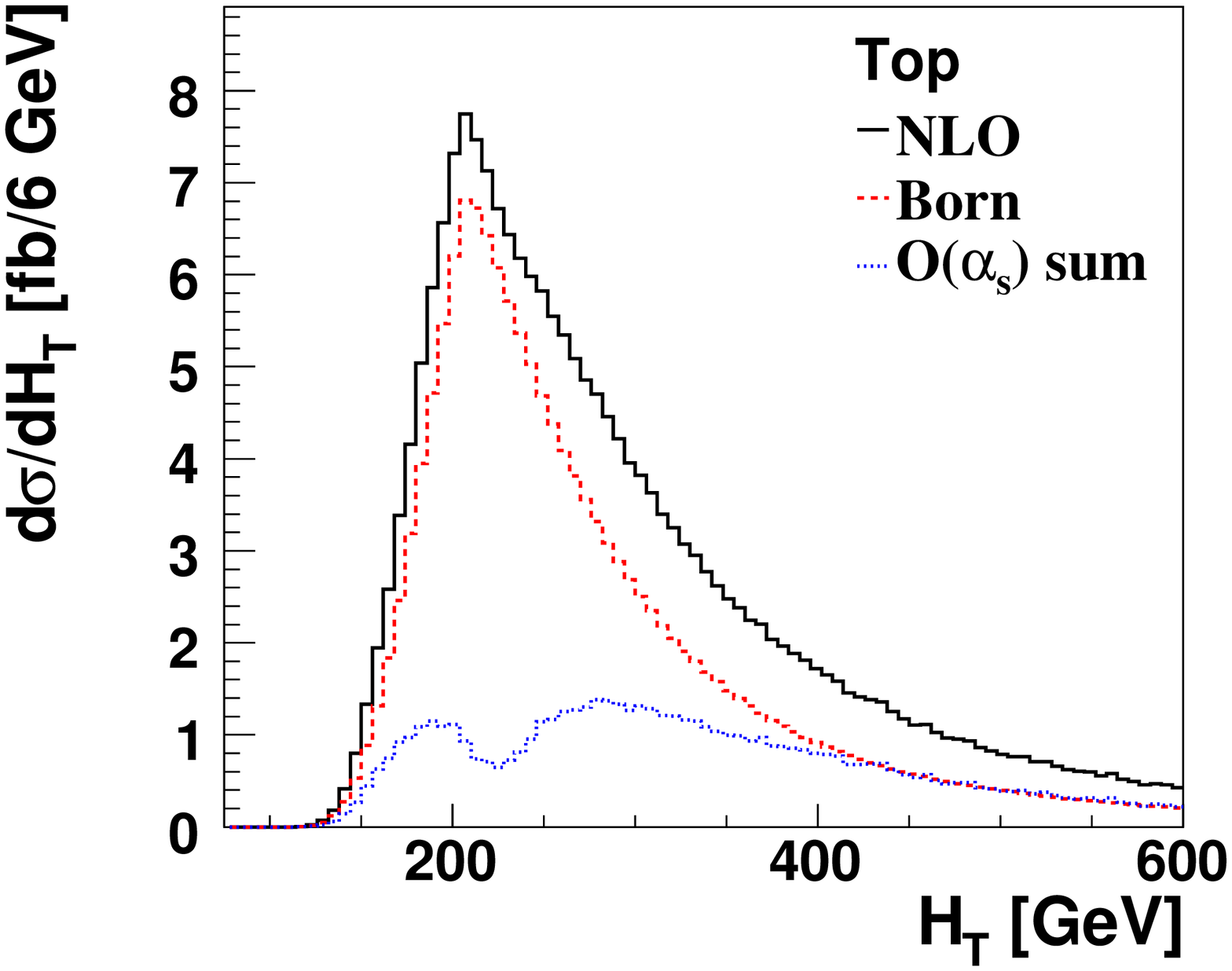}}%
\subfigure[]{\includegraphics[scale=0.3]%
{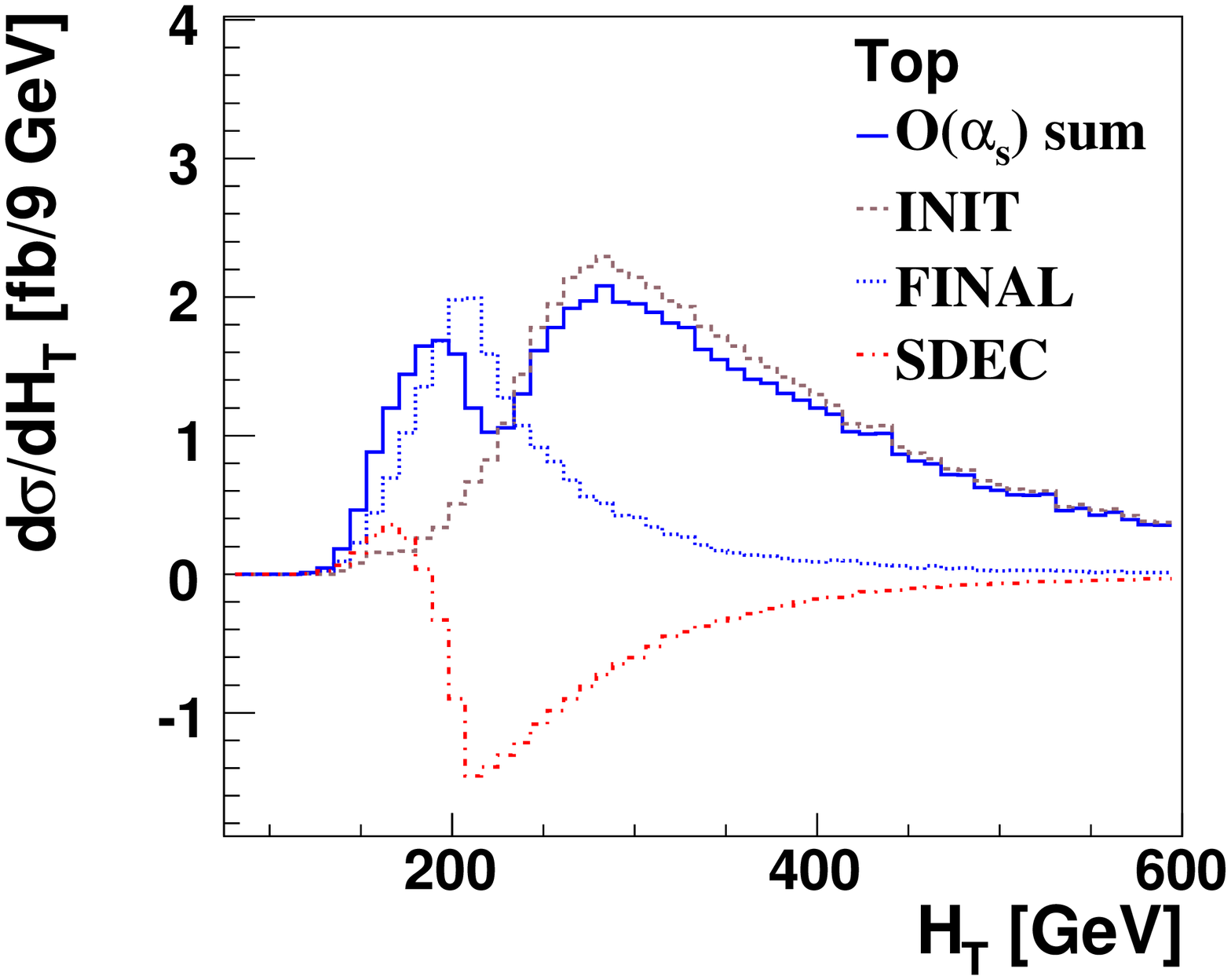}}

\caption{Comparison of the total transverse energy of single top quark events
after selection cuts between LO and NLO (a) and between the different
$O(\alpha_{s})$ corrections (b). 
\label{fig:ht}}
\end{figure*}

In order to distinguish the s-channel single top quark mode from dominant
backgrounds, it is important to know how the total transverse energy
$H_{T}$ changes with NLO corrections. In Fig.~\ref{fig:ht} we look
at the effect that $O(\alpha_{s})$ corrections have on the total
transverse energy $H_{T}$ of a single top quark event. 
$H_{T}$ is defined as
\begin{equation}
H_{T}=p_{T}^{lepton}+\met+\sum_{jets}p_{T}^{jet}.
\label{eq:HT}
\end{equation}
Clearly, the SDEC and FINAL $O(\alpha_{s})$ contributions shift
the total transverse energy down, while the INIT contribution shifts
it up. This again is due to the additional transverse energy that
a third jet in the initial state adds to the event. The shift to higher transverse energies is significant for single top quark measurements at the LHC, 
as $t\bar{t}$ events are generally a larger background for higher $H_T$ values.

\subsection{Event reconstruction}

One of the main reasons for studying single top quark events is 
to find out more about the properties of the top quark and its couplings. 
Furthermore, as single top quark production is
a weak interaction process, we expect a number of correlations between
the particles. 
It is therefore of interest to reconstruct the complete event,
including final state jets and intermediate particles.     

As explained in Refs.~\cite{Kane:1989vv,Cao:2004ap}, it is possible
to reconstruct the $W$ boson from the observed charged lepton and $\met$, 
where the unknown longitudinal momentum of the neutrino
$p_{z}^{\nu}$ is substituted by the restriction that the invariant
mass of the (charged lepton, neutrino) system has to equal the mass of the
$W$ boson. Of the two possible solutions we pick the one with the
smaller $\left|p_{z}^{\nu}\right|$. 
For the top quark reconstruction, we must then combine the reconstructed
$W$ boson with the $b_{dec}$ jet from the top quark decay. This
means we have to identify the correct jet as the $b_{dec}$ jet. There
are several different methods to select this jet. As both the $b_{dec}$ jet
and the $b_{fin}$ jet have high $p_{T}$ and are possibly b-tagged,
neither the method of choosing the leading jet (jet with the highest
$p_{T}$) nor the second jet (jet with the second highest $p_{T}$)
nor the b-tagged jet is very reliable in identifying the right jet.

A more effective algorithm is the so-called best-jet algorithm as
explained in Ref.~\cite{Abbott:2000pa} and more specifically for
our case in Ref.~\cite{Cao:2004ap}. Here, the $Wj$ or the $Wjj$
(only possible in cases where there is a third jet) combination, that
gives an invariant mass closest to the input $m_t$, is chosen
as the reconstructed top quark and the jet (or two-jet system), that
is thus identified, is called best-jet. Of the two possible other
jets, the one with larger $p_T$ is the so-called non-best-jet.

\begin{figure}
\includegraphics[scale=0.3]%
{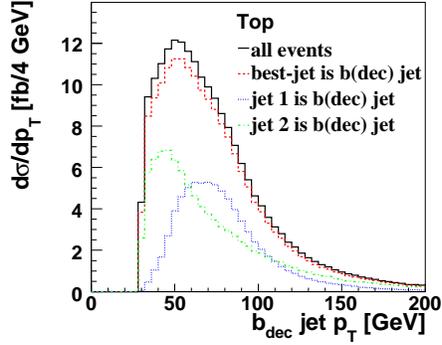} 

\caption{$p_{T}$ distribution of the $b_{dec}$ jet from the top quark decay,
for all events that pass the `loose' set of cuts (solid), only for those
events in which the $b_{dec}$ jet is the best-jet (dashed), only
for events in which the $b_{dec}$ jet is also the leading jet (dotted)
and only for events in which the $b_{dec}$ jet is the second jet
(dash dotted). 
\label{fig:beff}}
\end{figure}

Figure~\ref{fig:beff} shows the efficiencies of the best-jet, the
leading jet (jet 1) and the second jet (jet 2) algorithms. The leading
jet has an overall efficiency of 39\%. The second jet mainly corresponds
to the $b_{dec}$ jet for very high $p_{T}$ and has an overall efficiency
of about 49\%, while the best-jet algorithm has a high efficiency
for all momenta and identifies the $b_{dec}$ jet correctly in 93\% of all
events. Its effectiveness is mostly limited by the efficiency of the
$W$ boson identification; with a falsely reconstructed $W$ boson,
the identification of the $b_{dec}$ jet by the best-jet algorithm becomes
a random pick.

\begin{figure}
\includegraphics[scale=0.3]%
{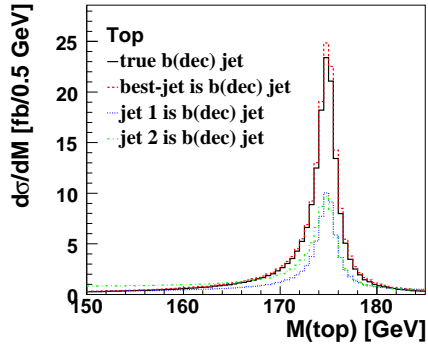} 

\caption{Invariant mass of the $Wj$ combination after selection cuts, where the jet $j$ is the real $b_{dec}$
jet (solid), the best-jet (dashed), the leading jet (dotted) and the
second jet (dash dotted). 
\label{fig:mtcomp}}
\end{figure}

The invariant mass of the $Wj$ combination is shown in Fig.~\ref{fig:mtcomp},
comparing the reconstruction using the best-jet, the leading jet and
the second jet algorithm to the reconstruction using the real $b_{dec}$
jet. For the identification of the real $b_{dec}$ jet, parton level
information is used and a possible third jet from the top quark decay is included. For all four curves, the $W$ boson is reconstructed
from the final state lepton and $\met$, so that all differences in shape
and height are due to the method of identifying the $b_{dec}$ jet.
As expected, the distribution using the best-jet fits the distribution
with the true $b_{dec}$ jet information much better than the distribution
using the leading jet or the second jet. It is important to notice
though, that this is due to two competing effects: (i) The best-jet
does not identify the correct $b_{dec}$ jet in 7\% of all events,
which reduces the height of the distribution in Fig.~\ref{fig:beff};
(ii) Due to the requirement to be as close as possible to the input $m_t$, 
mis-reconstructed events are shifted closer to 175 GeV, increasing the height of the
peak. The second effect is larger than the first, so that the peak
reconstructed with the best-jet algorithm is actually higher than
the peak reconstructed using the real $b_{dec}$ jet. 

\begin{figure*}
\subfigure[]{\includegraphics[scale=0.3]%
{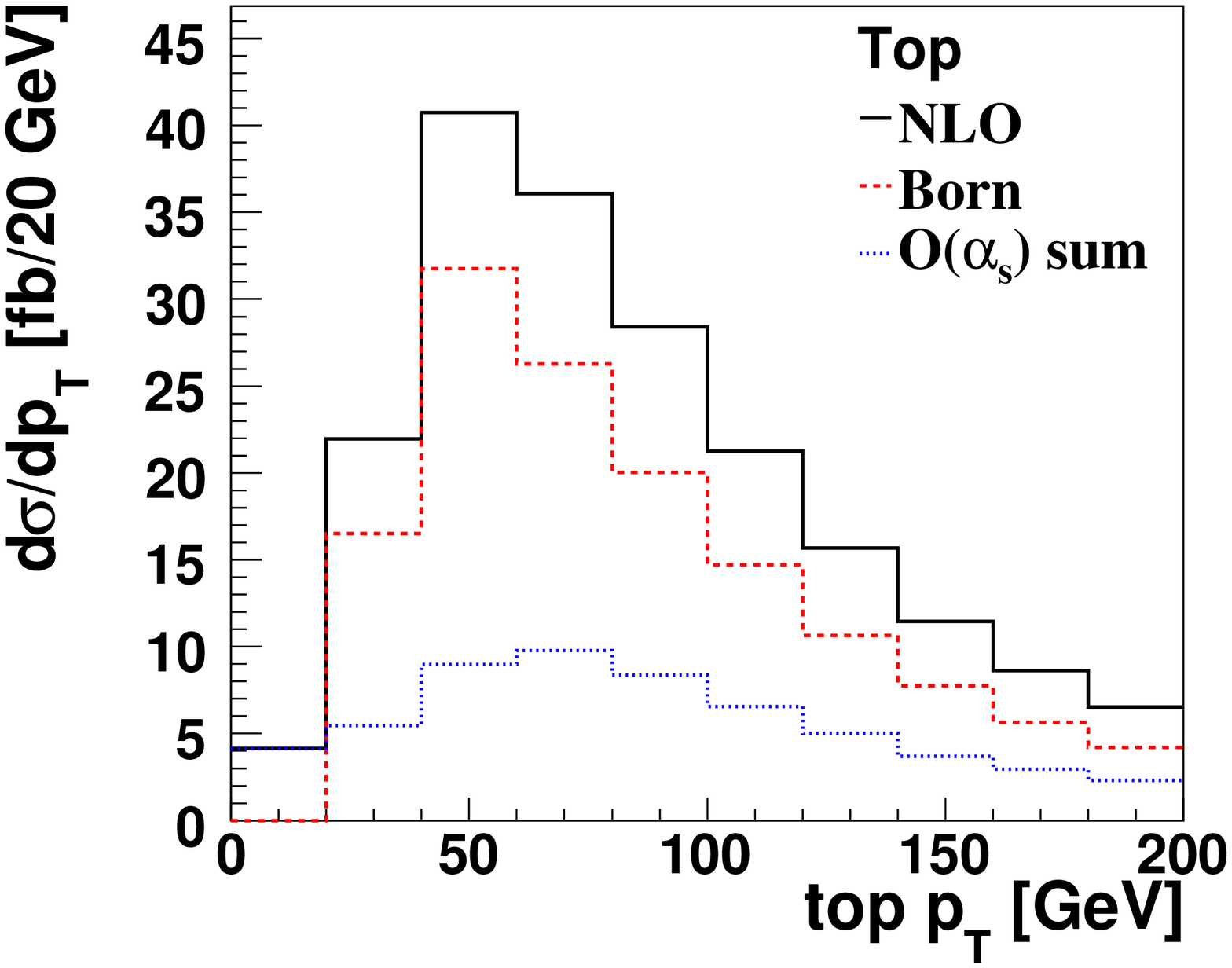}}%
\subfigure[]{\includegraphics[scale=0.3]%
{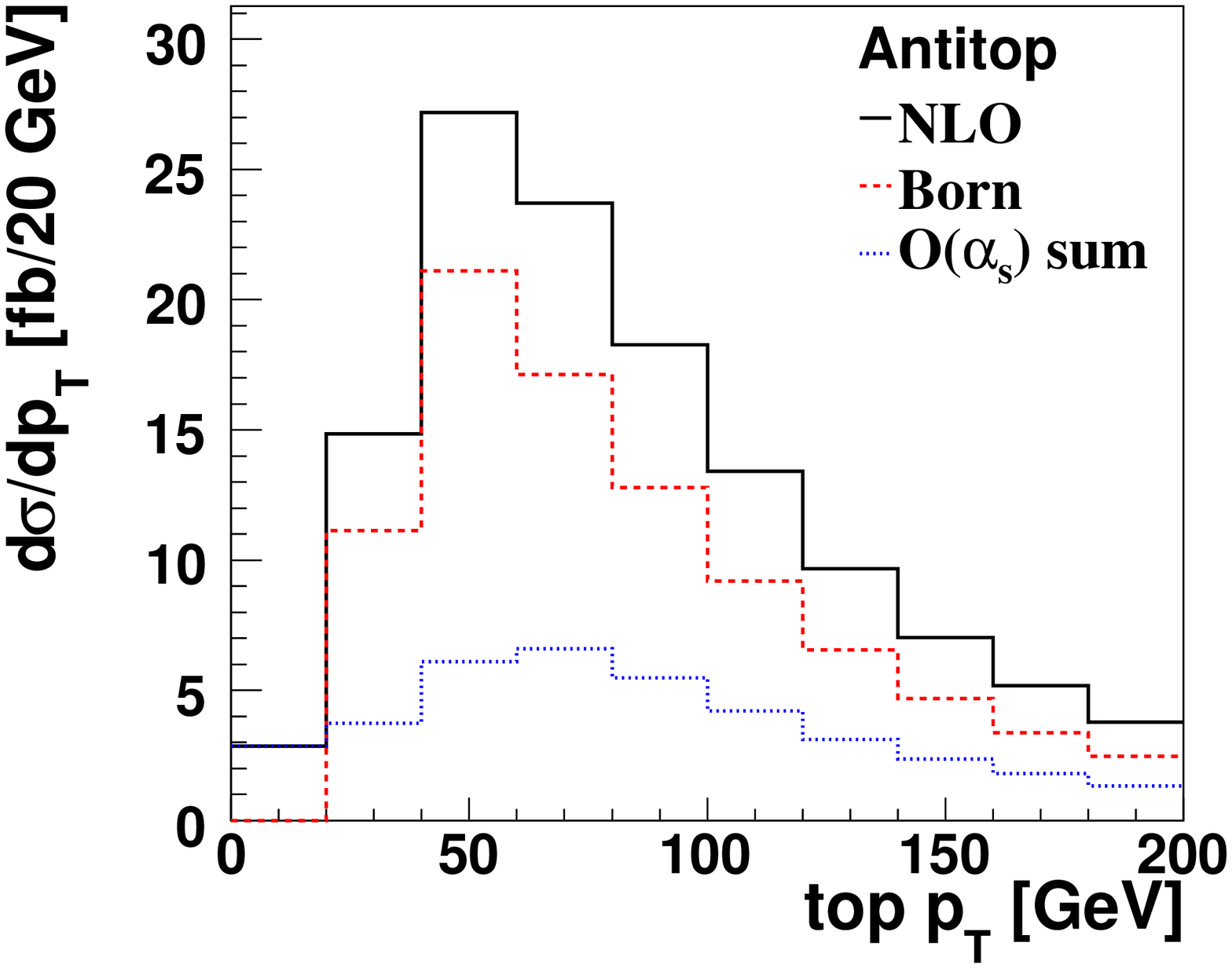}}%

\subfigure[]{\includegraphics[scale=0.3]%
{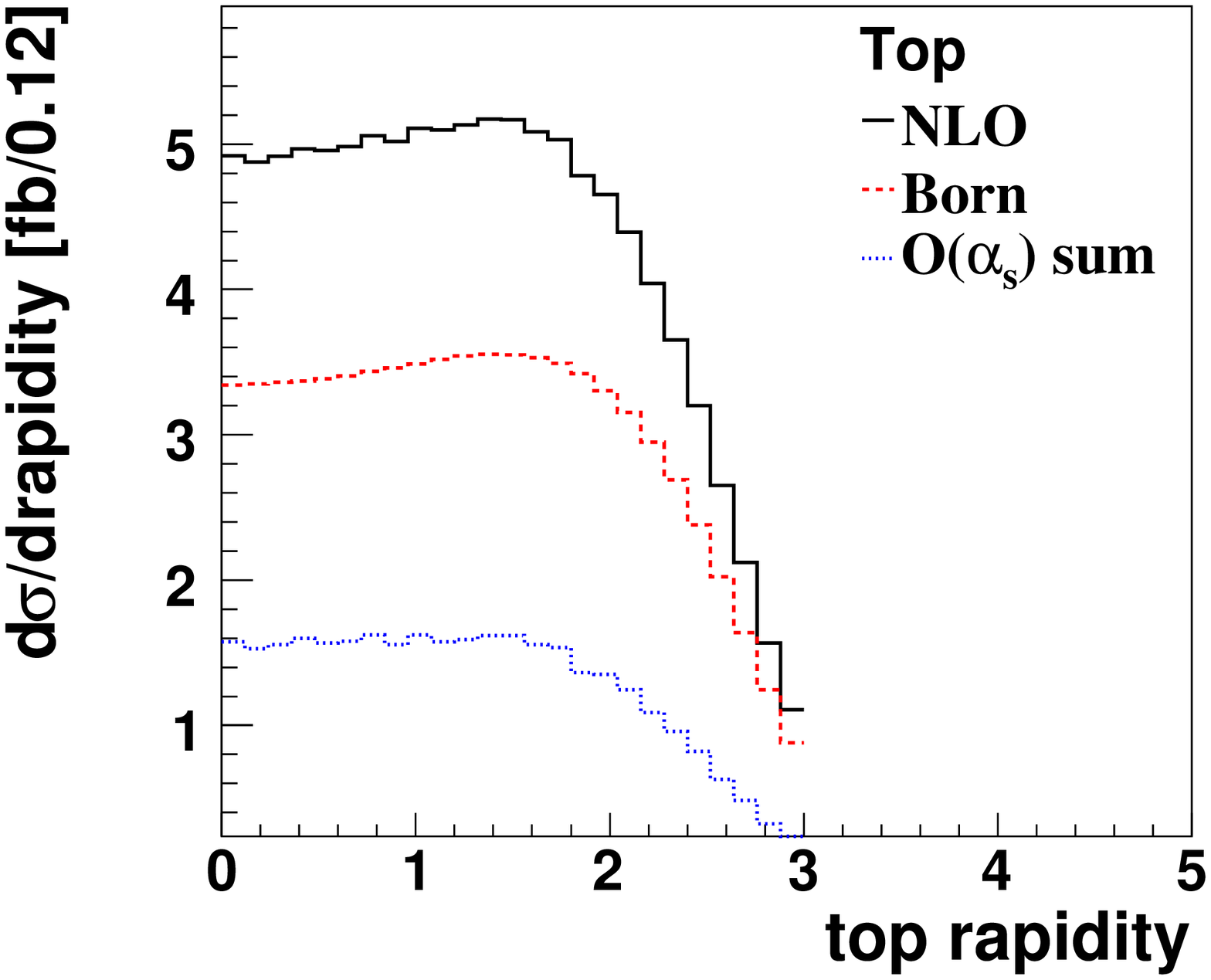}}
\subfigure[]{\includegraphics[scale=0.3]%
{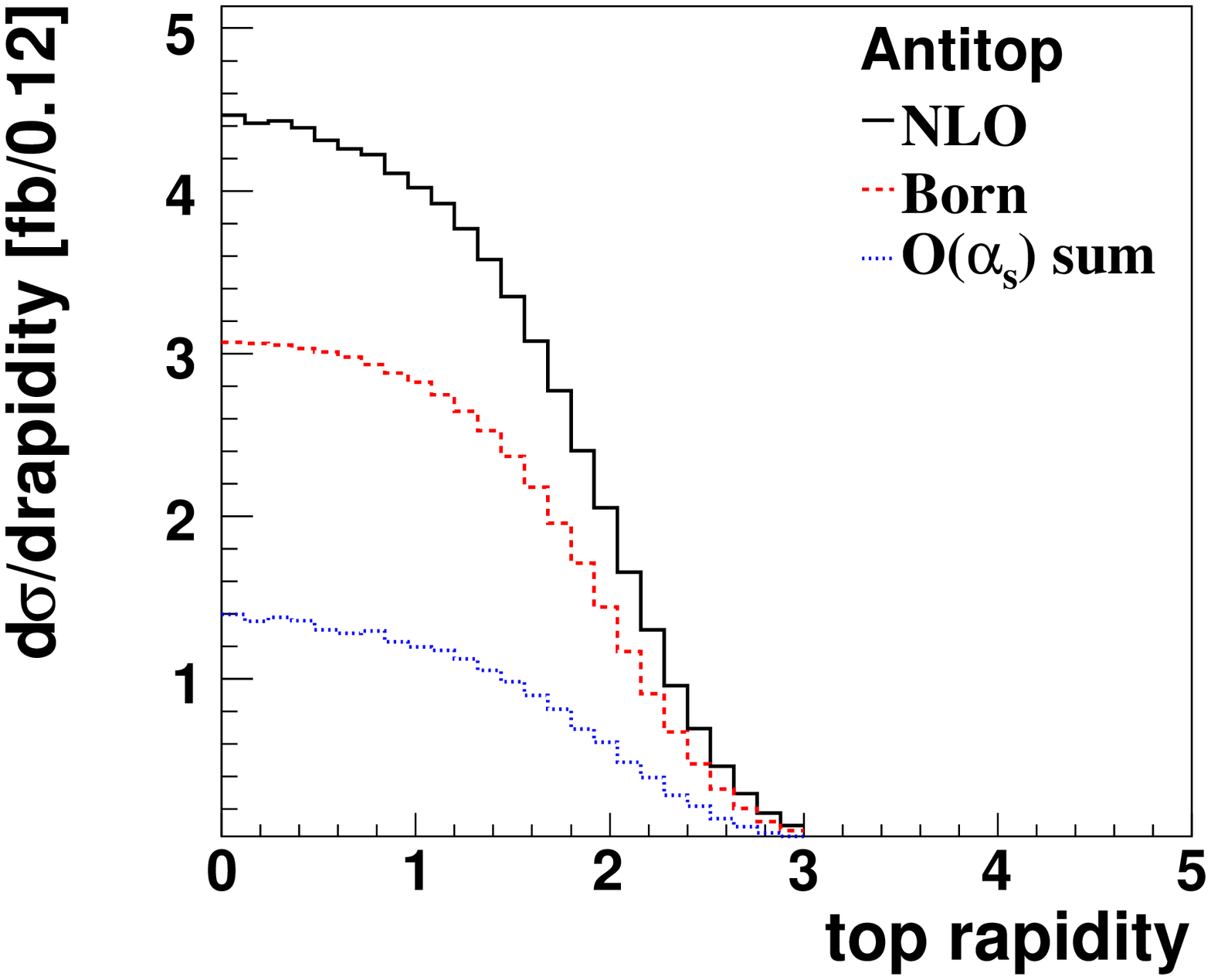}} 

\caption{Distributions of $p_{T}$ (a, b) and rapidity (c, d) of the top quark
(a, c) and of the antitop quark (b, d), both reconstructed with the
best-jet algorithm and after applying the `loose' set of selection cuts.
\label{fig:toppt}}
\end{figure*}

We reconstruct the top quark using the best-jet algorithm and are
now able to study some of its kinematic properties. In Fig.~\ref{fig:toppt},
the $p_{T}$ and rapidity distributions are shown for both top and antitop quark, reconstructed with the best-jet algorithm
and after applying the `loose' set of selection cuts. The shapes of
the $p_{T}$ distributions look similar, but the rapidity distribution
is wider for the top quark than for the antitop quark. This is a reflection
of the boosted kinematics of $W_{int}$ and the (anti)top quark itself (cf. Fig.~\ref{fig:spinboost}).
The NLO corrections do not significantly change the the shape of the
distributions.

\begin{figure*}
\subfigure[]{\includegraphics[scale=0.3]%
{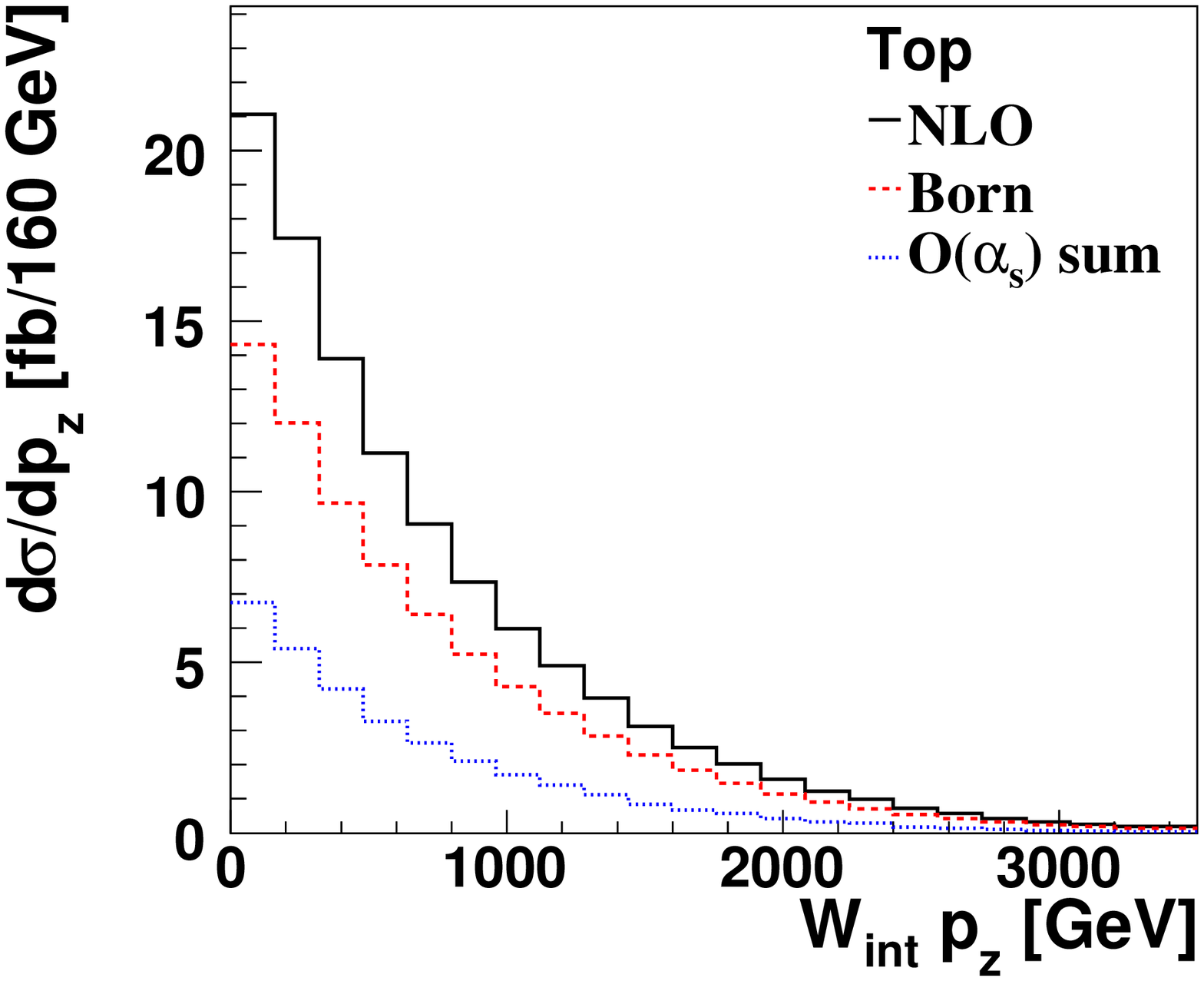}}%
\subfigure[]{\includegraphics[scale=0.3]%
{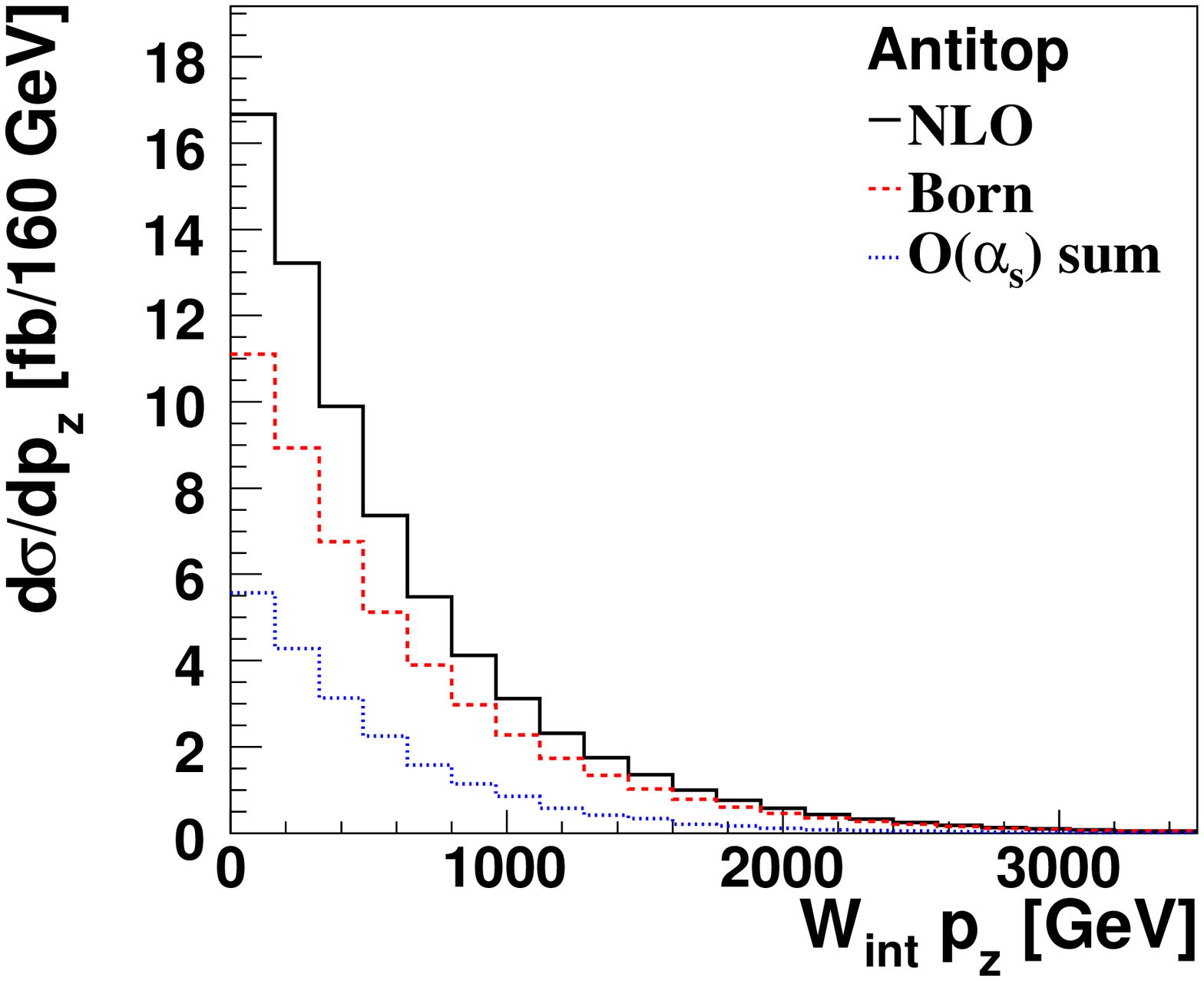}}
\subfigure[]{\includegraphics[scale=0.3]%
{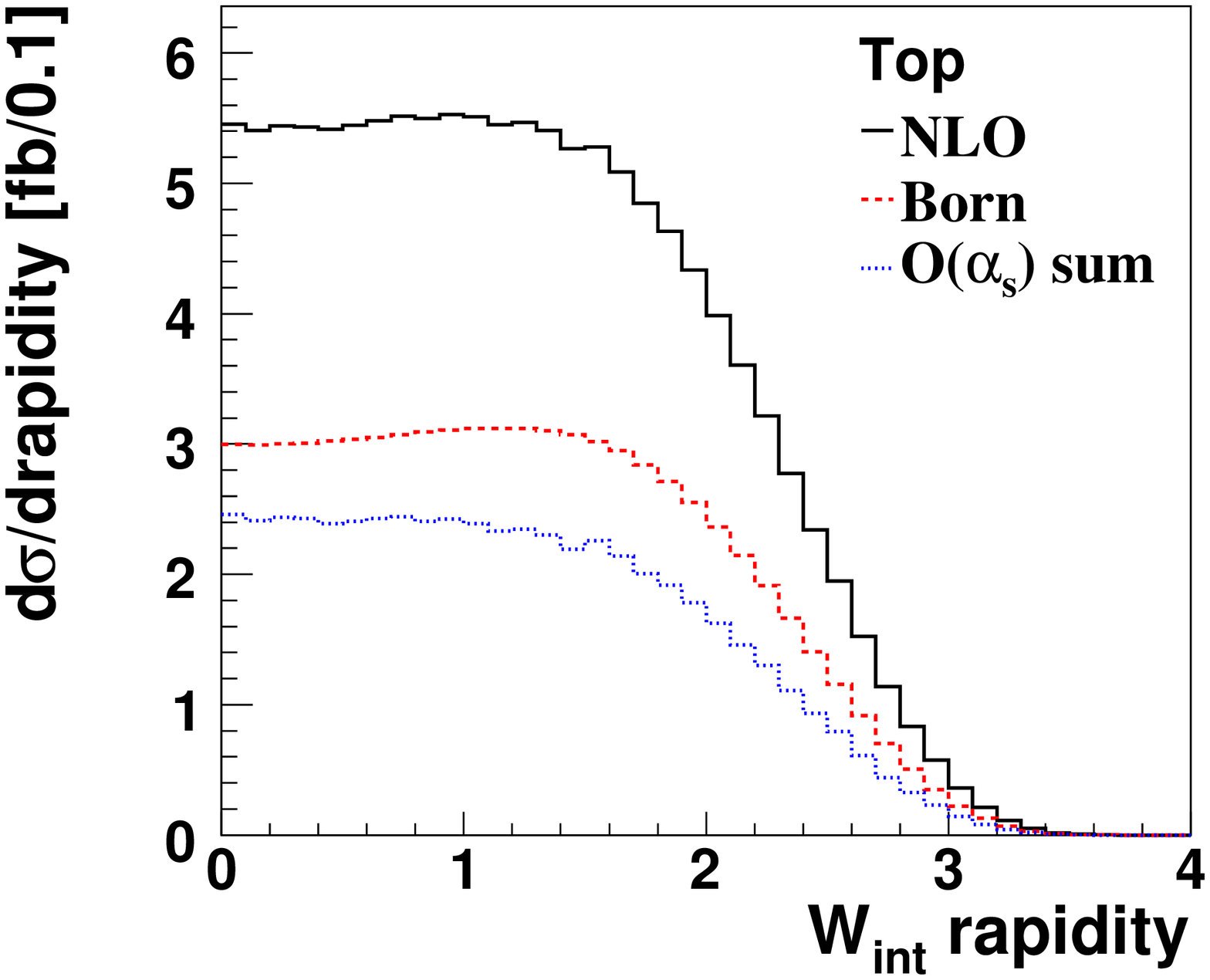}}%
\subfigure[]{\includegraphics[scale=0.3]%
{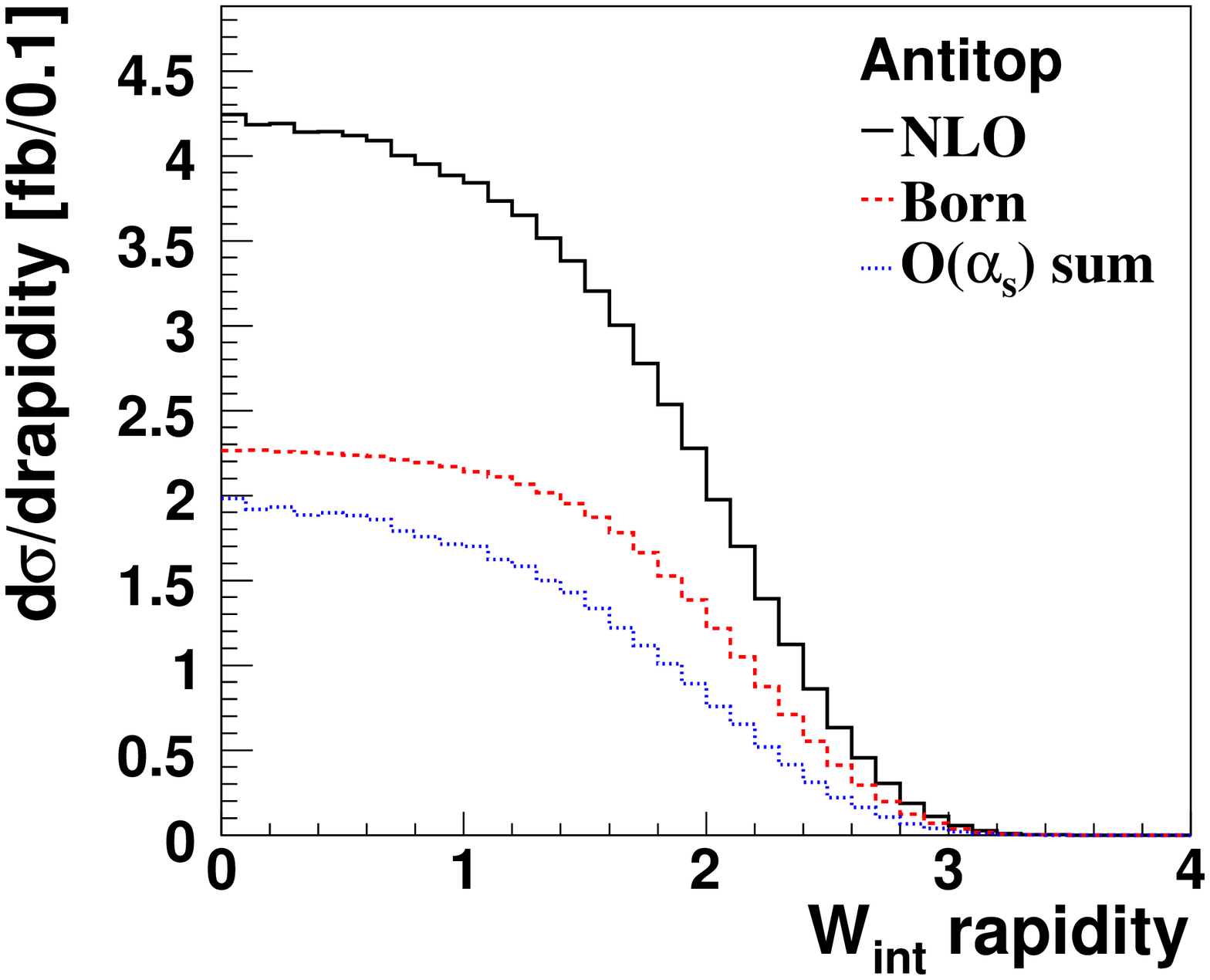}}
\subfigure[]{\includegraphics[scale=0.3]%
{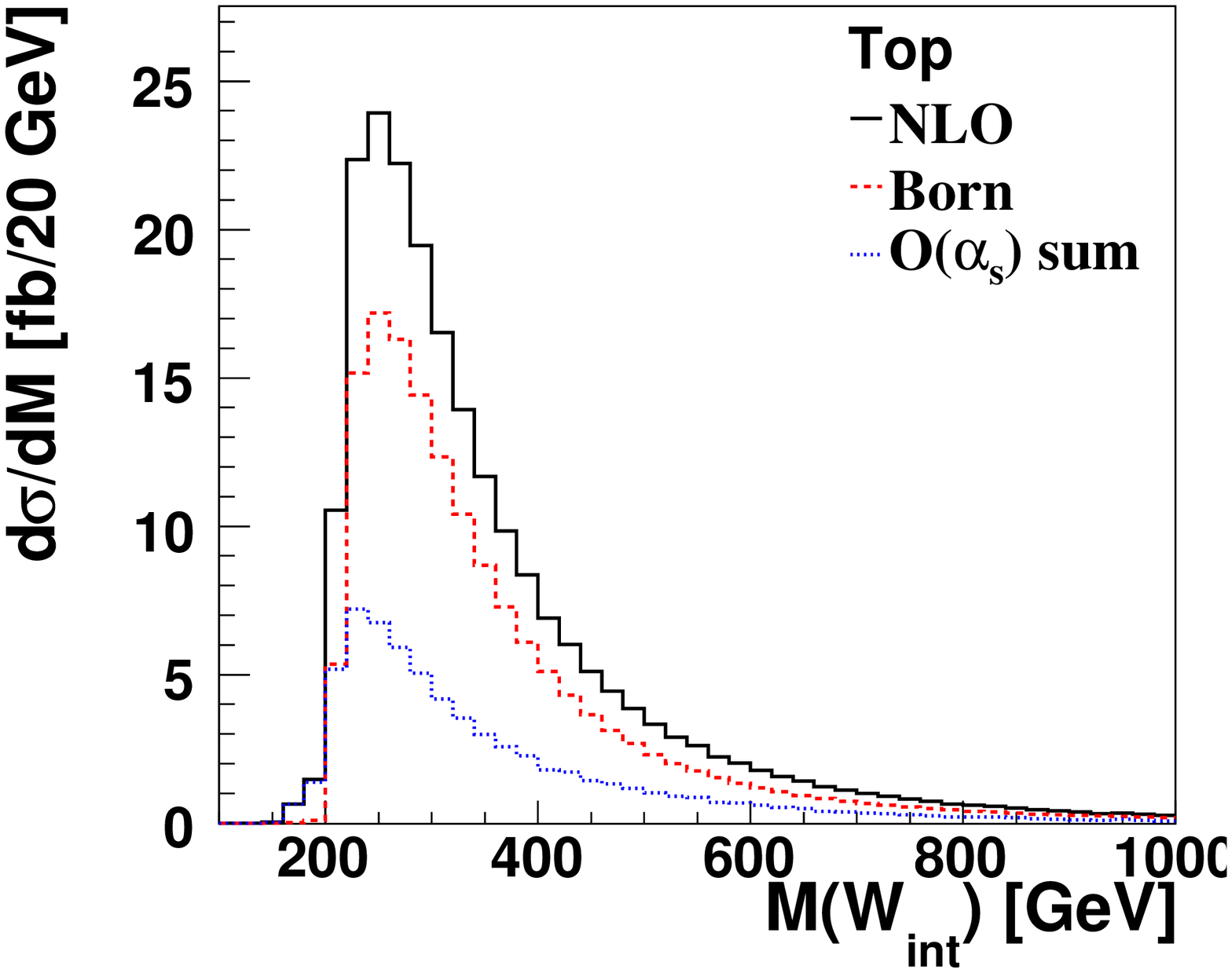}}%
\subfigure[]{\includegraphics[scale=0.3]%
{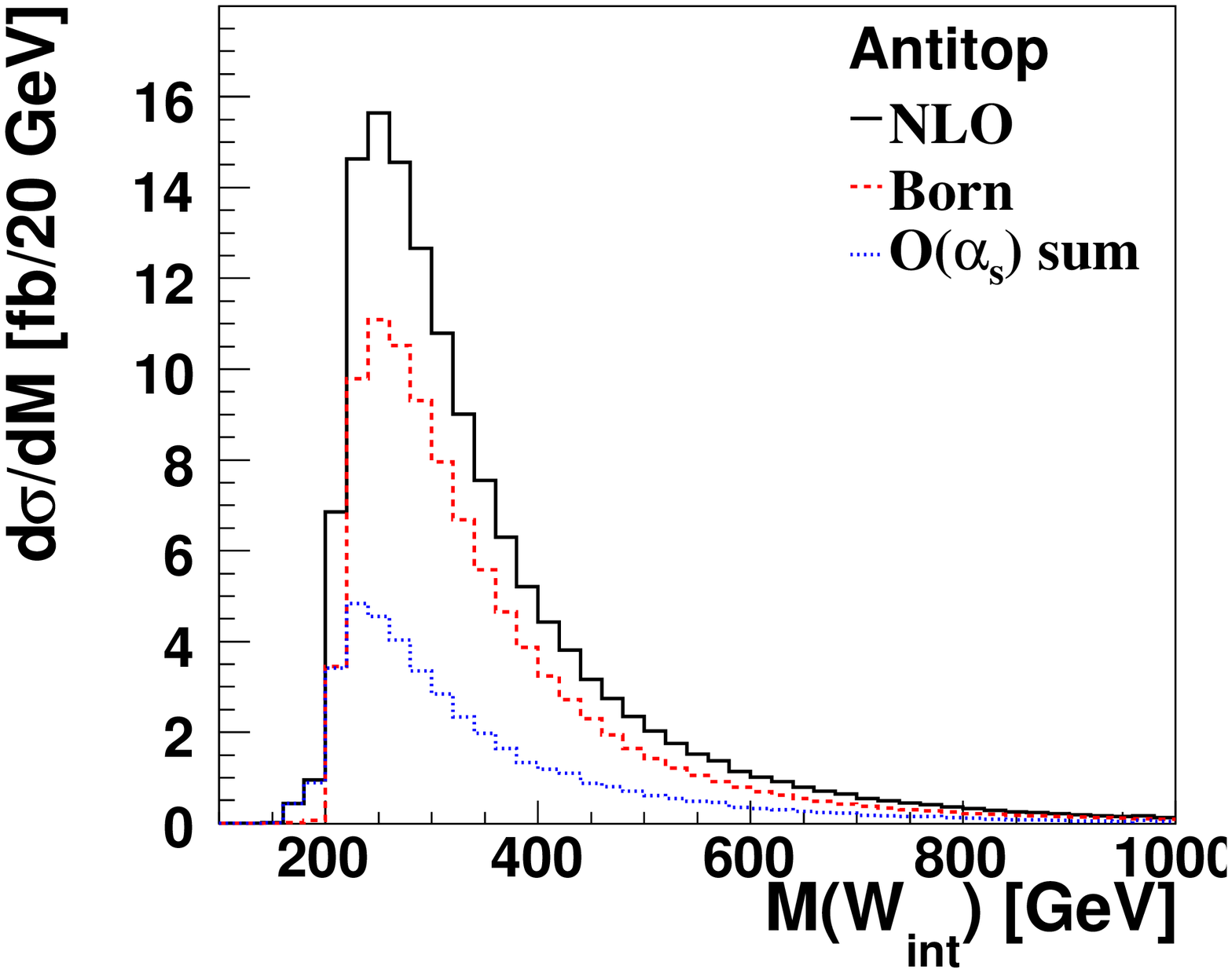}}
\caption{$p_z$ (a, b), rapidity (c, d) and invariant mass (e, f) distributions of the intermediate $W$ boson
after selection cuts. (a, c, e) show top quark production, (b, d, f) antitop quark production.The rapidity distribution only contains events with $p_T \neq 0$.
\label{fig:Wint}}
\end{figure*}

We reconstruct the virtual $W$ boson $W_{int}$ in single top quark production 
by combining the reconstructed top quark with the non-best-jet. This method is
exact if best-jet and non-best-jet are identified correctly and the
event does not contain a third jet from $O(\alpha_{s})$ FINAL corrections.
As discussed in Sec.~\ref{sub:LepMET}, at LO $W_{int}$ receives
a longitudinal boost, which is stronger for top than for 
antitop quark production. Figure~\ref{fig:Wint}
shows that the LO $p_{z}$ and rapidity distributions
are wider for $W_{int}$ in top than in antitop
quark production. 
Furthermore, the $O(\alpha_s)$ correction to the rapidity distribution of $W_{int}$ is narrower than 
the LO contribution. This is because at NLO 
the incoming partons could also be sea quarks or gluons which results in a less boosted 
$W_{int}$ system.

In order to search for new physics in the form of a $W'$ boson, it
is important to know the invariant mass distribution of the intermediate
SM $W$ boson. $W^\prime$ boson searches in the single top quark final state have been performed at the
Tevatron and have set lower $W'$ boson mass limits in the range 750 GeV to 800 GeV \cite{Abazov:2006aj,Abazov:2008vj,Aaltonen:2009qu}. The
LHC with its higher $E_{c.m.}$ will allow to probe up to
even larger boson masses. The invariant mass distribution of $W_{int}$ is 
shown in Fig.~\ref{fig:Wint} for SM single top and
single antitop quark production.
The $O(\alpha_{s})$ corrections,
more specifically the FINAL contribution, shift the invariant mass
peak to lower energies, as in our reconstruction of $W_{int}$ we do not include a possible third jet in
the final state.

\subsection{Kinematical and spin correlations\label{sub:SpinCorr}}

\subsubsection{Correlations between the $b$ jets}

\begin{figure*}
\subfigure[]{\includegraphics[scale=0.3]%
{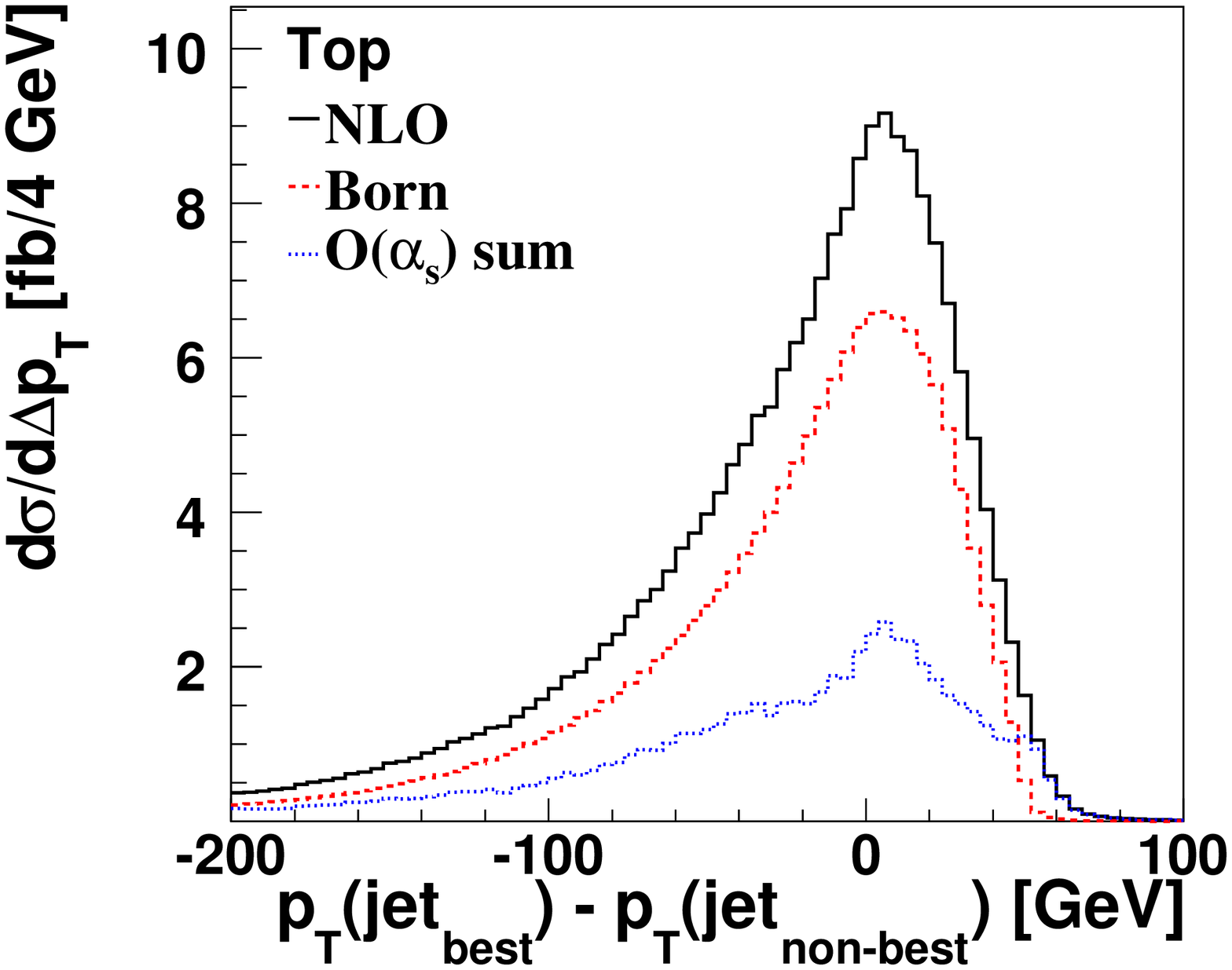}}%
\subfigure[]{\includegraphics[scale=0.3]%
{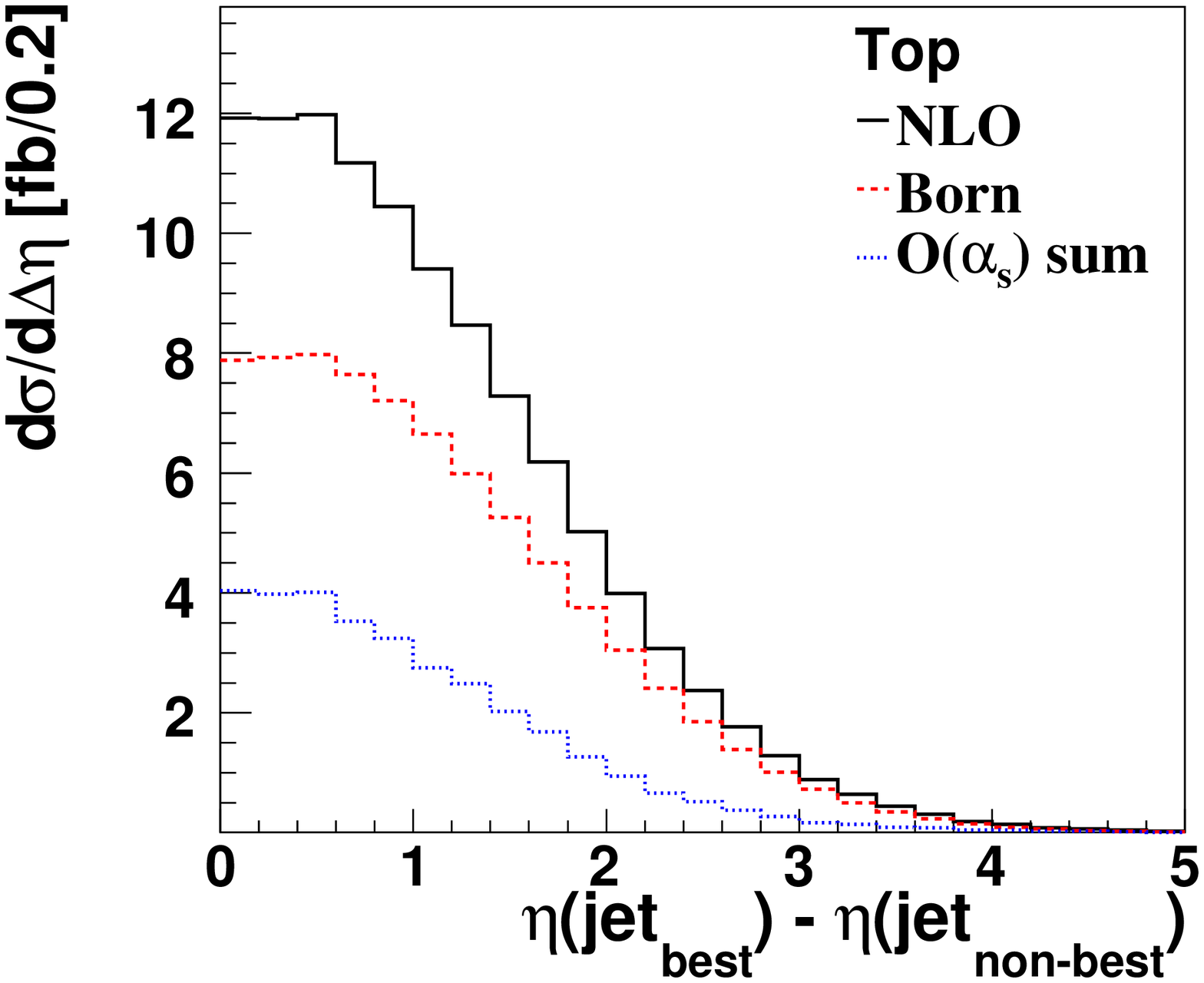}}

\caption{Difference in $p_{T}$ (a) and $\eta$ (b) between the best-jet and
the non-best-jet, after applying the `loose' set of kinematic cuts.\label{fig:PTDiffBg}}

\end{figure*}

Having reconstructed and identified the $W$ boson, the $b_{dec}$
jet and the $b_{fin}$ jet, it is now possible to study correlations
between objects in single top quark events. Figure~\ref{fig:PTDiffBg}
shows that the $b_{dec}$ jet (best-jet) and the $b_{fin}$ jet (non-best-jet)
in single top quark events are highly correlated. The correlations
are modified by the $O(\alpha_{s})$ corrections. The FINAL corrections
shift the $p_{T}$ difference between best-jet and non-best-jet
to more positive values, as their tendency to be collinear to the
boost direction lowers the $p_{T}$ of the $b_{fin}$ jet. For the
same reason, the $\eta$ correlation is shifted to slightly more central
values. 

\subsubsection{Top quark polarization}

\begin{figure}
\includegraphics[scale=0.6]%
{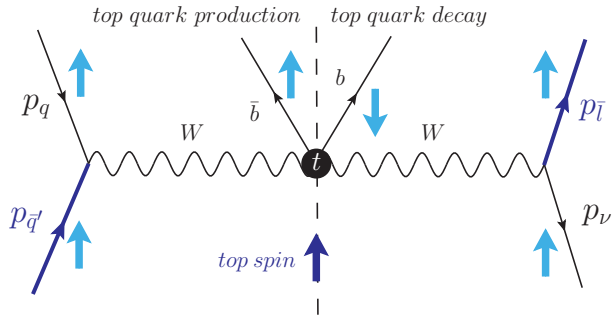} 

\caption{S-channel single top quark production from the perspective of a top
quark at rest. Thin arrows denote directions of momentum while bold arrows indicate spin directions.\label{fig:FeynmanSymm}}

\end{figure}

The SM predicts that single top quarks are highly polarized. Verifying this prediction can be used as a test
of the SM electroweak symmetry breaking mechanism and a check for
new physics \cite{Tait:2000sh,Chen:2005vr,Abazov:2008sz,Abazov:2009ky}. It is in principle possible to measure
the polarization of the single top quark, by making use of the fact
that the charged lepton from the top quark decay is maximally correlated
with the top quark spin \cite{Mahlon:1995zn,Parke:1996pr}, as illustrated in Fig.~\ref{fig:topdecay}. In the
following we therefore plot the angle between the charged lepton and
a reference axis in the rest frame of the top quark. Two different
choices for this reference frame have been used in the past: the helicity
basis and the so-called optimal basis \cite{Mahlon:1995zn,Mahlon:1996pn,Parke:1996pr,Tait:2000sh}.
In the helicity basis, the top quark spin is measured along the top
quark direction of motion in the center of mass (c.m.) frame of the
system. In the optimal basis, the top (antitop)
quark spin is measured along the direction of the incoming antiquark
(quark) in the c.m. frame of the top (antitop) quark. This reference
frame is called the optimal basis, as the top (antitop) quark produced
in s-channel single top quark processes is almost 100\% polarized
along (against) the direction of the incoming antiquark (quark). Figure~\ref{fig:FeynmanSymm} shows s-channel single top quark production from the perspective of a 
top quark at rest. It demonstrates
 how the optimal basis makes use of the momentum - spin correlations that are due to spin 
conservation and left-handedness of the weak interaction.

While at the Tevatron the antiquark comes predominantly from the antiproton,
the LHC is a $pp$ collider and there is no preferred direction for
the antiquark. We can however make use of the fact that the incoming
quark is more likely to be a valence quark, and has a larger longitudinal
momentum than the antiquark which comes always from the quark sea
of the proton. In most cases the direction of the longitudinal boost
of $W_{int}$ therefore indicates the direction of
the incoming quark. (This becomes a better approximation 
when the magnitude of the longitudinal momentum 
$\left|p_{z}\right|$ of $W_{int}$ is large.)
We choose the reference axis according to the
sign of $p_{z}(W_{int})$. After discussing
the helicity basis, we show how it is possible to enhance the
spin correlations for the optimal basis further by requiring a minimum
$p_{z}(W_{int})$.

For the helicity basis, the c.m. frame of the system has to be reconstructed in
order to define the top quark momentum. As discussed in Ref.~\cite{Cao:2005pq},
this is more complicated at NLO than at Born level, because of additional
jets. Ref.~\cite{Cao:2005pq} investigates two options for reconstructing
the c.m. frame: the $tb_{fin}(j)$-frame, which is the c.m. frame of
the incoming partons and the rest frame of all the final state objects
(reconstructed top quark and all other jets), and the $tb_{fin}$-frame,
which is the c.m. frame of the top quark and the non-best-jet. The
$tb_{fin}$-frame differs from the $tb_{fin}(j)$-frame only in exclusive
three-jet events. As shown in Table~\ref{tab:toppol3} and discussed
below, the degree of polarization is larger in the $tb_{fin}$-frame,
so we only show distributions for the top quark polarization
in this frame.

In the helicity basis, we examine the polarization of the top quark
by studying the angular distribution $\cos\theta_{hel}$ of the lepton
relative to the moving direction of the top quark, both in the c.m.
frame of the system, \begin{eqnarray}
\cos\theta_{hel}=\frac{\vec{p}_{t}\cdot\vec{p}_{\ell}^{*}}{|\vec{p}_{t}||\vec{p}_{\ell}^{*}|},\end{eqnarray}
 where $\vec{p}_{t}$ is the top quark three-momentum defined in the $tb_{fin}$- or the $tb_{fin}(j)$-frame, and $\vec{p}_{\ell}^{*}$
is the charged lepton three-momentum, after boosting it first into the c.m. frame of the system and then into the top quark rest frame. For a left-handed top quark, the angular correlation of the
lepton $\ell^{+}$ is given by $(1-\cos\theta_{hel})/2$, and for
a right-handed top quark, it is $(1+\cos\theta_{hel})/2$. For a right-handed
antitop quark, the angular correlation of the lepton $\ell^{-}$ is
given by $(1-\cos\theta_{hel})/2$, and for a left-handed antitop
quark, it is $(1+\cos\theta_{hel})/2$. %
\begin{figure*}
\subfigure[]{\includegraphics[scale=0.3]%
{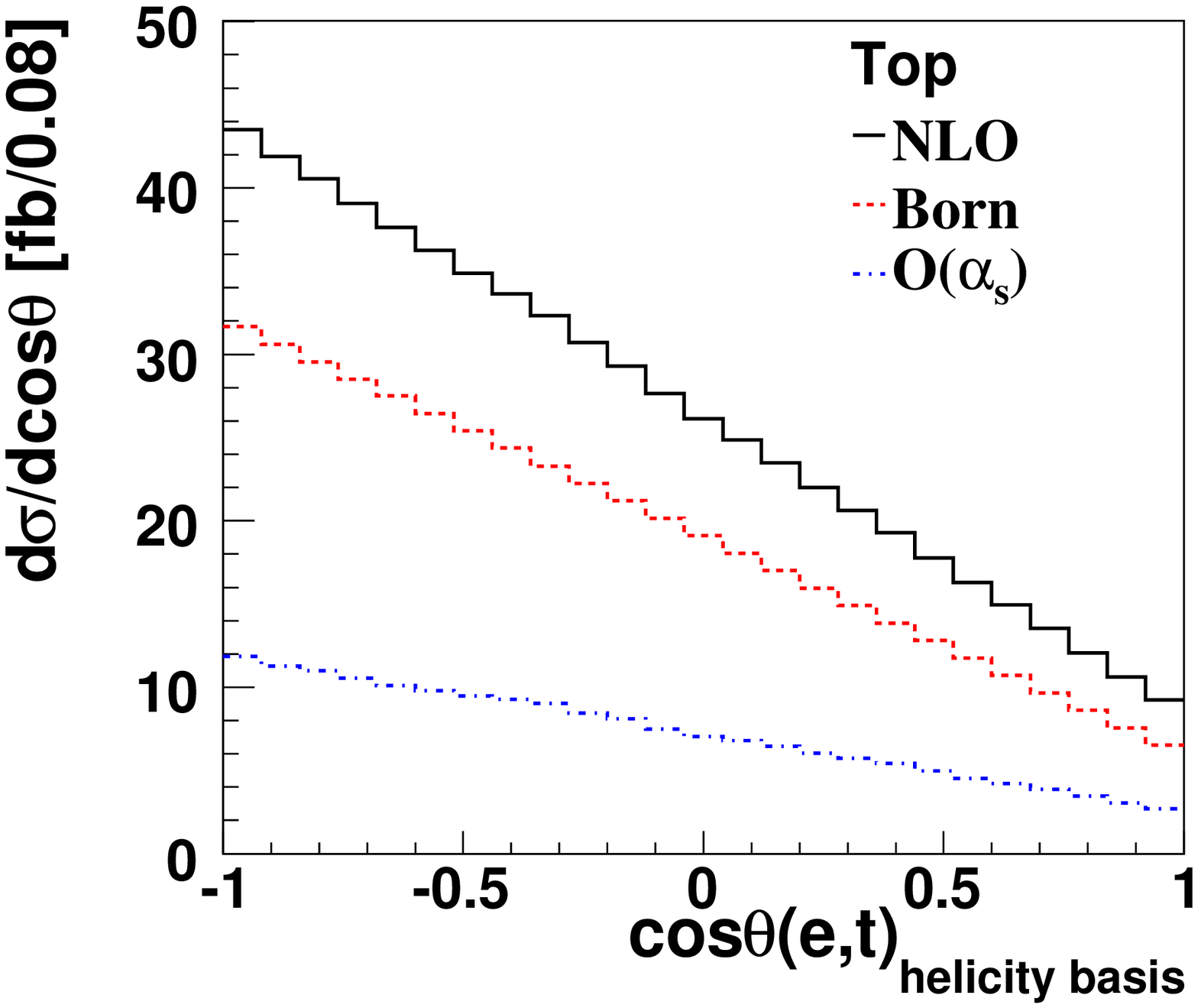}}%
\subfigure[]{\includegraphics[scale=0.3]%
{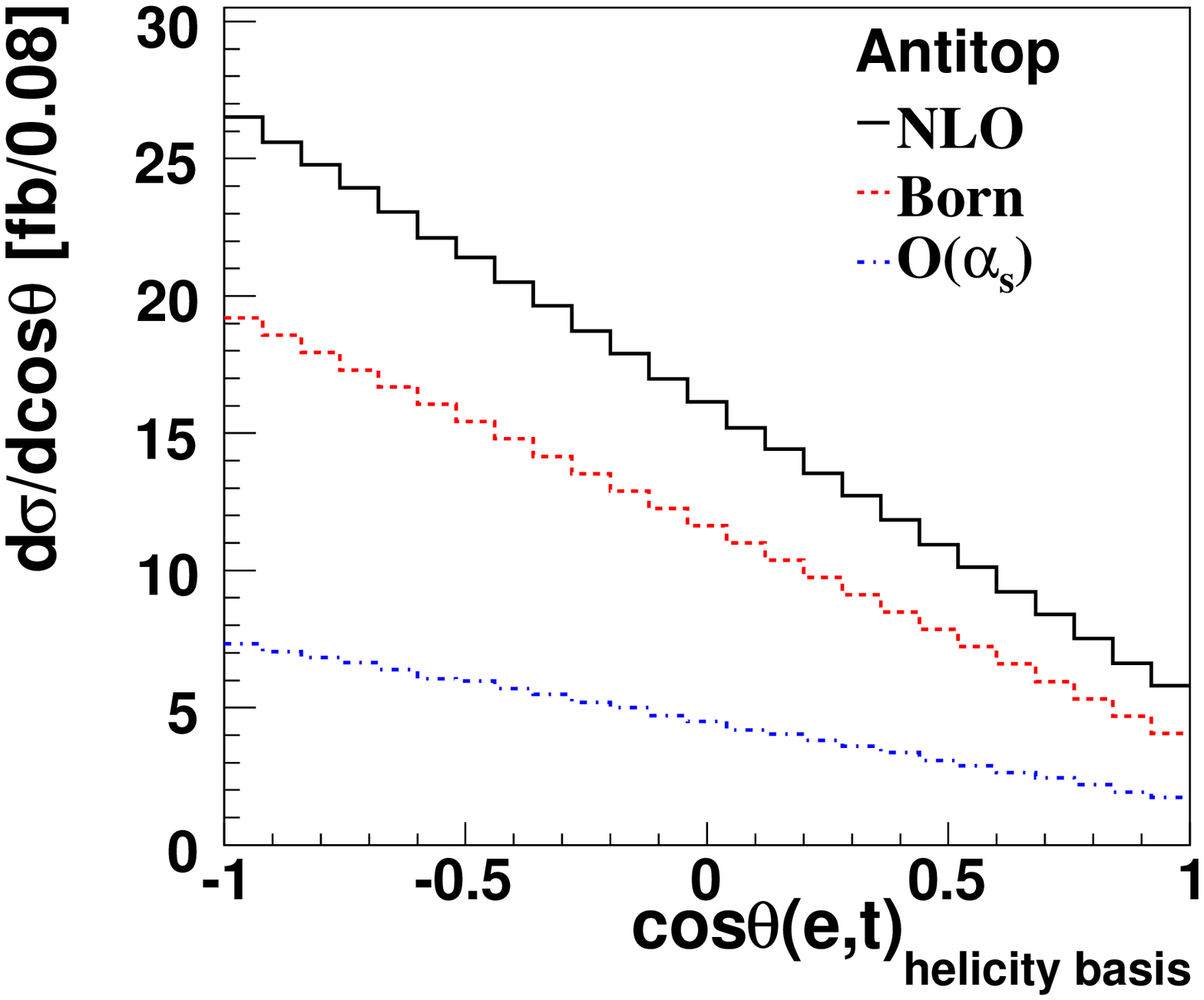}}%

\subfigure[]{\includegraphics[scale=0.3]%
{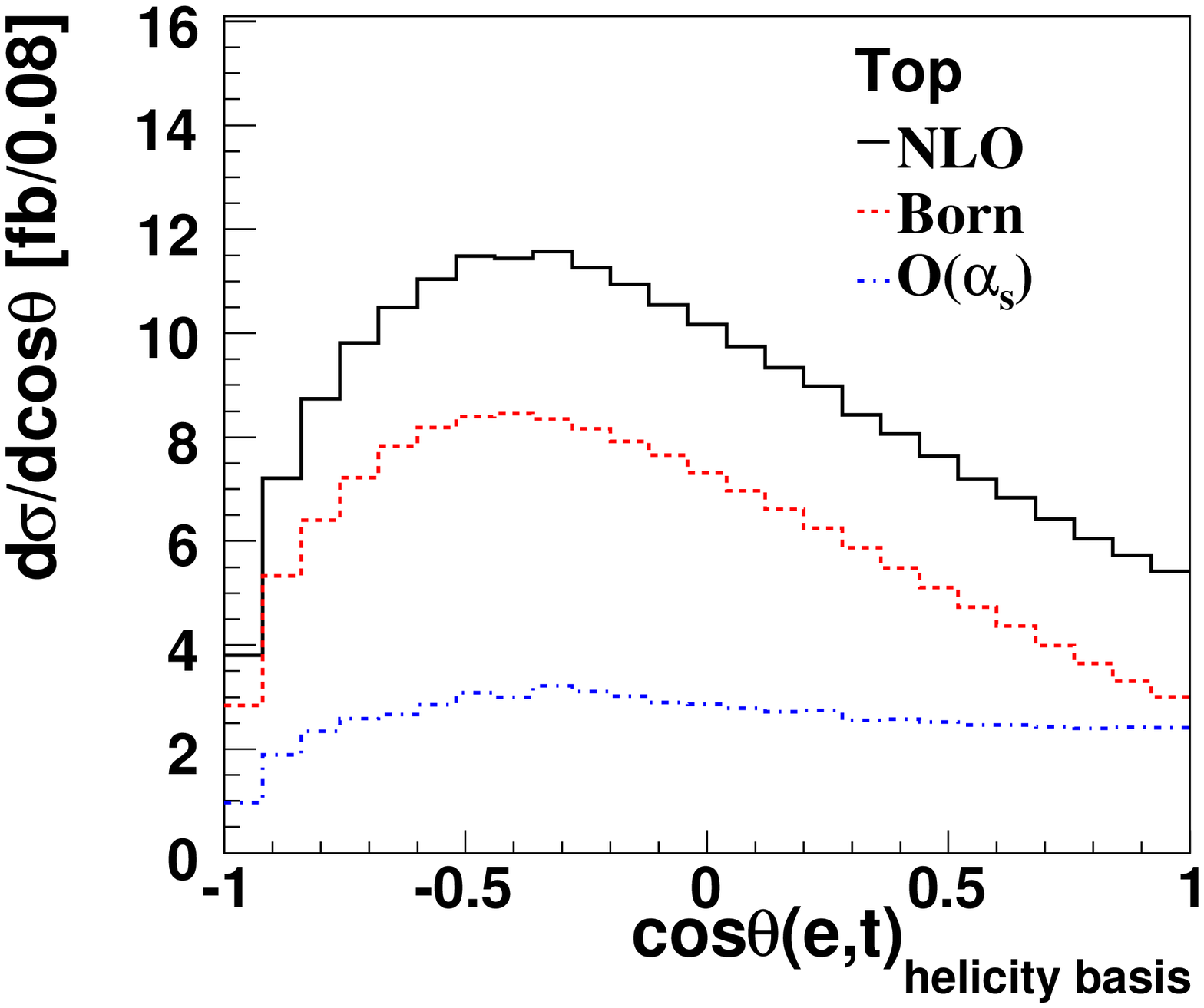}}
\subfigure[]{\includegraphics[scale=0.3]%
{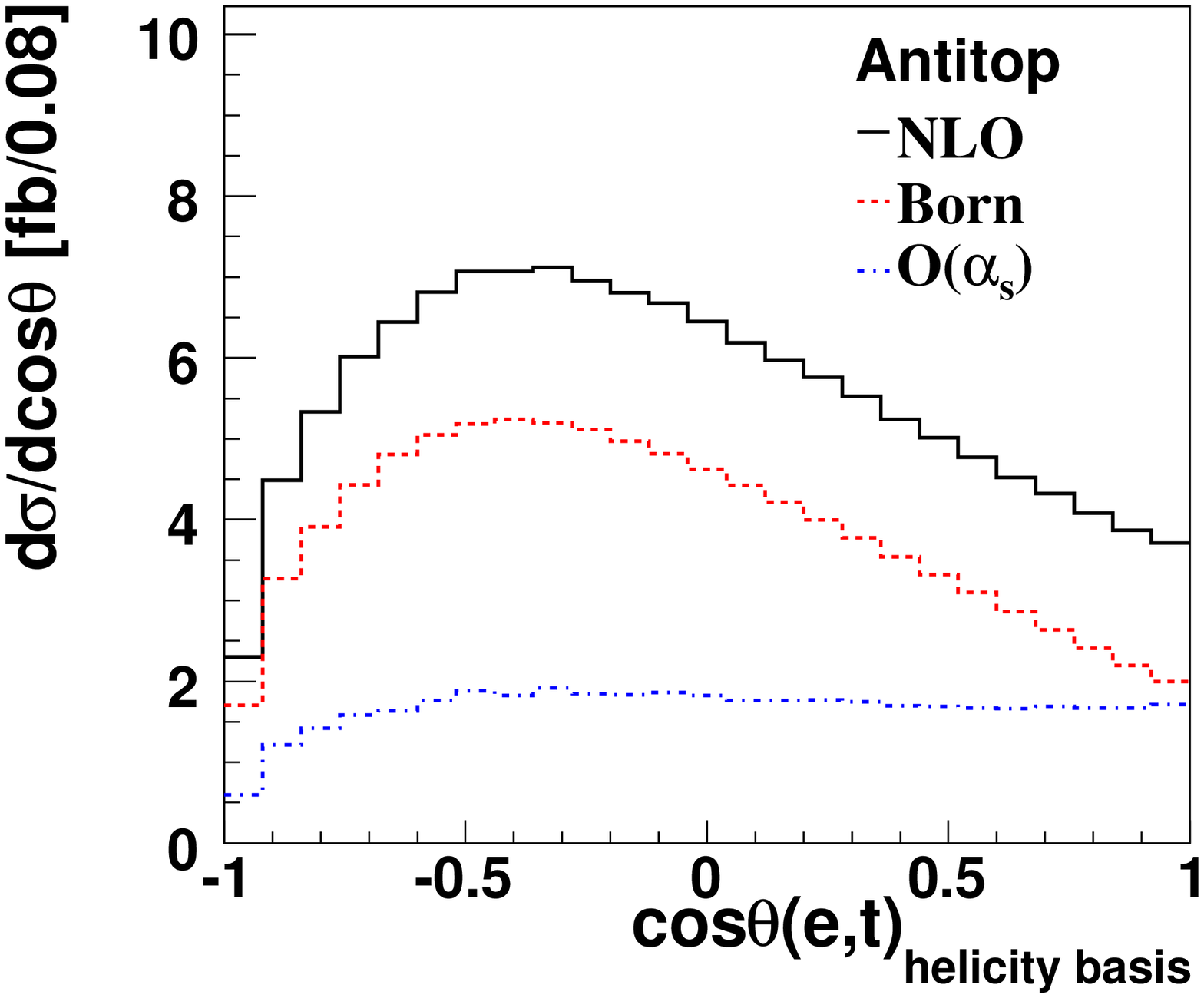}} 

\caption{Top quark polarization in the helicity basis at parton level (a, b)
and after event reconstruction and applying the `loose' set of selection
cuts (c, d), for top quark production (a, c) and antitop quark production
(b, d). The $tb_{fin}$-frame is chosen as the c.m. frame.\label{fig:hel}}

\end{figure*}

Figure~\ref{fig:hel} shows that this linear relationship for $\cos\theta_{hel}$
indeed describes s-channel single top quark events well at parton level.
It also shows that the top quark is not completely polarized in the
helicity basis, and that this polarization is weakened further at
NLO. 
After event reconstruction, the drop-off
close to $\cos\theta_{hel}$ = -1 is due to jet-lepton separation
cuts. 

\begin{figure*}
\subfigure[]{\includegraphics[scale=0.3]%
{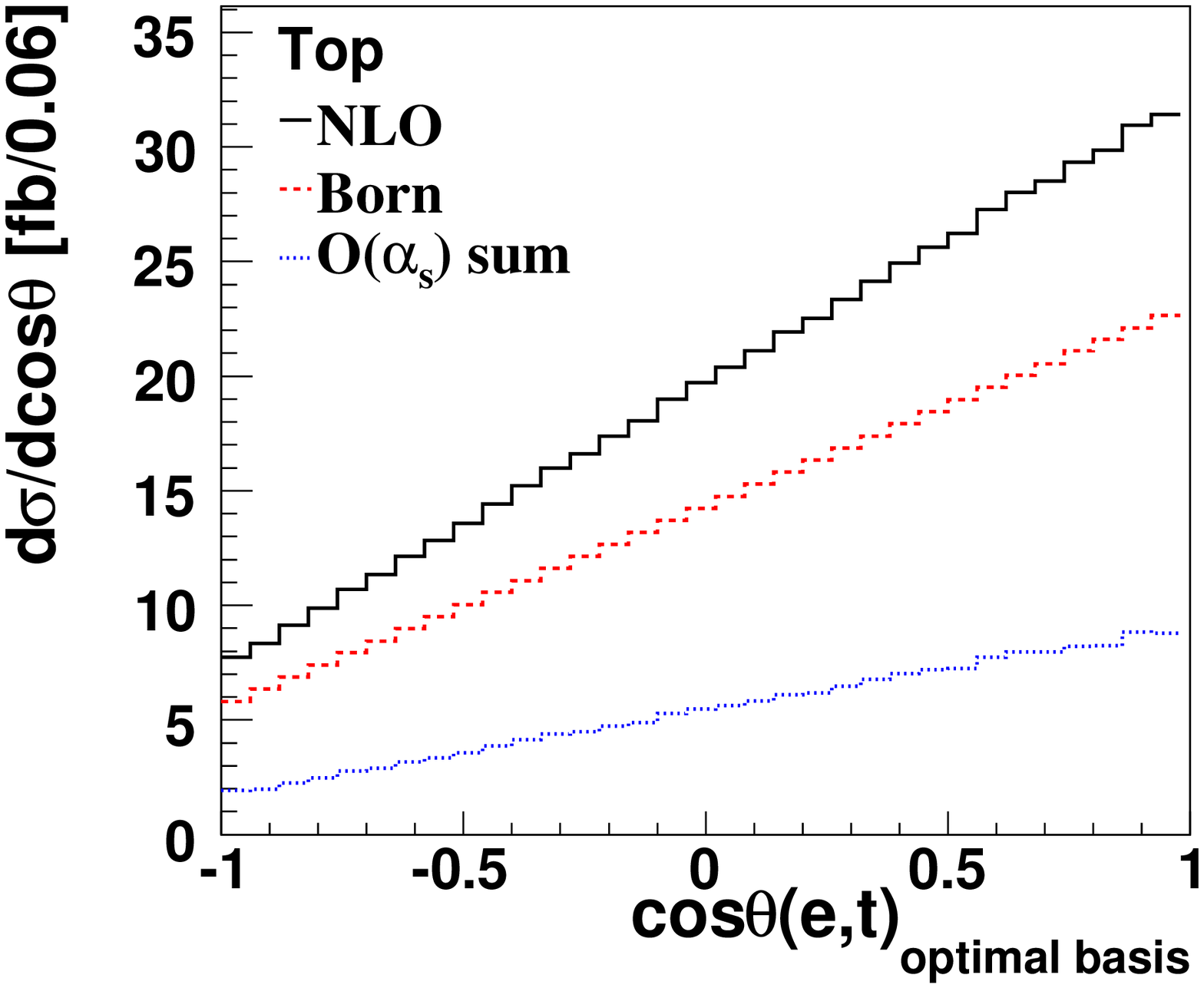}}%
\subfigure[]{\includegraphics[scale=0.3]%
{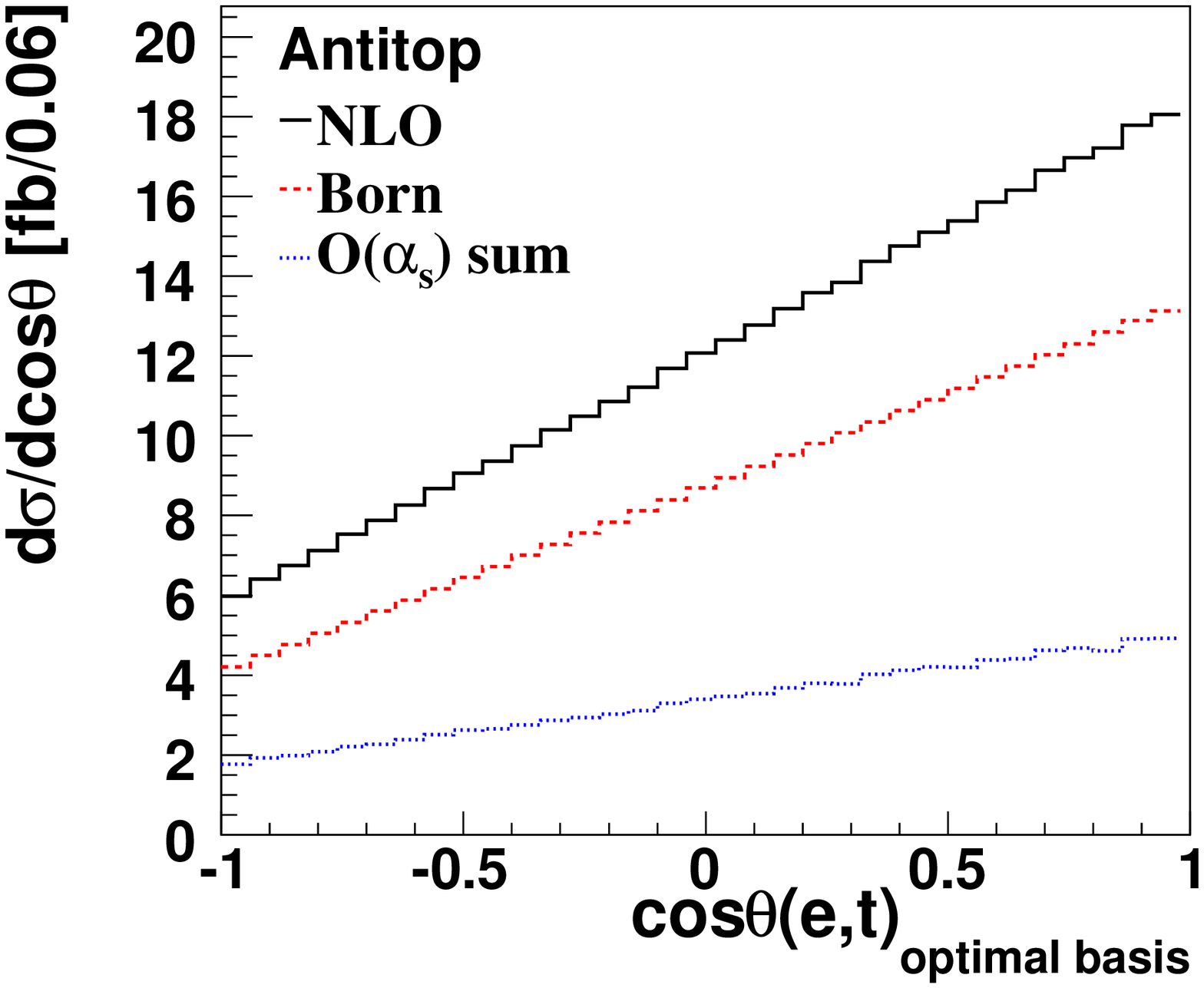}}%

\subfigure[]{\includegraphics[scale=0.3]%
{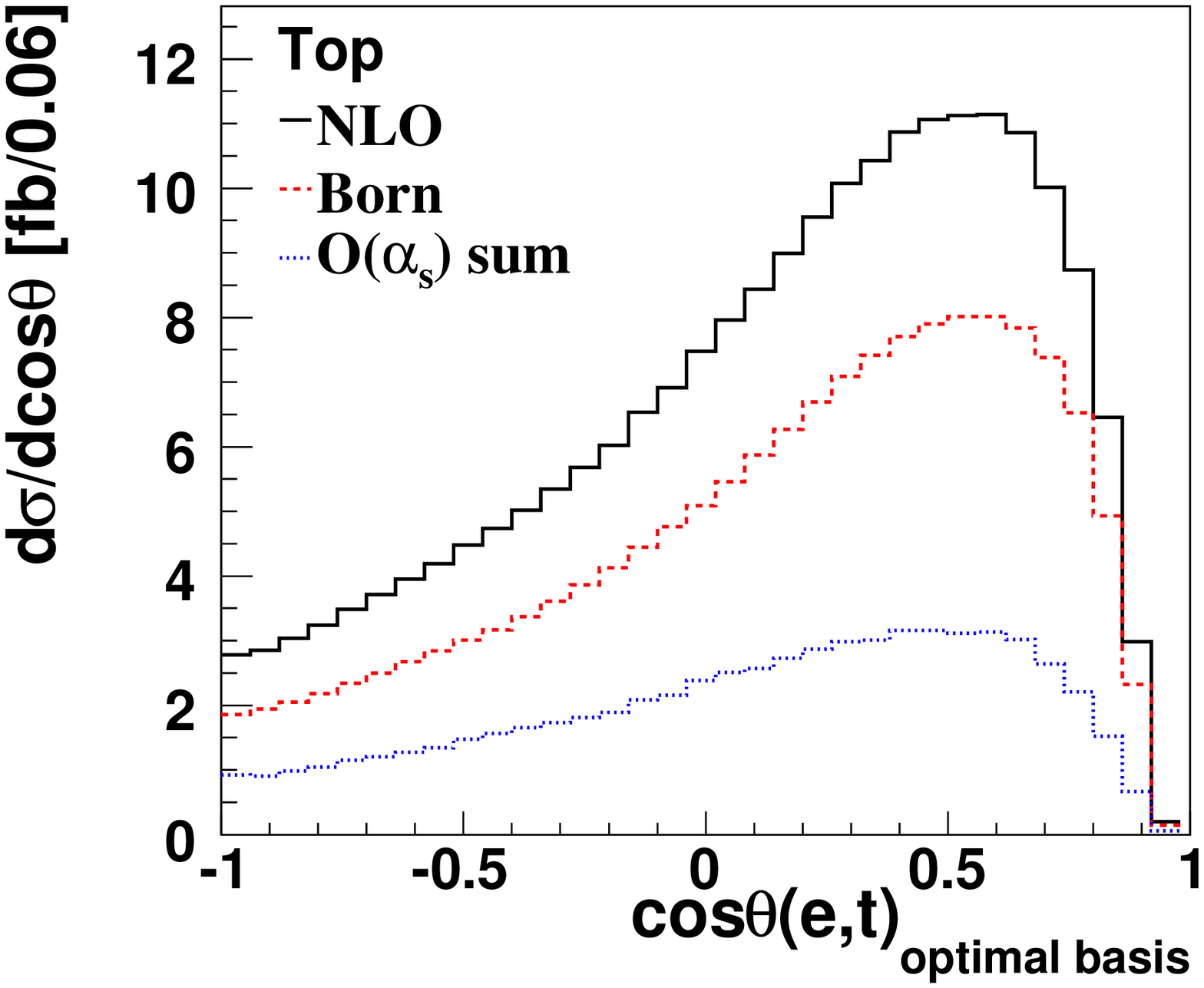}}
\subfigure[]{\includegraphics[scale=0.3]%
{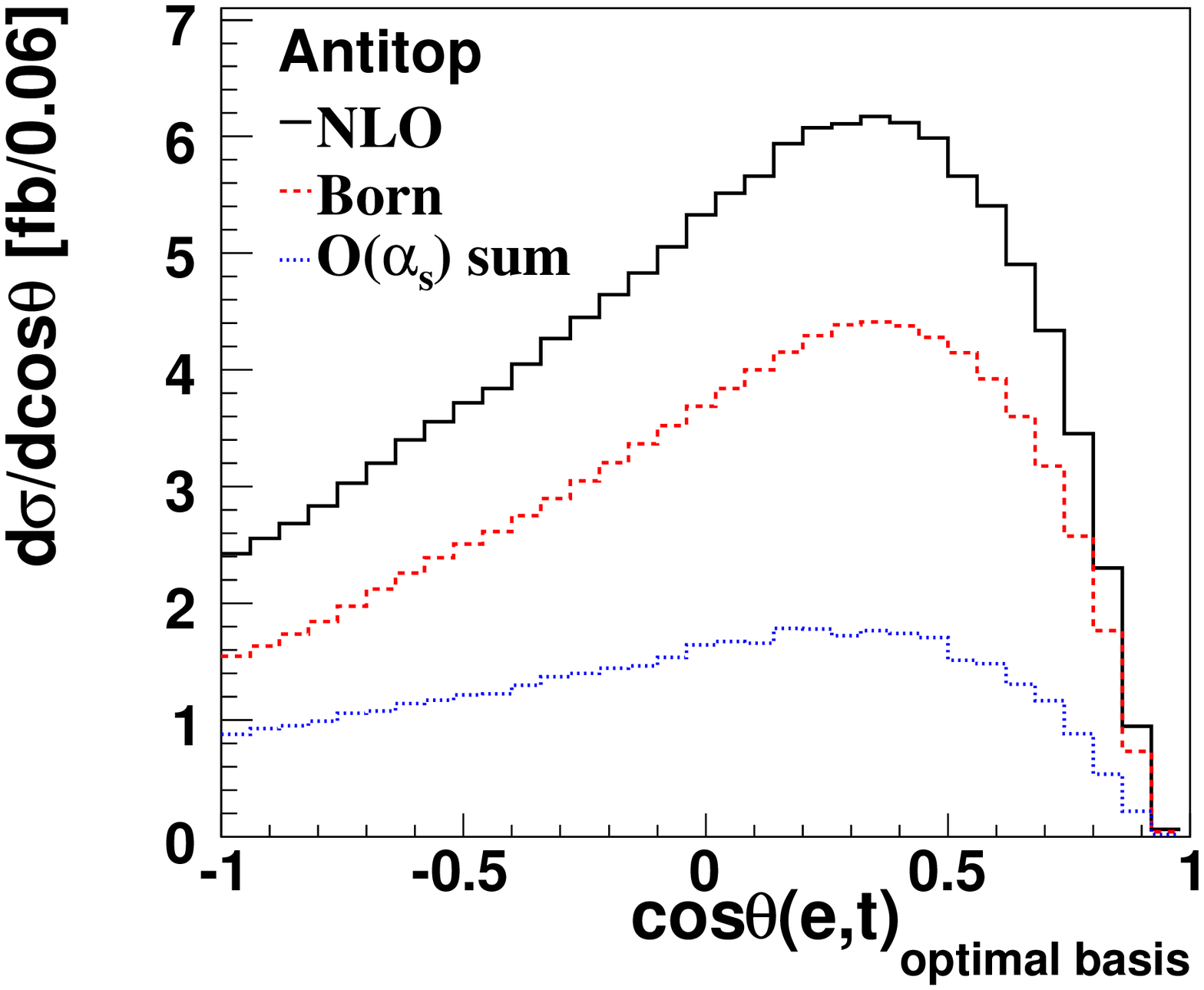}} 

\caption{Top quark polarization in the optimal basis at parton level (a,
b) and after event reconstruction and applying the `loose' set of selection
cuts (c, d), for top quark production (a, c) and antitop quark production
(b, d). \label{fig:opt}}

\end{figure*}

The corresponding distributions for the optimal basis can be seen
in Fig.~\ref{fig:opt}. Here, the relevant angular correlation is
\begin{eqnarray}
\cos\theta_{opt}=\frac{\vec{p}_{p_{1}}\cdot\vec{p}_{\ell}^{*}}{|\vec{p}_{p_{1}}||\vec{p}_{\ell}^{*}|}\,,\end{eqnarray}
where $\vec{p}_{p_{1}}$ is the three-momentum in the top quark rest
frame of the proton which travels (in the case of top quark production)
antiparallel to the longitudinal boost that $W_{int}$ receives, e.g.
in the $-\vec{p_{z}}(W_{int})$ direction.
In the case of antitop quark production the chosen proton travels
parallel to the longitudinal boost of $W_{int}$, e.g.
in the $\vec{p_{z}}(W_{int})$ direction.
$\vec{p}_{\ell}^{*}$ 
is the three-momentum of the lepton in the top quark rest frame. For
a(n) (anti)top quark polarized along (against) the moving direction
of the chosen proton, the angular distribution of the lepton $\ell^{+}$
($\ell^{-}$) is $(1+\cos\theta_{opt})/2$, while for a(n) (anti)top
quark polarized against (along) the moving direction of the chosen
proton, it is $(1-\cos\theta_{opt})/2$. Figure~\ref{fig:opt} shows
that there is indeed a linear relationship for $\cos\theta_{opt}$
at parton level, but that the top quark is not completely polarized in this basis.
After reconstruction and applying the `loose' set of cuts, there is
a cutoff at large $\cos\theta_{opt}$, due to the lepton $\eta$ cut.

\begin{figure*}
\subfigure[]{\includegraphics[scale=0.3]%
{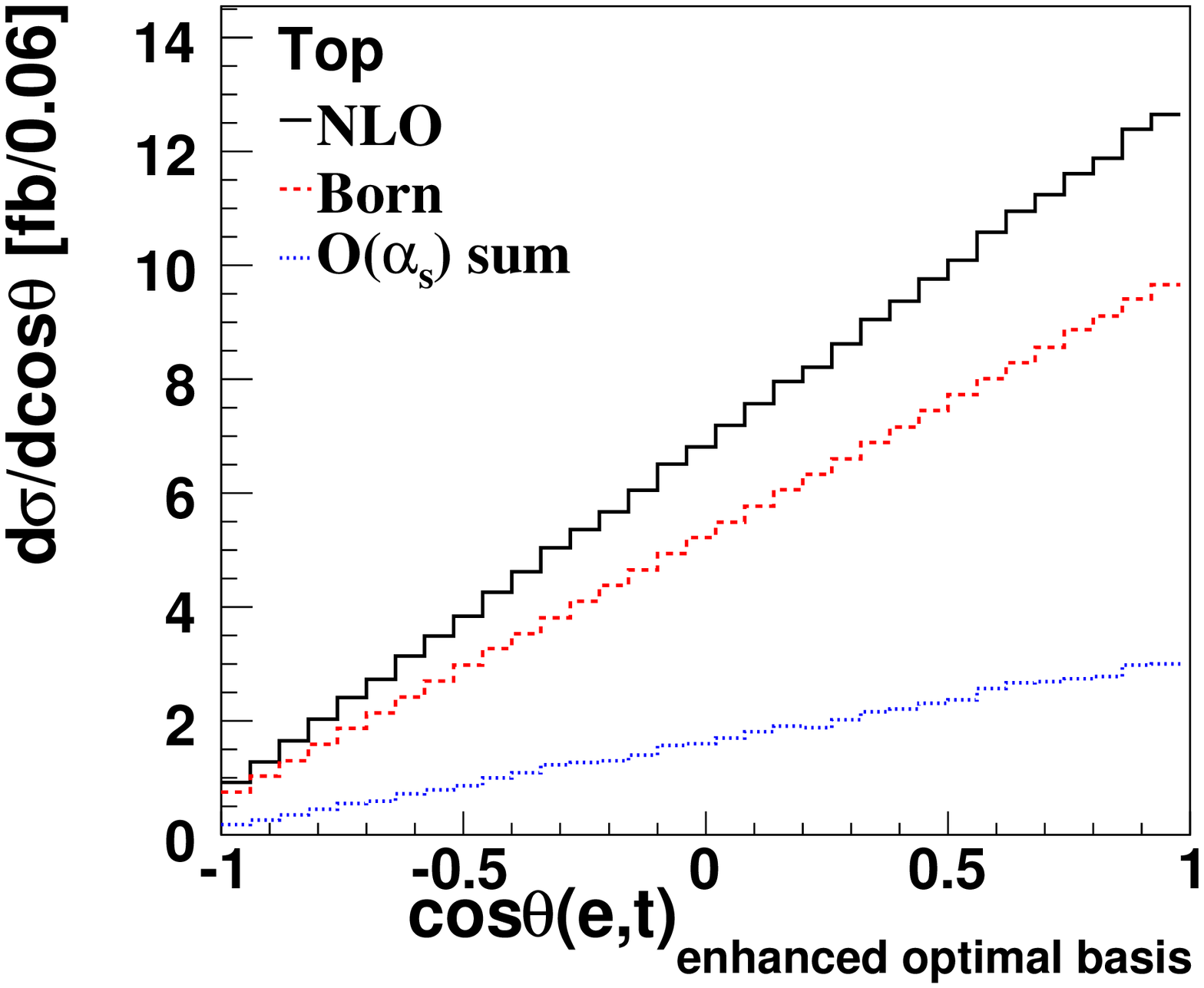}}%
\subfigure[]{\includegraphics[scale=0.3]%
{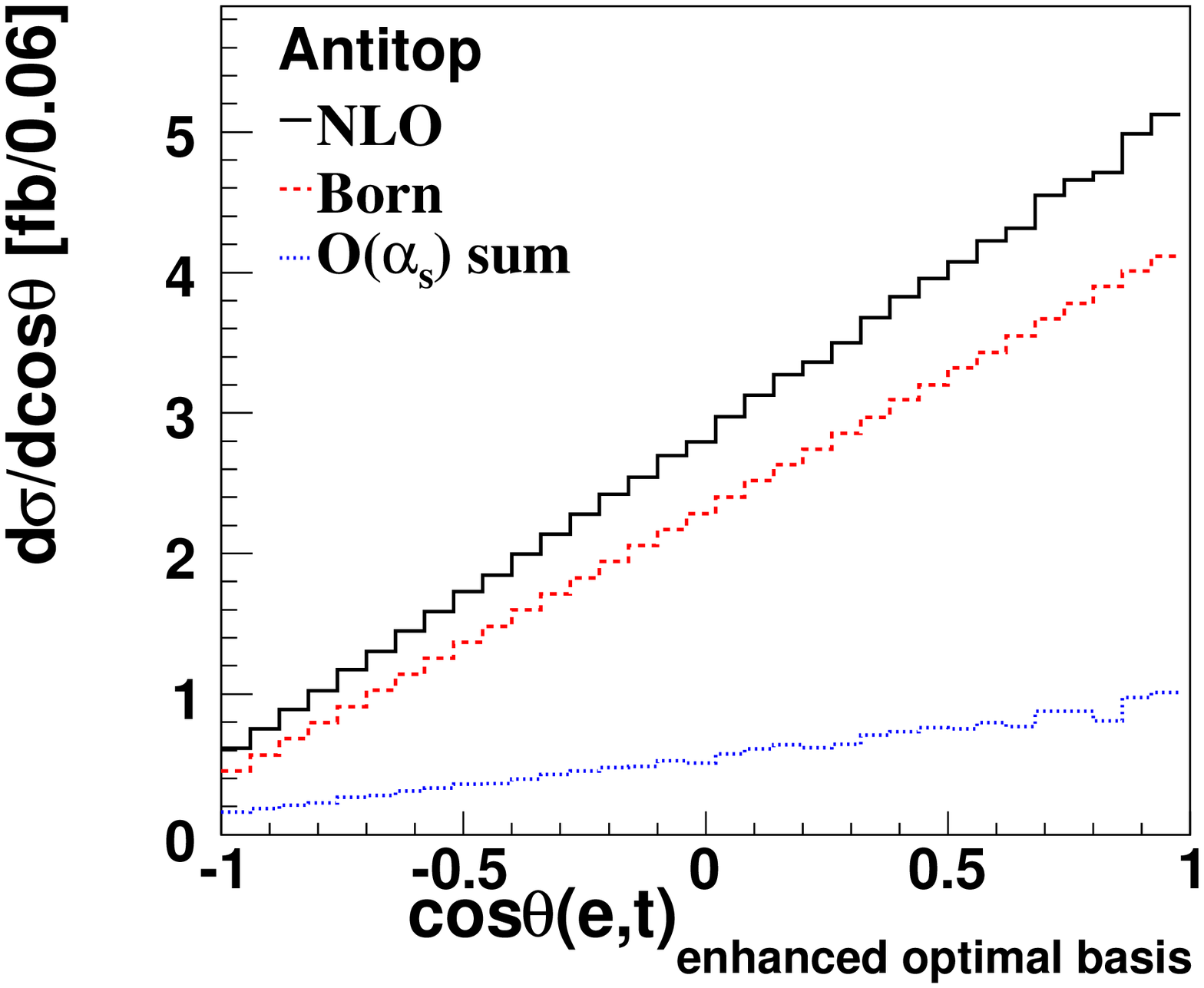}}%

\subfigure[]{\includegraphics[scale=0.3]%
{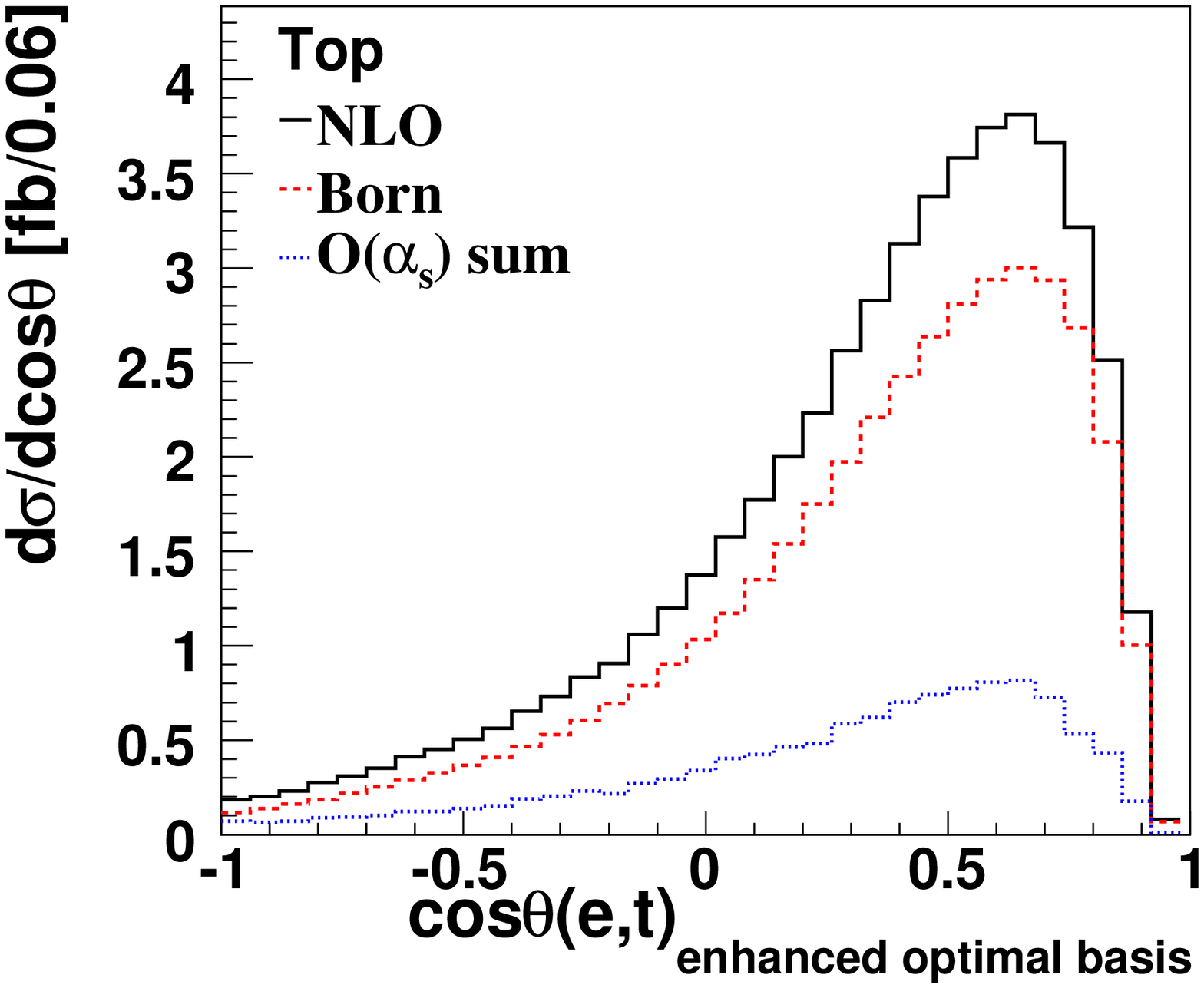}}
\subfigure[]{\includegraphics[scale=0.3]%
{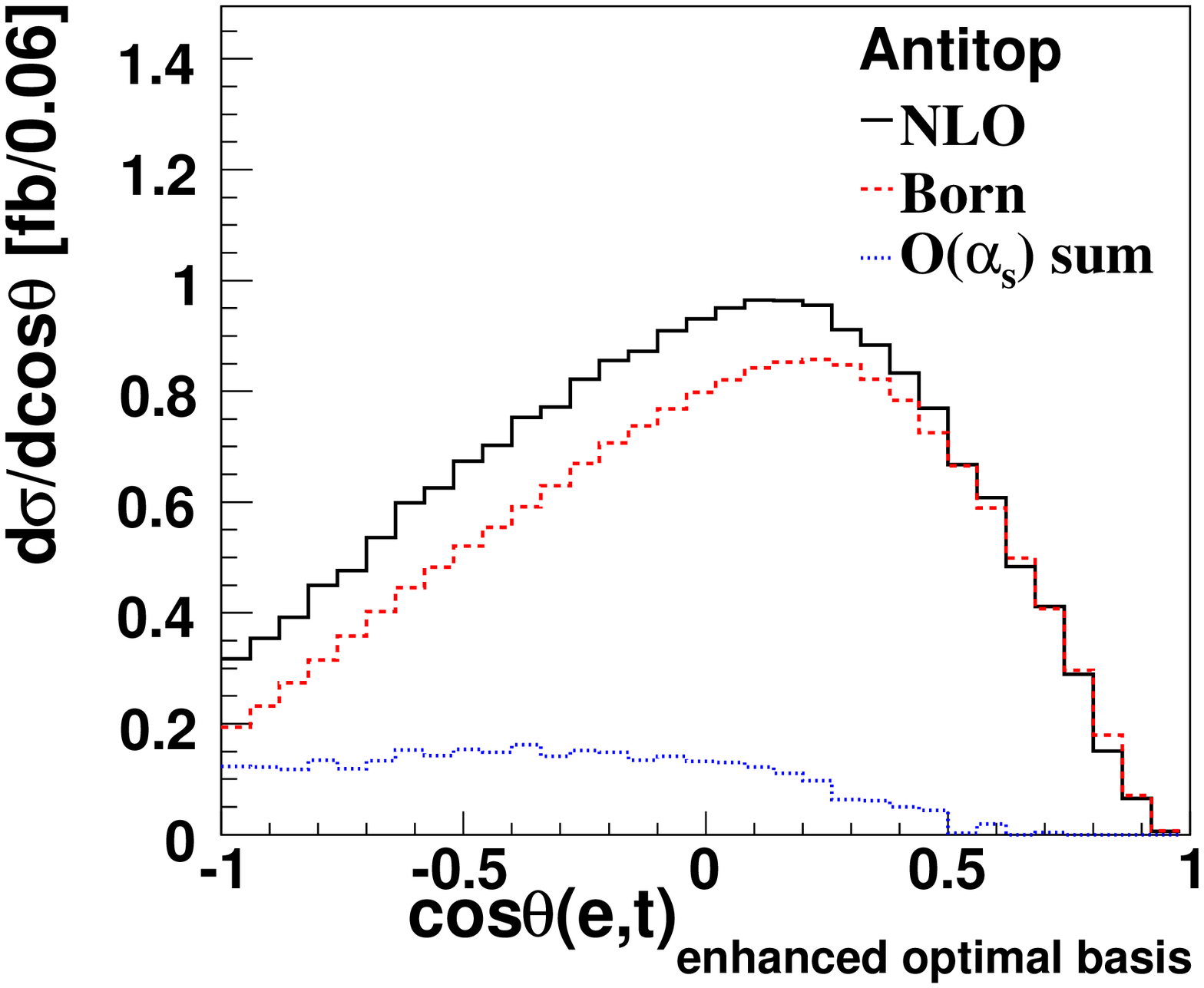}}

\caption{Top (a, c) and antitop (b, d) quark polarization in the enhanced
optimal basis, where $W_{int}$ has a longitudinal
momentum of at least 1000 GeV. (a, b) show the distributions before
selection cuts using all parton information, while (c, d) are the
distributions for reconstructed events after applying the `loose' set
of cuts.\label{fig:optenh}}

\end{figure*}

It is possible to enhance the performance of the optimal basis,
by using only those events in which $p_{z}(W_{int})$ is larger than a
certain threshold. For this study, thresholds of 500, 1000 and 2500 GeV have been tested.
 By imposing a cut on $p_{z}(W_{int})$, we are able to determine the direction of the incoming
antiquark correctly more often, as the momentum difference between
the incoming quarks increases. The efficiencies for the three different
thresholds 500, 1000 and 2500 GeV are 50\%, 25\% and 2\% respectively
(cf. Fig.~\ref{fig:Wint}). Figure~\ref{fig:optenh} shows that,
at parton level, the (anti)top quark is indeed highly polarized for
a $p_{z}(W_{int})$ cut of 1000 GeV. Reconstruction and selection
cuts change the distribution dramatically and differently for
top and antitop quarks. The cutoff close to $\cos\theta_{opt}$
= 1 is again due to the lepton $\eta$ cut. Due to spin correlations, the lepton $p_{T}$ cut
carves out a large number of events with $\cos\theta_{opt} <$ 0 for
the top quark, while for the antitop quark it removes mainly events
with $\cos\theta_{opt} > $ 0.

As in Ref.~\cite{Cao:2004ap}, we define the degree of polarization
$\mathcal{D}$ of the top quark as the ratio\[
\mathcal{D}=\frac{N_{-}-N_{+}}{N_{-}+N_{+}},\]
 where $N_{-}$ ($N_{+}$) is the number of left-hand (right-hand)
polarized top quarks in the helicity basis. Similarly, in the optimal
basis, $N_{-}$ ($N_{+}$) is the number of top quarks with polarization
against (along) the direction of the chosen proton three momentum
in the top quark rest frame $\vec{p}_{p_{1}}$. For the antitop quark
the relationships are the same.

\begin{table*}
\begin{centering}
\begin{ruledtabular}
\begin{tabular}{l|cc|cc|cc|cc|cc|cc}

 & \multicolumn{6}{c|}{Top} & \multicolumn{6}{c}{Antitop}\tabularnewline
\cline{2-13}
 & \multicolumn{2}{c|}{$\mathcal{D}$ } & \multicolumn{2}{c|}{\textbf{$\mathcal{F}$}} 
 & \multicolumn{2}{c|}{$\mathcal{A}$}  & \multicolumn{2}{c|}{$\mathcal{D}$ } 
 & \multicolumn{2}{c|}{\textbf{$\mathcal{F}$}} & \multicolumn{2}{c}{$\mathcal{A}$}\tabularnewline
 & LO & NLO & LO & NLO & LO & NLO  & LO & NLO & LO & NLO & LO & NLO\tabularnewline
\hline
Helicity ($tb_{fin}$-frame)& 0.69& 0.68& 0.84& 0.84& 0.34& 0.34& -0.68& -0.67& 0.84 & 0.83& -0.34& -0.33\tabularnewline
Optimal ($p_z(W_{int}) >$ 0) & -0.61 & -0.63 & 0.81 & 0.81 & -0.31 & -0.32 & 0.53 & 0.52 & 0.77  & 0.76 & 0.27 & 0.26\tabularnewline
Optimal ($p_z(W_{int}) >$ 500)  & -0.79 & -0.82 & 0.90 & 0.91 & -0.40 & -0.41 & 0.74 & 0.73 & 0.87  & 0.86 & 0.37 & 0.36\tabularnewline
Optimal ($p_z(W_{int}) >$ 1000)  & -0.88 & -0.89 & 0.94 & 0.95 & -0.44 & -0.45 & 0.83 & 0.81 & 0.91  & 0.91 & 0.41 & 0.41\tabularnewline
Optimal ($p_z(W_{int}) >$ 2500)  & -0.98 & -0.97 & 0.99 & 0.98 & -0.49 & -0.48 & 0.93 & 0.90 & 0.96  & 0.95 & 0.46 & 0.45\tabularnewline
\end{tabular}
\end{ruledtabular}
\par\end{centering}

\caption{Parton level degree of polarization $\mathcal{D}$, polarization fraction
$\mathcal{F}$, and asymmetry $\mathcal{A}$ for s-channel single top quark events before any cuts. Results are shown for both top (left) and antitop (right)
quark measured in the helicity basis ($tb_{fin}$-frame) and in the optimal basis with different $p_{z}(W_{int})$ thresholds. 
In this table, $\mathcal{F}$
corresponds to $\mathcal{F}_{-}$ in the helicity basis for left-handed
top quarks and to $\mathcal{F}_{+}$ in the optimal basis for top quarks with polarization along the chosen proton three-momentum.
For right-handed antitop quarks, 
$\mathcal{F}$ corresponds to $\mathcal{F}_{+}$ in the helicity basis and to $\mathcal{F}_{-}$ in 
the optimal basis for antitop quarks with polarization against the chosen proton three-momentum.\label{tab:toppol2}}

\end{table*}
\begin{table*}
\begin{centering}
\begin{ruledtabular}
\begin{tabular}{l|cc|cc|cc|cc|cc|cc}
 & \multicolumn{6}{c|}{Top} & \multicolumn{6}{c}{Antitop}\tabularnewline
\cline{2-13}
 & \multicolumn{2}{c|}{$\mathcal{D}$ } & \multicolumn{2}{c|}{\textbf{$\mathcal{F}$}} 
 & \multicolumn{2}{c|}{$\mathcal{A}$}  & \multicolumn{2}{c|}{$\mathcal{D}$ } 
 & \multicolumn{2}{c|}{\textbf{$\mathcal{F}$}} & \multicolumn{2}{c}{$\mathcal{A}$}\tabularnewline
 & LO & NLO & LO & NLO & LO & NLO  & LO & NLO & LO & NLO & LO & NLO\tabularnewline
\hline 
Helicity ($tb_{fin}(j)$, incl. 2-jet)& 0.74& 0.67&  0.87& 0.84& 0.37& 0.34& -0.74& -0.54& 0.87 & 0.77& -0.37 & -0.27\tabularnewline
Helicity ($tb_{fin}$, incl. 2-jet)& 0.74& 0.72& 0.87& 0.86& 0.37& 0.36& -0.74& -0.71& 0.87 & 0.86& -0.37& -0.35\tabularnewline
Helicity ($tb_{fin}(j)$, excl. 3-jet)&&0.76&& 0.88&& 0.38&& -0.65&& 0.83&& -0.33\tabularnewline
Helicity ($tb_{fin}$, excl. 3-jet)&& 0.81&& 0.91&& 0.41&& -0.80&& 0.90&& -0.40 \tabularnewline
Optimal (incl. 2-jet)& -0.66&-0.67  & 0.83& 0.83 &-0.33 &-0.33  & 0.59& 0.56 &0.79 &0.78 &0.29 & 0.28 \tabularnewline
Optimal (excl. 3-jet)&& -0.69 && 0.85 && -0.35 && 0.58 && 0.79 && 0.29 \tabularnewline
\end{tabular}
\end{ruledtabular}
\par\end{centering}

\caption{Degree of polarization $\mathcal{D}$, polarization fraction
$\mathcal{F}$, and asymmetry $\mathcal{A}$ for inclusive two-jet and exclusive three-jet
s-channel single top quark events after jet clustering ($p_{T}^{j}\ge 30$ GeV, 
$\left|\eta_{j}\right|\le 5$,
$\Delta R_{\ell j} = \Delta R_{jj}\ge 0.4$). Results are shown for both top (left) and antitop (right)
quark measured in the helicity basis comparing the two different
c.m. frames, and in the optimal basis. In this table, $\mathcal{F}$
corresponds to $\mathcal{F}_{-}$ in the helicity basis for left-handed
top quarks and to $\mathcal{F}_{+}$ in the optimal basis for top quarks with polarization along the chosen proton three-momentum.
For right-handed antitop quarks, 
$\mathcal{F}$ corresponds to $\mathcal{F}_{+}$ in the helicity basis and to $\mathcal{F}_{-}$ in 
the optimal basis for antitop quarks with polarization against the chosen proton three-momentum.\label{tab:toppol3}}

\end{table*}

The spin fractions $\mathcal{F}_{\pm}$ and the asymmetry ${\mathcal{A}}$
of the distribution are defined in Ref.~\cite{Cao:2004ap}. 
Without imposing any kinematic cuts, $D=2{\mathcal{A}}$, which can
indeed be seen in Table~\ref{tab:toppol2}. Furthermore, the ratio
of top quarks with spin along the basis direction will be $r_{\uparrow}=0.5-{\mathcal{A}}$
when no cuts are applied. However, when cuts are imposed, the two
relationships break down.

Table~\ref{tab:toppol2} shows $\mathcal{D}$, $\mathcal{F}$ and
$\mathcal{A}$ at parton level before any cuts.
It can be seen that in the helicity basis the polarization is slightly larger for
 top than for antitop quarks, as top quarks receive larger longitudinal boosts from $W_{int}$ and therefore have higher energies (cf. Sec.~\ref{sub:LepMET}).
The comparison between LO and NLO shows a slight decrease in $\left|\mathcal{D}\right|$ in the helicity basis for both top and antitop quark.
In the optimal basis, the larger longitudinal boost also increases the energy of the top quark, but more importantly makes it more likely that the chosen proton is the correct pick for the reference axis.
As shown in Table~\ref{tab:toppol2},
both quark types are more polarized in the helicity basis than in the optimal basis, if no $p_{z}$ cuts on $W_{int}$ are applied. With higher $p_{z}$ cuts on $W_{int}$,
$\left|\mathcal{D}\right|$ calculated in the optimal basis increases though, as higher cuts improve the chance of guessing the correct direction
of the initial antiquark.

The results after reconstructing the jets ($p_{T}^{j}\ge 30$ GeV, 
$\left|\eta_{j}\right|\le 5$,
$\Delta R_{\ell j} = \Delta R_{jj}\ge 0.4$) are listed in Table~\ref{tab:toppol3} for inclusive two-jet events as well as exclusive three-jet events. In comparison to the results before any cuts, $\left|\mathcal{D}\right|$ is generally larger here, as the $p_T$ cuts on the jets increase the energy of the (anti)top quark. Similarly, in both bases the polarization at NLO is higher for exclusive three-jet events than for inclusive two-jet events, because of the additional $p_T$ that a third jet in the initial state adds to the event, which again increases the energy of the (anti)top quark. Furthermore,
the table shows that for the helicity basis the $tb_{fin}$-frame
leads to higher polarizations than the $tb_{fin}(j)$-frame at NLO. 
This is because in most of the three-jet events the third jet
comes from INIT corrections and should not be included in the c.m.
system of the final state objects produced from $W_{int}$.

\begin{figure*}
\subfigure[]{\includegraphics[scale=0.3]%
{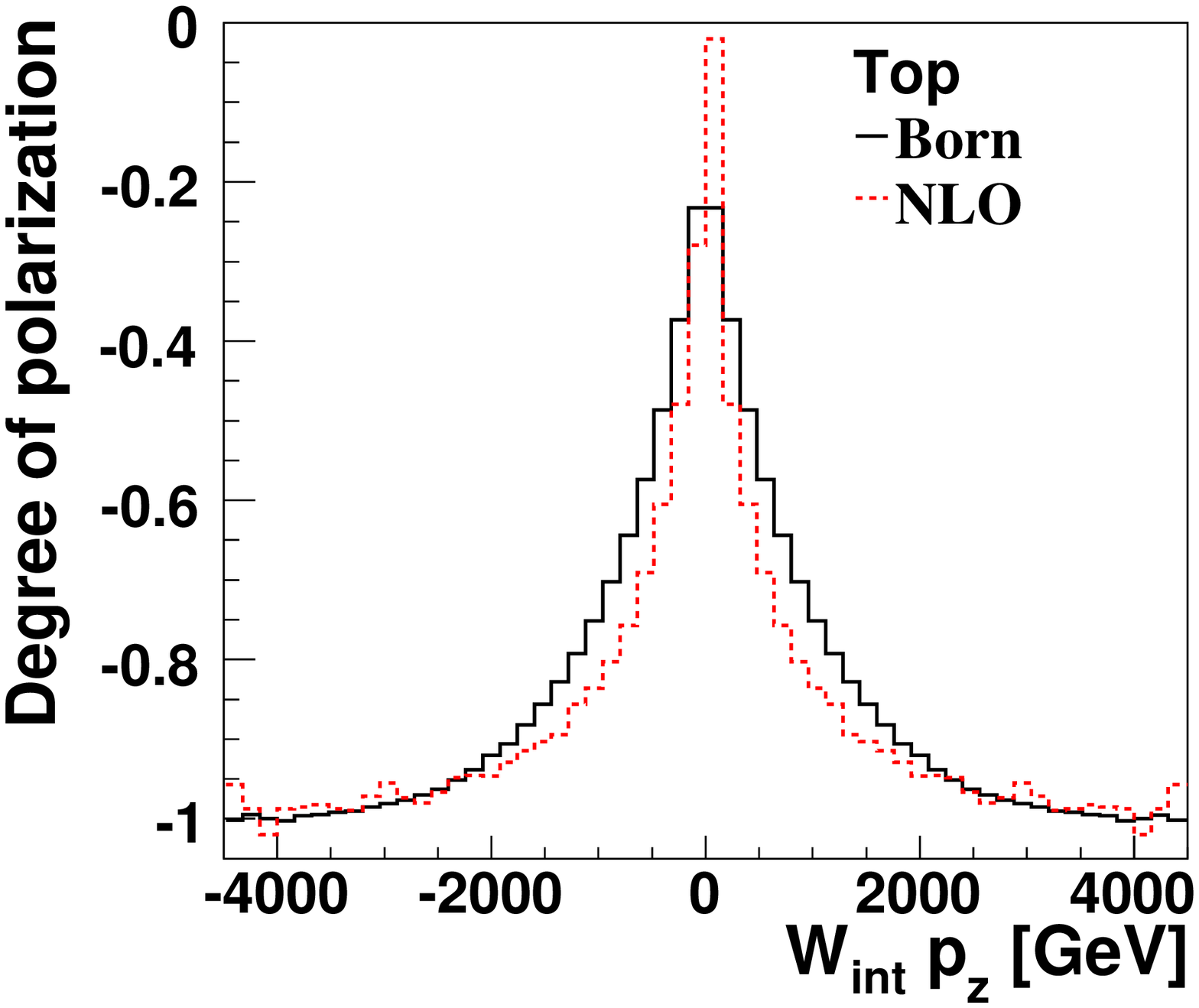}}%
\subfigure[]{\includegraphics[scale=0.3]%
{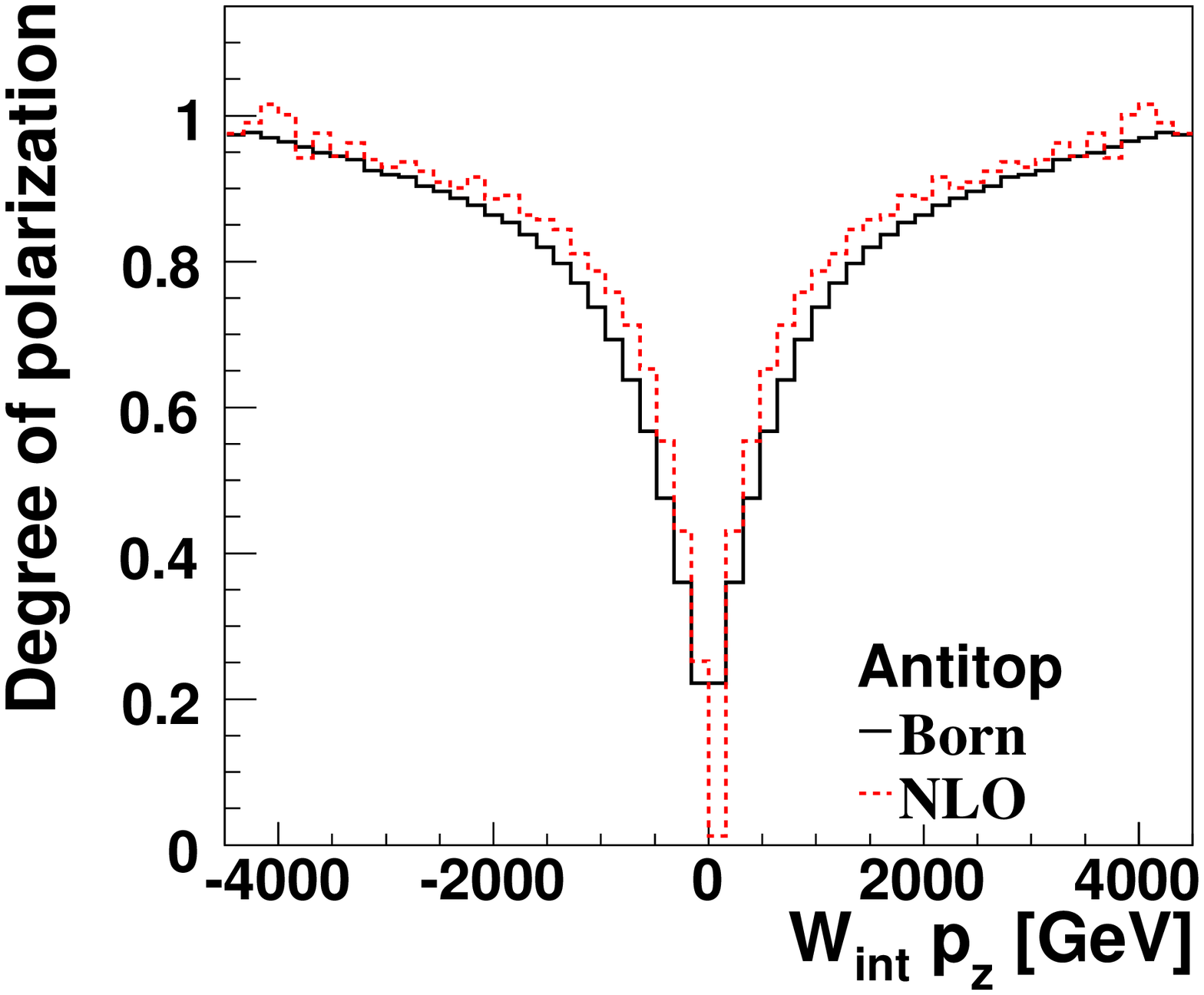}}
\caption{Degree of polarization $\mathcal{D}$ of top quark (a) and antitop quark (b) in the optimal basis,
for different values of $p_{z}(W_{int})$, 
at parton level before cuts. 
\label{fig:polWint}}
\end{figure*}

Figure~\ref{fig:polWint} shows the degree of polarization $\mathcal{D}$ of top and antitop quark
in the optimal basis, for different values of $p_{z}(W_{int})$
at parton level for top and antitop quark production. 

\subsection{Distributions for three-jet events}

Single top quark events at the LHC contain a large fraction of three-jet
events, which can be seen in Fig.~\ref{fig:jetPTEta}. It is therefore
of interest to discuss the kinematic properties of the third jet.
After selection cuts, this third jet corresponds to $O(\alpha_{s})$
corrections in about 80\% of the three-jet events. This means, for
most events, the emitted gluons/light quarks have a lower $p_{T}$ than the $b_{dec}$
and $b_{fin}$ jets, which is mainly due to the large amount
of initial state radiation that tends to be collinear to the beamline.

The emission of additional gluons/light quarks can be divided into production-stage
emission and decay-stage emission. Production-state emission includes
INIT and FINAL corrections and occurs before the top quark goes on-shell,
while decay-stage emission consists of the SDEC contribution and occurs
after the top quark goes on-shell. This classification of three-jet
events into production-stage or decay-stage is useful, but blurred
by the finite width of the top quark and, in experiments, by jet-energy
resolution and ambiguities in jet assignment.

\subsubsection{Kinematic distributions of the extra jet}

\begin{figure*}
\subfigure[]{\includegraphics[scale=0.3]%
{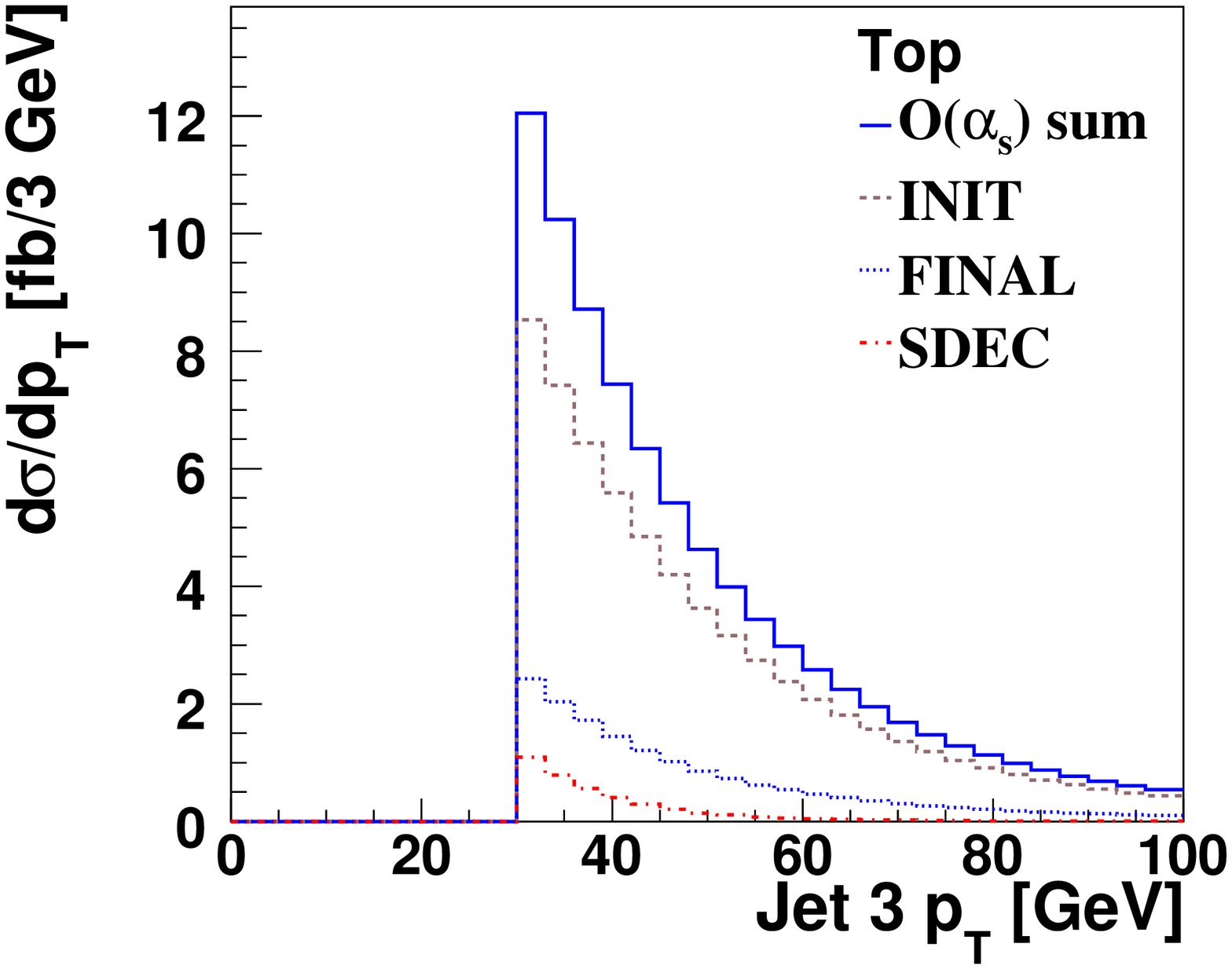}}%
\subfigure[]{\includegraphics[scale=0.3]%
{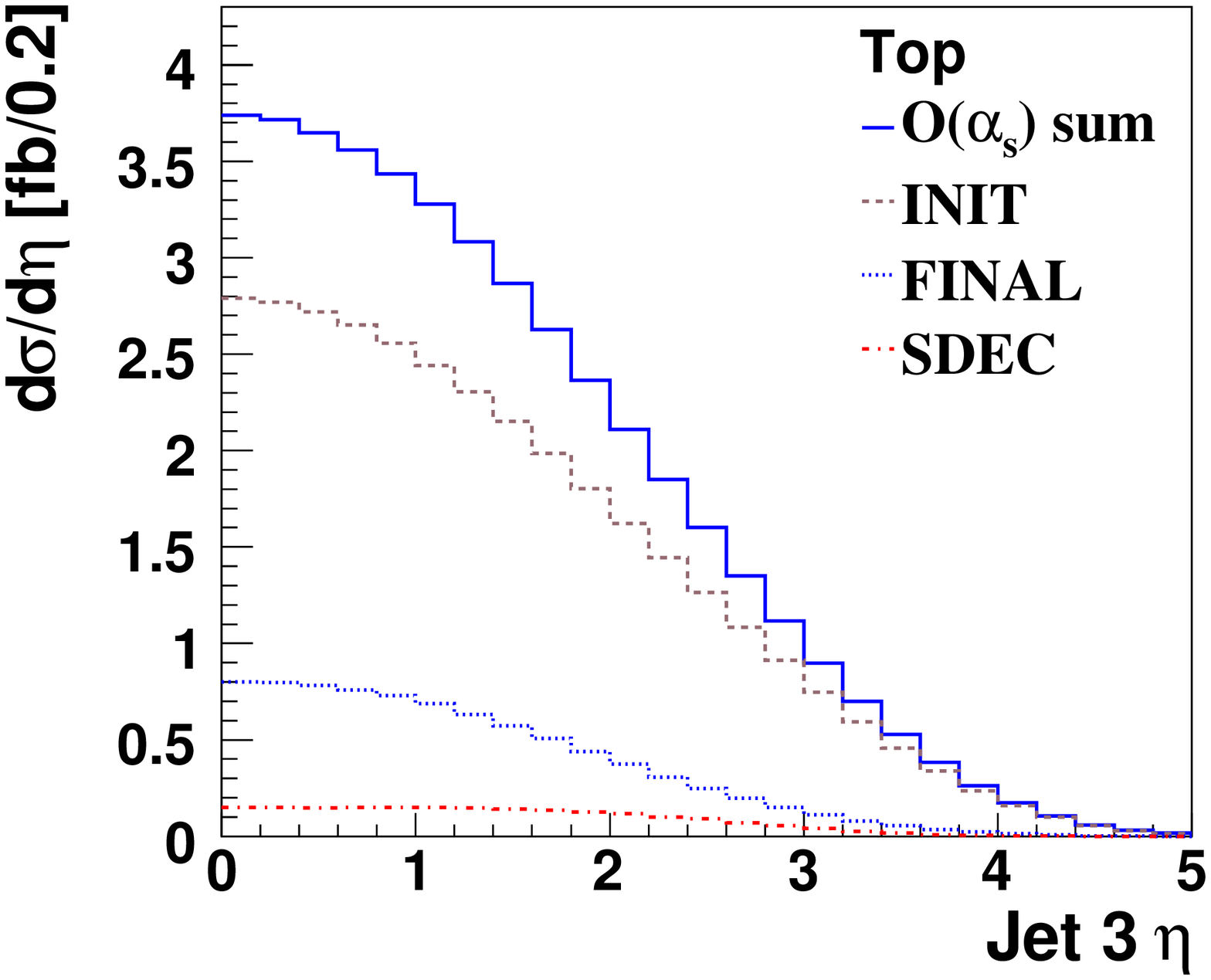}}

\caption{$p_{T}$ (a) and $\eta$ distribution (b) of the third jet after applying
the `loose' set of cuts for the different $O(\alpha_{s})$ corrections.\label{fig:jet3}}

\end{figure*}

The $p_{T}$ and $\eta$ distributions of the third jet for different
NLO corrections are presented in Fig.~\ref{fig:jet3}. The dominance
of the initial state corrections mentioned above is due to the collinear
enhancement of the incoming partons and can be seen in both distributions.
As it is determined by the collider energy, the INIT contribution
extends to far higher $p_{T}$ and larger $\eta$ than the other contributions.

\begin{figure*}
\subfigure[]{\includegraphics[scale=0.3]%
{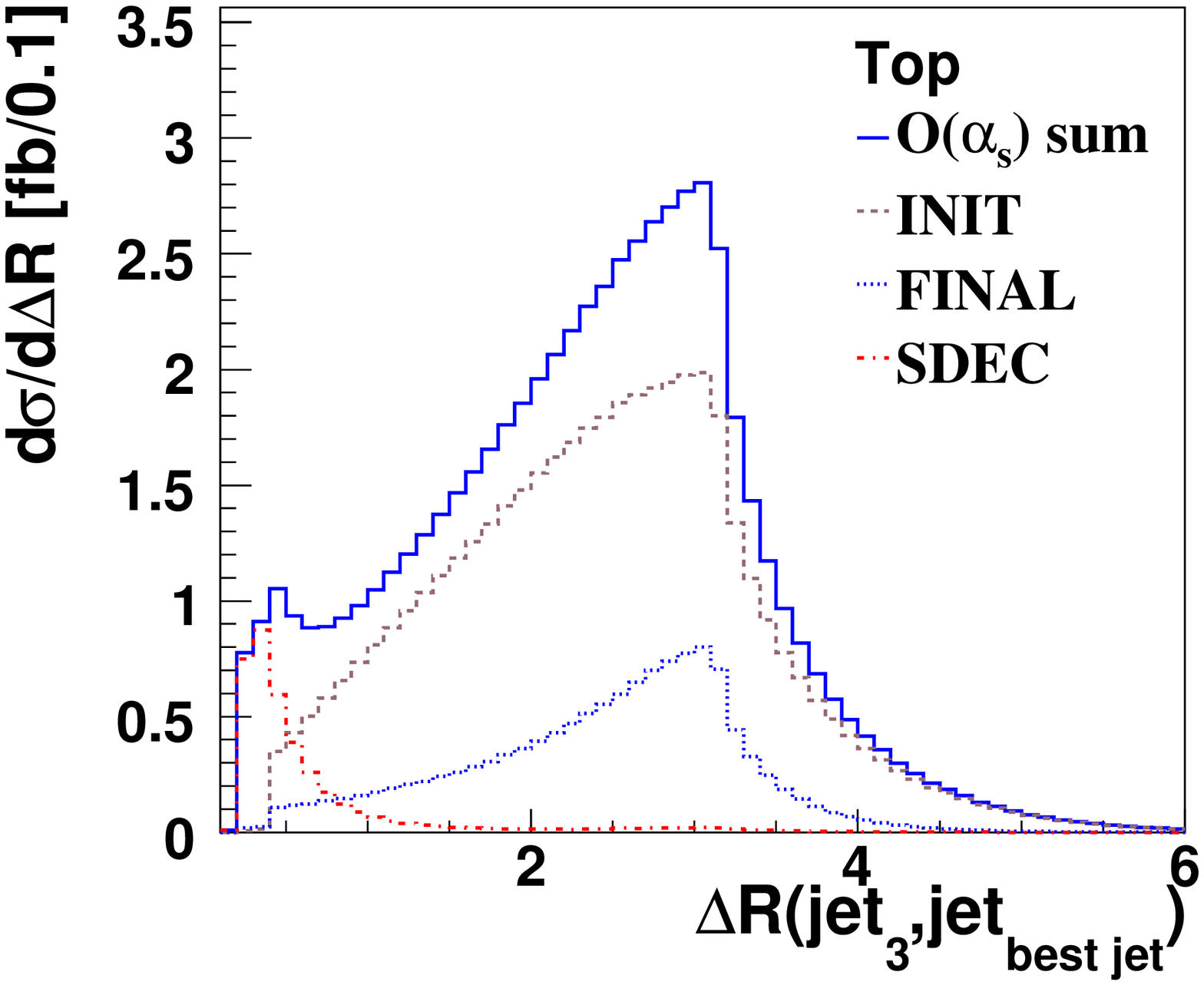}}%
\subfigure[]{\includegraphics[scale=0.3]%
{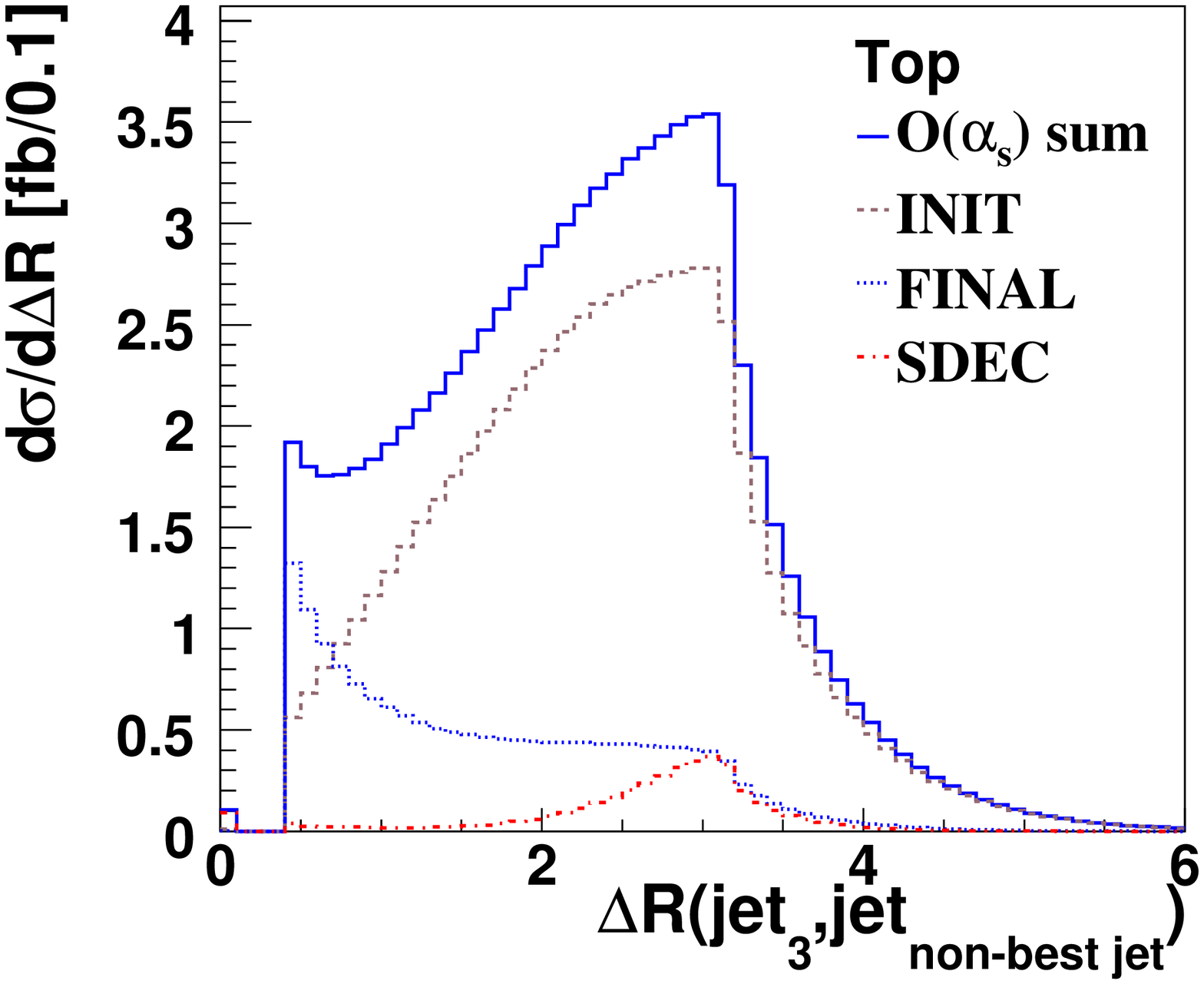}}

\caption{Difference in the $p_{T}$ between the third jet and the best-jet,
if the best-jet is not the third jet (a) and the third jet and the
non-best-jet (b), after applying the `loose' set of kinematic cuts.\label{fig:jet3dR}}

\end{figure*}

Figure~\ref{fig:jet3dR} shows that it is in principle possible to
identify the third jet as coming from the SDEC or the FINAL corrections,
by looking at the difference in the $p_{T}$ between the third jet
and the best-jet or the third jet and the non-best-jet respectively.
The $p_{T}$ difference between the third jet from the decay and the
best-jet has a peak close to zero for the decay contribution, while
the difference between the $p_{T}$ of the third jet from the FINAL
correction and the non-best-jet tends to be smaller than for the other
contributions. Unfortunately those peaks are not very high, and the
initial state correction is again dominant for slightly larger $p_{T}$
differences, so that an optimal cone size has to carefully balance
out the competing effects of desired jets falling outside of the cone
and unwanted initial state radiation being included. This is further complicated in experiments by hadronization and detector resolution
effects.

\subsubsection{Angular correlation between the extra jet and the best-jet}

\begin{figure*}
\begin{center}
\includegraphics[scale=0.3]%
{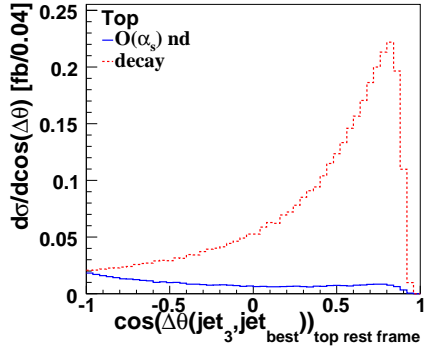}
\end{center}
\caption{Angular correlation between the third jet and the
best-jet after selection cuts. All NLO corrections except for
SDEC (solid), denoted as $O(\alpha_s)$nd, are compared to the SDEC contribution only (dotted).
Here, we only allow events in which exactly two jets are used to
form the best-jet.\label{fig:jet3cos}}

\end{figure*}

The best-jet algorithm can be used to distinguish production-stage
from decay-stage parton (gluon or quark) emission.  
In Fig.~\ref{fig:jet3cos}, we show the angular correlation 
between the third jet and the best-jet after event reconstruction.
Here, we include only events in which exactly two jets are used to form
the best-jet. 
There is a clear separation between the SDEC contribution only 
and the rest of the NLO corrections, 
even after event reconstruction with kinematic cuts imposed.

\section{Conclusions\label{sec:Conclusions}}

We have presented a study of s-channel single top and 
antitop quark production at the LHC based on the full NLO calculations.
We have studied the NLO QCD corrections to the production and decay
of the single top quark and shown their effect on the inclusive cross
section as well as on kinematical distributions, including top quark
polarization measurements. For this, we have divided the higher order
corrections into three Gauge invariant sets. The inclusion of NLO
corrections allows more precise predictions of the properties of single
top quark events, which is mandatory for using single top quark events
to test the SM and search for new physics.

The NLO corrections increase the s-channel inclusive cross section
significantly. Due to the PDF of the proton, the cross section is
larger for single top quark production than for single antitop quark
production and depends on the value of $m_t$. Simple kinematic cuts,
as they are used by the ATLAS and the CMS experiments, reduce the
acceptance considerably, while the percentage of three-jet events
remains high, especially due to collinear enhancement of the initial
state corrections.

The LO kinematical distributions of final and intermediate state objects
in single top quark events are dominated by spin correlations and 
the momentum difference between the two incoming partons. These effects are generally smaller
for antitop quark production than for top quark production and reduced
by $O(\alpha_{s})$ corrections in both cases. This is because 
the NLO corrections weaken spin correlations
and, in some cases, lower the momentum difference between the incoming
partons.

The total transverse energy $H_{T}$, the invariant mass of the ($b_{dec}$
jet, $b_{fin}$ jet) system, and the invariant mass of the ($b_{dec}$
jet, lepton) system
are examples of distributions that are characteristic of single top
quark events. They change significantly if NLO corrections are included.
The invariant mass distribution of the SM $W_{int}$
could be a useful discriminator for $W'$ boson searches at the LHC.

For the reconstruction of the top quark it is important to identify
the correct jet as the $b_{dec}$ jet, that is the jet which is produced
when the top quark decays. The most efficient method for this identification
is found to be the best-jet algorithm, which picks the $Wj$ or $Wjj$
combination that gives an invariant mass closest to the input $m_t$.

The spin correlations and the fact that the top quark decays before
it can hadronize makes it possible to measure its polarization. We
identify appropriate frames for top quark spin correlation
measurements and find that an additional cut on $p_z(W_{int})$ 
significantly increases the measured
spin correlations in the optimal basis.
\clearpage{}
\begin{acknowledgments}
S.~H. and R.~S. are supported in part by the U.S. National Science 
Foundation under Grant No. PHY-0757741. 
Q.~H.~C. is supported in part by the Argonne National Laboratory and
University of Chicago Joint Theory Institute (JTI) Grant No. 03921-07-137, 
and by the U.S. Department of Energy under Grants No. DE-AC02-06CH11357
and No. DE-FG02-90ER40560. C.~P.~Y. acknowledges the support of 
the U.S. National Science Foundation under Grant No. PHY-0555545 and PHY-0855561. 
C.~P.~Y. would also like to thank the hospitality of National Center
for Theoretical Sciences in Taiwan and Center for High Energy Physics, 
Peking University, in China, where part of this work was done.
\end{acknowledgments}
\bibliographystyle{apsrev}


\end{document}